\documentclass[11pt, a4paper]{article}

\RequirePackage{amsthm,amsmath,amsfonts,amssymb}
\usepackage[round]{natbib} 
\RequirePackage[colorlinks,citecolor=blue,urlcolor=blue]{hyperref}
\RequirePackage{graphicx}
\RequirePackage{bm, dsfont}
\usepackage{listings}
\usepackage[margin=1in,headheight=13.6pt]{geometry}
\usepackage{authblk}
\usepackage{caption}
\usepackage{dsfont}
\usepackage{hyperref}
\usepackage{booktabs}
\usepackage{rotating}
\usepackage{enumitem}
\usepackage{chngcntr}
\counterwithout{figure}{section}

\DeclareMathOperator{\E}{\mathbb{E}}

\DeclareMathOperator{\x}{\mathbf{x}}

\DeclareMathOperator{\g}{\mathbf{g}}
\DeclareMathOperator{\w}{\mathbf{w}}
\DeclareMathOperator{\thetaf}{\bm{\theta}}
\DeclareMathOperator{\betaf}{\bm{\beta}}

\DeclareMathOperator{\tf}{\mathbf{t}}
\DeclareMathOperator{\Vvec}{  \mathbf{v}  }

\DeclareMathOperator{\yf}{\mathbf{y}}

\DeclareMathOperator*{\Df}{\bm{\mathcal{D}}}
\DeclareMathOperator*{\D}{\mathcal{D}}
\newcommand\independent{\protect\mathpalette{\protect\independenT}{\perp}}
\def\independenT#1#2{\mathrel{\rlap{$#1#2$}\mkern2mu{#1#2}}}

\begin{document}
	
	\title{A Bayesian prevalence-incidence mixture model for screening outcomes with misclassification\thanks{\textbf{Acknowledgments:} We gratefully acknowledge the provision of the familial risk CRC EHR data used in the application section, which were provided by the Dutch Familial Colorectal Cancer Surveillance (FACTS) randomized controlled trial \citep{hennink_randomized_2015} and the Dutch Radboudumc Nijmegen and Rijnstate Arnhem hospitals. We specifically thank Tanya M. Bisseling, Marjolein Greuter, Simone D. Hennink, Nicoline Hoogerbrugge, Monique E. van Leerdam, and Marcel B. W. Spanier  for their contributions.}}
	\author[1]{Thomas Klausch (\href{t.klausch@amsterdamumc.nl}{t.klausch@amsterdamumc.nl})}
	\author[1]{Birgit I. Lissenberg-Witte}
	\author[1]{Veerle M.H. Coup\'e}
	\affil[1]{Amsterdam University Medical Center, Department of Epidemiology and Data Science, 
		Amsterdam, The Netherlands} 
	
	\maketitle
 
\begin{abstract}
Screening and surveillance programs for cancer, such as colorectal cancer (CRC), often yield electronic health records (EHR) of screening time, test results, and covariates. We consider EHR from CRC surveillance of individuals who have a high cancer risk due to their family history. These individuals, therefore, receive regular colonoscopies with the goal of finding and removing adenomas, precursor lesions to CRC. Our objective is to estimate time to adenoma incidence and explore associations with covariates. However, in doing so, several challenges of the CRC surveillance EHR have to be addressed. Importantly, the adenoma events are interval-censored, meaning the exact event times are unknown and only fall within intervals defined by colonoscopy visits. Furthermore, colonoscopies can miss adenomas due to human or technical error, leading to misclassification of individuals with adenomas as adenoma-free. Finally, the EHR data include individuals with adenomas at baseline, termed prevalent cases. This prevalence status may be unobserved if the baseline colonoscopy is missing or fails to detect existing adenomas. To address these challenges in the CRC EHR, and screening data in general, we develop a new  Prevalence-Incidence Mixture model (PIM) with a Bayesian estimation back-end through data augmentation and regularization priors. We show how to fit the model, estimate cumulative incidence functions, and evaluate model fit using information criteria as well as a non-parametric estimator. In extensive simulations, we show good performance of the model when informative priors on the test sensitivity are provided, which is usually possible. An implementation in the \texttt{R} package \texttt{BayesPIM} is provided. \\

\end{abstract}

\section{Introduction} \label{sec:introduction}
Screening and surveillance programs aim to detect diseases, such as cancer, at an early stage to improve treatment outcomes. Electronic health records (EHR) from these programs provide valuable insights into disease progression and can be used to optimize screening schedules, such as personalizing test frequency and timing. This study focuses on EHR from colorectal cancer (CRC) surveillance among individuals with elevated risk due to family history (Section \ref{sec:dataintro}). These individuals undergo regular colonoscopies to detect and remove adenomas, which are precursors to CRC. Since individuals in this high risk group may develop adenomas more frequently and rapidly than the general population, estimating adenoma incidence time is critical to optimize surveillance. To this end, we develop a modelling framework with an accompanying \texttt{R} package called \texttt{BayesPIM}. Our model is a type of so-called prevalence-incidence mixture model (PIM) with a Bayesian estimation back-end. The class of PIM was first suggested by \citet{cheung_mixture_2017} and \citet{hyun_flexible_2017}. The primary goal of their PIM was to model time to incidence when the disease status is ascertained at irregularly spaced discrete points in time (interval censoring) and a part of the population can have the (pre-state) disease already at baseline (i.e., the point in time when an individual is first included in the study), which is called prevalence.  This observation process was also present in the CRC EHR, because the CRC screening process led to interval censoring and individuals may have had adenomas already at inclusion into study; see Section \ref{sec:dataintro}. \\

Prevalence complicates the estimation of the interval-censored incidence model, in particular if the prevalence status is not observed (latent) for some or all individuals in the EHR. In the Cheung-Hyun PIM, the prevalence status is unobserved if no test is administered at baseline for at least a subset of the population and hence it is unknown for these individuals whether they have prevalent disease or not.   Ignoring the issue of latent prevalence in standard interval-censored survival models, such as the Accelerated Failure Time (AFT) model or the Cox model, causes underestimation of the time to incidence, because latent prevalent cases can be discovered after baseline and are then treated as incident cases. The Cheung-Hyun PIM model solves this issue through putting point probability mass on incidence at baseline (time zero), denoting effectively immediate transition for prevalent cases. This modelling decision is similar to so-called cure models that allow a proportion of the population to never progress, which is achieved by putting incidence probability mass on infinity (for a review of cure models, see \citet{amico_cure_2018}). The Cheung-Hyun PIM jointly estimates the incidence and the prevalence model by an expectation maximization (EM) algorithm, implemented in the \texttt{R} package \texttt{PIMixture} \citep{PIMixture} that is currently suggested as a principal modelling approach for screening data on the US National Cancer Institute (NCI) website \citep{NCIwebsite}. The \texttt{PIMixture} methodology has been applied in numerous epidemiological studies, such as estimating time to CIN2/3 lesions with prevalence at baseline using cervical cancer screening EHR  \citep{clarke_five-year_2019,saraiya_risk_2021}. The model has also been recently extended to take competing events into account \citep{hyun_sample-weighted_2020}. \\

Building on the PIM framework, our model \texttt{BayesPIM} extends the approach to account for imperfect test sensitivity (less than one), where sensitivity is the probability to find the (pre-state) disease when it is truly present. This enhancement is motivated by the CRC EHR, where colonoscopies can miss adenomas due to technical or human error at baseline or during follow-up (misclassification). The test sensitivity of a colonoscopy for a small to medium-sized adenoma varies depending on the study between approximately 0.65 and 0.92 \citep{van_rijn_polyp_2006}. As \texttt{PIMixture} and standard models for interval-censored survival data \citep{boruvka_cox-aalen_2015, anderson-bergman_icenreg_2017} have been developed under the assumption of perfect test sensitivity, their estimates are prone to bias. Importantly, with imperfect tests, latent prevalence due to misclassification at baseline may occur even when all subjects receive a baseline test, while \texttt{PIMixture} assumes latent prevalence can only occur in subjects that do not receive a baseline test. \texttt{BayesPIM} handles latent prevalence regardless of whether its cause is an omitted baseline test or misclassification and can co-estimate the test sensitivity. As we show, including prior information on the sensitivity is an advantage of the Bayesian approach, as it stabilizes estimation and incorporates prior uncertainty. While the model accommodates imperfect sensitivity, we assume perfect specificity, a reasonable assumption in CRC screening where initial findings are usually confirmed via pathology which corrects falsely positive tests. \\

\texttt{BayesPIM} allows for parametric survival distributions and uses model selection criteria to identify the best fit. In addition, we propose a joint non-parametric estimator of the cumulative incidence function (CIF) and prevalence which can be used to validate model fit visually. Our procedure is based on the  non-parametric CIF estimator  \texttt{em\_mixed} developed by \citet{witte_em_2017}, originally intended for interval-censored screening data with misclassification but without baseline prevalence. Using a recoding step similar to an approach described by \citet{cheung_mixture_2017} to adapt the \cite{turnbull_empirical_1976} non-parametric maximum likelihood (NPMLE) estimator to their PIM setting, we adapt \texttt{em\_mixed} to the PIM setting with misclassification and prevalence. \\ 

Misclassification and prevalence have also been discussed in the context of multi-state models. Specifically, continuous-time hidden Markov models (HMM) allow modelling transition processes between multiple (disease) states, while specifying a probability distribution relating the observed states to underlying true states which may be used to account for imperfect test sensitivity \citep{jackson_multistate_2003, jackson_multi-state_2011}.  Bayesian HMM have been suggested that can additionally deal with prevalence \citep{luo_bayesian_2021, raffa_multivariate_2015}. These models rely on the Markov assumption, meaning that the transition probability to a state does not depend on the time spent in the state which is akin to assuming an exponential transition time distribution (constant hazard function). We believe that this is too restrictive for many cancer screening settings, where hazards of (pre-state) disease events may change across time. Semi-Markov models have been suggested to account for the time-dependence of transition rates \citep{aastveit_new_2023, barone_bayesian_2022}. AFT models, applied in the present approach, are very similar to semi-Markov models in this regard; for example, a Weibull AFT specification allows for non-constant hazards of transition. Recently, \citet{klausch_bayesian_2023} used AFT models for three-state screening data, demonstrating the robustness of Bayesian estimation when using regularization priors, similar as in \texttt{BayesPIM} (Section \ref{sec:priors}). However, these models do not take prevalence and misclassification into account. \\

This article is structured as follows. Section \ref{sec:dataintro} introduces the Dutch CRC EHR as the motivating case of our methodology. Subsequently, sections \ref{sec:DGM} and \ref{sec:estimation} describe the data-generating mechanism and estimation methodology, respectively. Section \ref{sec:simulation_synth} evaluates \texttt{BayesPIM} through simulations including settings created using resampling of the real-world Dutch CRC data. Finally, Section \ref{sec:application} presents the application to the Dutch CRC EHR and Section \ref{sec:discussion} the discussion.

\section{Motivating case study: The Dutch CRC EHR} \label{sec:dataintro}
Our motivating case involves CRC surveillance through colonoscopy in individuals with an elevated CRC risk due to family history. Specifically, we analyze a merged dataset ($n = 810$) of EHR from two sources (Table \ref{tab:baseline_char}). The primary dataset comes from the Dutch Familial Colorectal Cancer Surveillance (FACTS) randomized controlled trial \citep{hennink_randomized_2015}, which experimentally compared two surveillance protocols in this high-risk group (three- vs. six-year intervals between colonoscopies). This was supplemented with observational EHR from a similar high-risk population undergoing CRC surveillance according to current guidelines. These additional data were collected through ongoing research at two Dutch hospitals, Radboud UMC Nijmegen and Rijnstate Arnhem. Eligible participants for this case study were those with at least one successful colonoscopy, with baseline defined as the time of the first colonoscopy. The primary objective was to estimate the time from baseline to the first adenoma. However, given the progressive nature of CRC, a cancer may occasionally be detected instead of an adenoma during surveillance. In this study, the number of cancers was negligible ($n = 8$, 1\%). Since we consider the time until the first adenoma, in this study, follow-up was censored at the first positive test if an adenoma was found. \\

\begin{table}[t]
	\centering
	\caption{Baseline characteristics of the Dutch CRC EHR screening data (time is measured in years).}
	\label{tab:baseline_char}
	\begin{tabular}{lrrr}
		\multicolumn{2}{c}{} & \multicolumn{2}{c}{Event in follow up} \\
		\cline{3-4}
		Statistic & Total & Adenoma & Right censored \\ \hline
		Sample size n & 810 & 330 & 480 \\ 
		Confirmed prevalent (column \%)& 20.4 & 50.0 & 0 \\
		Baseline test unsuccessful (column \%)& 6.5 & 6.7 & 6.5 \\
		Censoring time (min/median/max) & 1.2/6.0/30.1 & 1.2/6.3/22.1 & 1.4/6.0/30.1 \\
		No. of screenings (min/median/max) & 1/1.5/7 & 1/1.5/7 & 1/2/7 \\
		Interval length (min/median/max) & 1.0/3.1/20.3 & 1/3.3/13.7 & 1/3.1/20.3 \\
		Age at baseline (min/median/max) & 24.3/53.2/86.3 & 27.6/54.7/86.3 & 24.26/52.3/73.5 \\
		Gender is female (column  \%) & 55.8 & 53.0 & 57.8 \\
		\hline
	\end{tabular}
\end{table}

At baseline, 20.4\% ($n = 165$) of individuals were found to have an adenoma. Since any finding during colonoscopy is verified by pathology we assume that the combined test has perfect specificity (cf. Section \ref{sec:introduction}). Hence, 20.4\% of individuals have confirmed (i.e., known) prevalence (Table \ref{tab:baseline_char}). However, in the remaining 79.6\% ($n = 645$), the prevalence status was unknown (latent) due to two possible reasons. First, although every individual received a colonoscopy at baseline, in 6.5\%, the baseline colonoscopy was not completed successfully, due to, for example, incomplete visualization of the colon. These test outcomes are categorized as missing. In the remaining cases, while the baseline colonoscopy was successfully performed and returned a negative result, it may have missed an adenoma (misclassification). Thus, although the baseline non-completion rate was relatively low (6.5\%), substantial uncertainty regarding baseline prevalence remains due to the potential for missed adenomas. \\

There were six types of surveillance visit patterns present in the CRC EHR that are processed by \texttt{BayesPIM} (Table \ref{tab:visit_times_example}). Table \ref{tab:visit_times_example} gives an example data vector for all types using four screening moments (i.e., time points $v_1$ to $v_4$). For convenience, we always set $v_1=0$ denoting baseline; whether the result of a colonoscopy is available at baseline is indicated by $r_i$, with $r_i=1$ if yes and $r_i=0$ if not. The result may be missing because no colonoscopy was conducted at baseline (so called "no-show") or because the colonoscopy was unsuccessful (e.g. incomplete imaging of the colon). We address all resulting  pattern types in turn. First, incident cases in the follow-up can occur after a successful baseline colonoscopy (type 1, 18\%). The data example shows that incidence was found at $v_3=6$ years. The adenoma thus surely occurred before 6 years, and, more specifically, it occurred in the $(3.0,6.0]$ interval if it was not missed at the colonoscopy at $v_2=3$ years, or it occurred in $(0, 3.0]$ if it was not missed at the $v_1=0$ baseline colonoscopy. Whether the adenoma was missed at any test occasion is unknown. Second, right censoring (loss to follow-up) can occur after a successful baseline colonoscopy (type 2, 47\%). Here, we adopt the common notation that the last screening time is set to infinity in case of right censoring. The data example then suggests that right censoring  occurred after the last colonoscopy had been performed at $v_3=6.3$ years. Note that due to misclassification right censoring may have occurred even if an adenoma had been present before $v_3=6.3$ including the possibility of a latent prevalent adenoma at baseline. To the contrary, type 3 denotes prevalence found at baseline, as indicated by $v_1=0$ without further screening moments (20.4\%; cf. Table \ref{tab:baseline_char}). It is important to distinguish the prevalent case from type 4 that denotes a negative baseline test with loss to follow up before the first surveillance moment in the regular follow-up, as indicated by $(0,\infty)$ in the data example (8.9\%). Again, type 4 also might denote a prevalent case with falsely negative baseline test. Furthermore, note that case types 3 and 4 do not provide information on incidence, but they do inform the estimation of the prevalence model. The final two types 5 (2.7\%) and 6 (3.8\%), concern cases without baseline test ($r_i=0$) and an event in the follow-up (until $v_3=8.3$ years) or loss to follow-up (at $v_2=5.9$ years), respectively. 

\begin{table}[t]
	\centering
	\caption{Typology of screening cases in the Dutch CRC EHR with observed proportion per type (n=810; note: the data example uses up to four visits for illustration, but the real CRC EHR contained up to seven visits, see Table \ref{tab:baseline_char}).}
	\label{tab:visit_times_example}
	\begin{tabular}{llcccccc}
		\multicolumn{3}{c}{} & \multicolumn{5}{c}{Data example} \\
		\cline{4-8}
		Type & Description & \% observed & $v_1$ & $v_2$ & $v_3$ & $v_4$ & $r_i$ \\
		\hline
		1 & Incident in follow-up with baseline test & 17.7 & 0 & 3.0 & 6.0 & & 1 \\
		2 & Right censored with baseline test & 46.5 & 0 & 2.9 & 6.3 & $\infty$ & 1 \\
		3 & Observed prevalent at baseline test & 20.4 & 0 & & & & 1 \\
		4 & Right censored with only a baseline test & 8.9 & 0 & $\infty$ & & & 1 \\
		5 & Incident in follow-up without a baseline test & 2.7 & 0 & 6.2 & 8.3 & & 0 \\
		6 & Right censored without a baseline test & 3.8 & 0 & 5.9 & $\infty$ & & 0 \\
		\hline
	\end{tabular}
\end{table}

\section{\texttt{BayesPIM}: data generating mechanism} \label{sec:DGM}

We first introduce notation (Section \ref{sec:notation}) and give an overview of the hierarchical model structure of \texttt{BayesPIM} (Section \ref{sec:hierarchical}). Subsequently, we give details on the prevalence-incidence mixture model (Section \ref{sec:AFTmixturemodel}) and the prior assumptions (Section \ref{sec:priors}). 

\subsection{Notation} \label{sec:notation}

Every individual $i=1,\dots,n$ has an observed vector of ordered screening times $\Vvec_i= (v_{i1},v_{i2},\dots,v_{ic_i})$, with $v_{i1}<v_{i2}<\dots<v_{ic_i}$ of length $c_i$, where we set $v_{i1} = 0$ to denote baseline and $v_{ic_i}=\infty$ if the screening series is right censored, so that $v_{ic_i}<\infty$ only if the last screening time corresponds to a positive test (Table \ref{tab:visit_times_example}). The latent time of right censoring (loss to follow-up) is denoted $s_i$. Right censoring occurs when $v_{ij+1}$ would be greater than $s_i$; see Section \ref{sec:hierarchical}. The associated test outcomes at $\Vvec_i$ are $\yf_i=(y_{i1},y_{i2},\dots,y_{ic_i})$, where $y_{ij} = 1$ denotes a positive test and $y_{ij} = 0$ a negative test. Since we assume perfect test specificity (cf. Section \ref{sec:introduction}) and are interested in the the time until the first event/adenoma (cf. Section \ref{sec:dataintro}), a positive test at $v_{ic_i} < \infty$ implies that $\yf_i=(0,0,\dots,0,1)$ (i.e., a vector of length $c_i$ with zeros in all except the last position). For right censoring, we have that $\yf_i=(0,0,\dots,0,0)$ (i.e., a vector of length $c_i$ with zeros in all positions). Disease testing is administered at each scheduled screening time, including baseline when indicator $r_i=1$. If $r_i=0$, $y_{i1}$ is unavailable (missing) regardless of cause (e.g. missing test, unsuccessful test). Furthermore, every individual has covariates $\x_i$ of dimension $p \times 1$, where $\x_i$ is understood to have a unit entry in the first position (intercept). We write $\mathbf{X}$ for the $n \times p$ matrix of covariates. The observed data for individual $i$ are thus $\D_i = \{r_i, \Vvec_i,\x_i, \yf_i \}$ wrapped as set $\Df = \{\D_1,\dots,\D_n\}$. \\

The latent transition time $t_i$ (incidence) is interval-censored by $\Vvec_i$. The time $t_i$ is measured from baseline (i.e.\ time zero). Its units are implied by the unit of measurement of the screening times $\Vvec_i$ (e.g.\ days, months, or years). We model the time $t_i$ through an Accelerated Failure Time (AFT) model using $\x_i$ (Section \ref{sec:AFTmixturemodel}), parametrized by a coefficient vector $\betaf \in \mathbb{R}^{p}$ and a scale parameter $\sigma  \in \mathbb{R}^{+}$ (Section \ref{sec:AFTmixturemodel}). We write $\tf = (t_1,\dots,t_n)$ for the vector of all transition times.  Furthermore, we use latent variable $g_i$ to indicate the prevalence status at baseline, with $g_i = 1$ indicating prevalence and $g_i=0$ indicating non-prevalence. We model the prevalence status using covariates $\x_i$, through a probit model (Section \ref{sec:AFTmixturemodel}). The probit model is parametrized by coefficients $\thetaf \in \mathbb{R}^{p}$ and its linear term is denoted $\mu_i = \x_{i}^T \thetaf$. We write $\bm{g}= (g_1, \dots, g_n)$ for the vector of all prevalence status indicators. We note that while we generically refer to the covariates in the incidence and prevalence model as $\x_i$, \texttt{BayesPIM} allows including different, possibly overlapping subsets of covariates in the two models. For notational ease, we use the short-hand notation $\x_i$ denoting that the same covariates are entered in both models. Furthermore, when $g_i=1$ (prevalence) the time $t_i$ is defined as a counterfactual. This means that $t_i$ is the time when transition to disease would occur if $i$, counter to fact, were non-prevalent. This conceptualization enables joint Gibbs sampling of the latent (unobserved) $g_i$ and $t_i$; see Section \ref{sec:estimation}. \\

We use latent variable $y^*_{ij}$ to indicate disease presence, taking value 1 if the disease is present at occasion $j$ and 0 otherwise, where
\begin{align} \label{eq:def_y_star}
	y^*_{ij}(g_i, t_i,v_{ij})	=
	\begin{cases}
		\mathds{1}_{\{v_{ij} \ge t_i\}} \quad &\text{if} \ g_i=0 \quad \text{(non-prevalent)}\\
		1 \quad &\text{if} \ g_i=1 \quad \text{(prevalent)}.
	\end{cases}
\end{align}

Using (\ref{eq:def_y_star}), the test sensitivity is defined as $\kappa = \Pr(y_{ij} = 1 \mid y^*_{ij} = 1)$. We note that $g_i \equiv y^*_{i1}$ but for notational clarity, we keep $g_i$ as the prevalence mixture indicator that is modelled. A special case emerges when the baseline test is positive ($y_{i1}=1$). Due to the assumption of perfect test specificity (Section \ref{sec:introduction}), it is then known that $g_i= y^*_{i1}= 1$. However, $g_i$ remains latent when $y_{i1}=0$ (possibly falsely negative test) or when the baseline test is missing ($r_i=0$).\\

We denote probability density functions (PDF) and probability mass functions (PMF) with $f$; for example $f_t$ gives the PDF of $t_i$ and $f_g$ gives the PMF of the Bernoulli variable $g_i$. The cumulative distribution function (CDF) of $t_i$, also called the cumulative incidence function (CIF, defined as one minus the survival function), is denoted $F_t$. The CDF of the normal distribution is denoted $\Phi$. Furthermore, the prior PDFs are denoted $\pi$ with prior parameters $\tau_{\beta}$ (prior of $\betaf$), $\lambda$  (prior of $\sigma$), $\tau_{\theta}$ (prior of $\thetaf$), and $\bm{\alpha}$ (prior of $\kappa$); see Section \ref{sec:priors}.

\subsection{Hierarchical model structure and assumptions} \label{sec:hierarchical}
\begin{figure}[t]
	\centering
	\includegraphics[trim = 50 200 450 20, clip, scale = 0.5]{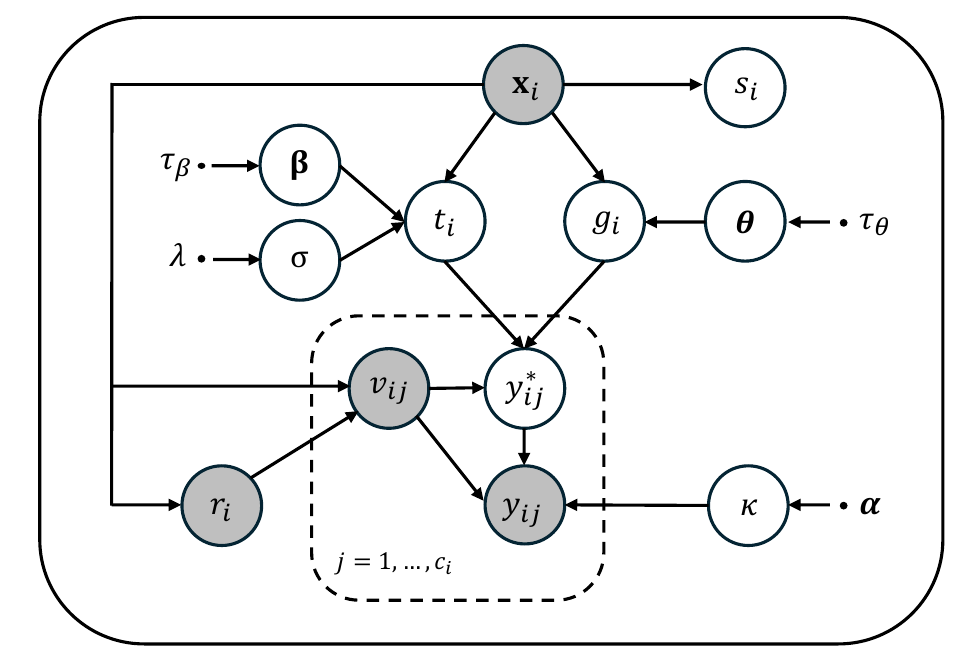}
	\caption{DAG illustrating the hierarchical model structure of \texttt{BayesPIM} using plate notation. White circles denote unobserved variables, grey circles denote observed variables, dots denote fixed parameters, and arrows denote the direction of dependence.  For clarity, dependence of $v_{ij}$ on history $\bar{\bm{v}}_{ij}$ is suppressed and $v_{i1}=0$ (baseline time) is implied. }
	\label{fig:dag}
\end{figure}

We assume the following hierarchical model structure illustrated by a Directed Acyclic Graph (DAG) in Figure \ref{fig:dag}. The parameters are generated from the prior (see Section~\ref{sec:priors})
\begin{align*}
	\betaf, \sigma, \thetaf, \kappa \mid \tau_{\beta}, \lambda, \tau_{\theta}, \bm{\alpha} \ &\sim \ \pi(\betaf, \sigma, \thetaf, \kappa \mid \tau_{\beta}, \lambda, \tau_{\theta}, \bm{\alpha}).
\end{align*}

For $i=1, \dots ,n$ and fixed vector $\x_i$,  the incidence time $t_i$, prevalence status $g_i$, baseline test decision $r_i$, and right censoring time $s_i$ are then generated as:
\begin{align}
	t_i \mid \x_i, \betaf, \sigma \ &\sim \ f_t(t_i \mid \x_i, \betaf, \sigma) \nonumber \\
	g_i \mid \x_i, \thetaf \ &\sim \ \text{Bernoulli}(\Phi(\mu_i)) \nonumber \\
	r_i \mid \x_i \ &\sim \ f_r(r_i \mid \x_i ) \nonumber \\
	s_i \mid \x_i \ &\sim \ f_s(s_i \mid \x_i ) \nonumber
\end{align}

 In addition, we allow for the boundary cases $r_i=0$ for all $i$ (nobody gets a baseline test)  and $r_i=1$ for all $i$ (everybody gets a baseline test), in which case $r_i$ is not random. The incidence time $(t_i)$ and prevalence $(g_i)$ models are discussed in more detail in Section~\ref{sec:AFTmixturemodel}.   \\
 
Subsequently, for each $i$, the screening times $\Vvec_i$ and test outcomes $\yf_i$ are generated recursively (i.e., the observation process). Specifically, at baseline, indicator $r_i$ decides if a baseline test is done. For $r_i=0$, $y_{i1}$ is missing (latent) and screening continues as described below. Else, the baseline test outcome is drawn (recall $g_i = y_{i1}^*$ the prevalence status) as
\begin{align*}
y_{i1}\mid y_{i1}^*,\kappa &\sim \text{Bernoulli}(\kappa \ y_{i1}^* ).
\end{align*}

Hence, the baseline test cannot be positive in case of non-prevalence ($\Pr(y_{i1} =1\mid y_{i1}^*=0,\kappa ) = 0$) and else has probability $\kappa$ to be positive. If the baseline test is positive, screening terminates and we set $\Vvec_i = (v_{i1}) = (0)$ with $\yf_i = (y_{i1}) = (1)$ and $c_i=1$ (cf. Table \ref{tab:visit_times_example}, Type 3). Else, the following steps are repeated for $j=2,3,\dots$ until an event (positive test) or right censoring; then we set $c_i=j$:
\begin{align}
	v_{ij}\mid r_i,\bar{\bm{v}}_{ij},\x_i &\sim f_v(v_{ij}\mid r_i, \bar{\bm{v}}_{ij},\x_i)  \label{eq:schedule_mechanism}\\
	y_{ij}\mid v_{ij}, y_{ij}^*,\kappa &\sim \text{Bernoulli}(\kappa\,y_{ij}^*  \mathds{1}_{\{v_{ij}<\infty\}}), \label{eq:obsprocess}
\end{align}
 with vector $\bar{\bm{v}}_{ij}=(v_{i1},\dots,v_{i,j-1})$ for $j \ge 2$ the screening times history before $v_{ij}$. Specifically, at any screening moment $j'\ge 2$ we draw a candidate screening time $v_{ij'}$ according to (\ref{eq:schedule_mechanism}). If the screening time $v_{ij'}$ would exceed the time of right censoring $s_i$ ($v_{ij'} > s_i$), we set $c_i = j'$ and $v_{ic_i}=\infty$ to indicate right censoring and we also set $y_{ic_i}=0$ with probability one; see (\ref{eq:obsprocess}). Else, we draw (\ref{eq:obsprocess}) and if the test is positive, $y_{ij'}=1$, we set $c_i=j'$, so that $v_{ic_i}<\infty$ is the last observed screening time. If $y_{ij'}=0$ (negative test), screening continues with drawing a new screening time according to (\ref{eq:schedule_mechanism}). \\
 
 For model estimation, it is not necessary to parametrize the distributions of $r_i$, $s_i$ and $v_{ij}$; see Section \ref{sec:estimation}. However, in Simulations 1 and 2 (Section \ref{sec:simulation_synth}) we give examples illustrating how all latent and observed model variables can be generated in simulation studies.\\
  
In summary, our model encodes the following key (conditional) independence assumptions ($\independent$ denoting independence):
\begin{enumerate}[label=(\alph*)]
	\item $t_i \independent g_i \mid \x_i$ (independence of prevalence-incidence components)
	\item $r_i \independent (g_i, t_i) \mid \x_i$ (missing at random baseline test outcomes)
	\item $v_{ij} \independent (g_i, t_i) \mid (r_i, \bar{\bm{v}}_{ij}, \x_i)$ (uninformative scheduling of screening)
	\item $s_i \independent (g_i, t_i) \mid \x_i$ (uninformative censoring)
	\item $y_{ij} \independent ( r_i, \bar{\bm{v}}_{ij}, \x_i, \bar{\bm{y}}_{ij}) \mid (v_{ij}, y_{ij}^*, \kappa)$ (stable test performance)
\end{enumerate}
where $\bar{\bm{y}}_{ij}= (y_{i1},\dots,y_{ij-1})$ is the history of screening outcomes. Assumptions (a) and (b) are similar to the missing at random (MAR) assumption given $\x_i$ \citep{little_statistical_2002} while assumption (c) denotes uninformative scheduling of screening tests and (d) uninformative right censoring. Assumption (e) implies that the screening test cannot be impacted by the timing or outcomes of past screenings and works equivalently for all individuals regardless of covariates $\x_i$. We return to these assumptions in the Discussion (Section \ref{sec:discussion}).

\subsection{Prevalence-incidence mixture model} \label{sec:AFTmixturemodel}
The latent incidence time variable $t_i$ follows an AFT model 
\begin{align} \label{eq::mod_x}
	\log t_i &= \x_i^T \betaf  + \sigma \epsilon_i.
\end{align}

The specific distribution chosen for the residuals $\epsilon_i$ induces the distribution $f_t(t_i \mid \x_i, \betaf, \sigma)$ through the change of variables defined by (\ref{eq::mod_x}). For example, if $\epsilon_i$ is standard extreme value, logistic, or normal distributed, time $t_i$ is, respectively, Weibull, log-logistical or log-normal distributed (conditionally on $\x_i$). These models allow time-dependency of hazards of an event and also are called semi-Markov models. An exponential (Markov-type) model with constant hazards over time is obtained by choosing a Weibull distribution and constraining $\sigma=1$.  \\

In addition, we model disease prevalence at baseline $(g_i)$ by a probit model where 
\begin{align} \label{eq::mod_g}
	\Pr(g_i = 1 \mid \x_i, \thetaf) = \Phi( \x_i^T \thetaf  ) = \Phi( \mu_i  ).
\end{align}

The probit model usually gives similar probability estimates as the equally common logistic model used by \citet{cheung_mixture_2017} and \citet{hyun_flexible_2017}, where the two approaches differ primarily on the computational end. Specifically, a probit model is chosen because it allows conjugate normal parameter updates during model estimation which is fast (Section \ref{sec:estimation}). For this, \texttt{BayesPIM} internally employs the latent variable formulation of the probit model, where a latent variable $w_i = \mu_i + \phi_i$, with $\phi_i \ \sim \ N(0,1)$ and $g_i = \mathds{1}_{\{w_i>0\}}$, so that $\Pr(g_i=1\mid  \x_i, \thetaf)=\Pr(w_i>0\mid \x_i, \thetaf)=\Phi(\mu_i)$. The model residuals $\epsilon_i$ and $\phi_i$ are independent which induces the assumption of "independence of prevalence-incidence components" conditional on $\x_i$; see Section \ref{sec:hierarchical}.    \\

The CIF of time $t$ conditional on non-prevalence at baseline and a new user-specified covariate vector $\tilde{\x}$ is given by $F_t(t\mid g=0,\tilde{\x}, \betaf, \sigma)$. This CIF is interpreted as the cumulative risk of disease for healthy individuals at baseline. Furthermore, to describe the CIF of the mixture of prevalent and non-prevalent cases we need to set the event time of prevalent cases to a value, where zero is the obvious choice, i.e.
\begin{align*} 
	t^* := (1-g) t.
\end{align*}

Subsequently, we estimate the mixture CIF
\begin{align*}
	F_{t^*}(t \mid \tilde{\x}, \betaf,\sigma,\thetaf,) = \Phi( \tilde{\x}^T \thetaf)
	+ \bigl(1-\Phi(\tilde{\x}^T \thetaf)\bigr)
	F_t(t \mid g = 0, \tilde{\x}, \betaf, \sigma).
\end{align*}

This CIF is similar to the CIF defined by \citet{cheung_mixture_2017} who also put point probability mass at zero for prevalent cases. Furthermore, when we marginalize $F_t$ and $F_{t^*}$ over the distribution of the covariates we obtain marginal (population-averaged) CIFs. Section \ref{sec:est_ppd} and the Supplemental Material (Section \ref{sec:supplement_sampling_ppdcif}) give further details on inference on these CIFs.

\subsection{Prior assumptions} \label{sec:priors}
The model parameters $(\betaf, \sigma, \thetaf, \kappa)$ are random variables with independent prior distributions
\begin{align*} 
	\pi(\betaf, \sigma, \thetaf,\kappa\mid \tau_{\beta}, \lambda,  \tau_{\theta}, \bm{\alpha}) = \bigg[ \prod_{j=1}^{p} \pi(\beta_{j} \mid \tau_{\beta}) \bigg] \bigg[ \prod_{k=1}^{p}  \pi(\theta_{k} \mid \tau_{\theta}) \bigg] \pi(\sigma\mid\lambda) \pi(\kappa\mid\bm{\alpha}) .
\end{align*}

Specifically, we choose zero-centred standard normal priors for the regression parameters, i.e. $\beta_{j} \sim N(0,\tau_{\beta})$ and $\theta_{j} \sim N(0,\tau_{\theta})$ with $\tau_{\beta}=\tau_{\theta}=1$. Furthermore, we specify a half-normal prior with variance $1$ for $\sigma$,
i.e. $\sigma \sim N^+(0,\lambda)$ with $\lambda=1$. These prior choices are called weakly informative, as they do not provide strong information on the location of the parameters, but do regularize parameter estimation. Regularization facilitates parameter estimation in sparse data settings that result from the interval and right censored observation process, the fact that in cancer screening there are typically few events, and latent prevalence \citep{klausch_bayesian_2023}. Finally, for the sensitivity parameter $\kappa$ we use a $\text{Beta}(\alpha_1,\alpha_2)$ prior, which can be chosen informatively. Alternatively, $\kappa$ can be fixed at a known value (a point prior). We evaluate the performance of uninformative and informative prior choices for $\kappa$ in simulations (Section \ref{sec:simulation_synth}). For an example, see the application (Section \ref{sec:application}).

\section{Model estimation} \label{sec:estimation}

In this section, we explain the estimation back-end of \texttt{BayesPIM}. We begin by deriving the observed data likelihood, followed by details on the Gibbs sampling estimation procedure, posterior distribution of CIFs, and   model fit evaluation.

\subsection{Observed-data likelihood}
As we derive in detail in the Supplemental Material (Section \ref{sec:supplement_proofs_obsll}), the observed-data likelihood $\mathcal{L}( \betaf, \sigma, \thetaf, \kappa  \mid  \Df )$ is proportional to
\begin{align}\label{eq::obs_data_ll}
	\prod_{i \in \mathcal{I}_0} &\bigg[ (1 - \Phi( \mu_i) )  \kappa^{y_{ic_i}} \sum_{j=1}^{c_i-1} (1-\kappa)^{(c_i-j-1)} \big[ F_t(v_{ij+1} \mid \x_i,\betaf,\sigma) - F_t(v_{ij} \mid \x_i,\betaf,\sigma) \big] \nonumber \\ 
	&\quad + \Phi( \mu_i)  \kappa^{y_{ic_i}}(1-\kappa)^{(c_i+r_i-2)} \bigg] \times  \prod_{i \in \mathcal{I}_1} \Phi( \mu_i)  \ \kappa,
\end{align}

where $\mathcal{I}_0$ denotes the set of all individuals $i$ with a negative baseline test ($r_i=1$ and $y_{i1}=0$) or a missing baseline test ($r_i=0$) and $\mathcal{I}_1$ denotes the set of all individuals $i$ with a positive baseline test ($r_i=1$ and $y_{i1}=1$). As described in Section \ref{sec:notation}, a positive baseline test implies that prevalence is known with certainty ($g_i=1$). Therefore, known prevalent cases ($\mathcal{I}_1$) inform the prevalence model through likelihood contribution $\Phi( \mu_i) \kappa$ only and are not part of the mixture structure that applies to negative or missing baseline tests ($\mathcal{I}_0$). This mixture form emerges, because the latent variables $g_i$ and $t_i$ are integrated out of the complete-data likelihood. \\

Here, we give a brief summary of the results from the Supplemental Material (Section \ref{sec:supplement_proofs_obsll}). In particular, the observed-data likelihood is given by the complete-data likelihood marginalized over latent $g_i$ and $t_i$ which, due to the factorization implied by the hierarchical model described in Section \ref{sec:hierarchical} and the DAG (Figure \ref{fig:dag}), is proportional to
\begin{align}\label{eq:LL_contributions}
	\prod_{i=1}^n \sum_{l = 0}^{1}  \Phi(\mu_i)^{l} (1-\Phi(\mu_i))^{(1-l)} \int_0^\infty \Pr(\yf_i \mid g_i = l, r_i, t_i, \Vvec_i, \kappa ) \ f_t(t_i\mid \x_i, \betaf, \sigma) dt_i. 
\end{align}

Our goal is to obtain a closed-form solution of the integral in (\ref{eq:LL_contributions}). We first note that the probability term $\Pr(\cdot)$ denotes the likelihood contribution of the test outcomes of individual $i$ conditional on the latent $g_i$ and $t_i$, given by
\begin{align} \label{eq:prob_bdelta}
	\Pr(\yf_i \mid g_i, r_i, t_i, \Vvec_i, \kappa ) 
	&=
	\begin{cases}
		0 \quad &\text{if}\ g_i = 0 \ \text{and}\ t_i > v_{ic_i} \\
		\kappa^{y_{ic_i}}(1-\kappa)^{m_i} \quad &\text{if}\ g_i = 0 \ \text{and}\ t_i \le v_{ic_i} \\
		\kappa^{y_{ic_i}}(1-\kappa)^{(c_i+r_i-2)} \quad &\text{if}\ g_i = 1,
	\end{cases}
\end{align}

with $m_i =  \sum_{j=1}^{c_i-1} y_{ij}^*(g_i, t_i, v_{ij})$ the number of falsely negative tests. We note that, if $g_i=0$, likelihood (\ref{eq:prob_bdelta}) does not depend on the presence of a baseline test ($r_i$), because the disease is not present at baseline and we assume the test specificity to be perfect (Section \ref{sec:introduction}). Also due to perfect specificity, the case $t_i > v_{ic_i}$ (if $g_i=0$) has a likelihood of zero whenever the last observed test is positive (i.e., $v_{ic_i}<\infty$) because then incidence must satisfy $t_i \le v_{ic_i}$. Under right censoring we have $v_{ic_i}=\infty$, so no such upper bound applies. Technically, this zero-probability acts as a constraint on the latent $t_i$ in the integral in (\ref{eq:LL_contributions}). In particular, when $v_{ic_i}<\infty$ the $l=0$ term is effectively integrated over $(0, v_{ic_i})$ rather than $(0, \infty)$.\\

To obtain a closed-form solution for (\ref{eq:LL_contributions}), we first note that (\ref{eq:prob_bdelta}) can also be written as a sum if $g_i=0$, i.e.
\begin{align} \label{eq::prob_bdelta_g0}
	\Pr(\yf_i \mid g_i=0, r_i, t_i, \Vvec_i, \kappa )  = \kappa^{y_{ic_i}} \sum_{j=1}^{c_i-1} (1-\kappa)^{(c_i-j-1)} \mathds{1} _{ \{v_{ij} < t_i \le v_{ij+1}\} }.
\end{align}

Here, $(c_i-j-1)$ is the number of falsely negative tests applying if $t_i$ lies in the interval $(v_{ij}, v_{ij+1}]$. Hence, (\ref{eq::prob_bdelta_g0}) shows a computationally effective way of determining $m_i$ in (\ref{eq:prob_bdelta}). Furthermore, substituting (\ref{eq::prob_bdelta_g0}) into (\ref{eq:LL_contributions}) allows solving the integral and yields the final likelihood expression (\ref{eq::obs_data_ll}); for details see the Supplemental Material. The likelihood of the test outcomes (\ref{eq:prob_bdelta}) and its sum representation (\ref{eq::prob_bdelta_g0}) also play a central role in the derivation of the full conditional distributions used by the Gibbs sampler, discussed next.

\subsection{Gibbs sampler} \label{sec:Gibbs}

We are now interested in posterior inference on $(\betaf, \sigma, \thetaf, \kappa)$ based on the posterior
\begin{align*}
	f_{\betaf, \sigma, \thetaf, \kappa}(\betaf, \sigma, \thetaf, \kappa \mid \Df ) \propto \mathcal{L}(\betaf, \sigma, \thetaf, \kappa \mid \Df ) \ \pi(\betaf, \sigma, \thetaf, \kappa \mid \tau_{\beta}, \tau_{\theta}, \lambda,  \bm{\alpha}).
\end{align*}

However, due to the mixture structure of the likelihood (\ref{eq::obs_data_ll}), direct sampling from the posterior is not feasible in closed form. Therefore, \texttt{BayesPIM} employs a Metropolis-within-Gibbs sampler with data augmentation. Equations (\ref{eq::fc.g})--(\ref{eq::fc.kappa}) give the full conditional sampling steps for the random model variables and the parameters that are run repeatedly in sequence for $k=1,\dots,K$ updates until convergence (see sections \ref{sec:simulation_synth} and \ref{sec:application} for details on the convergence criteria we applied). The conditional independences implied by the hierarchical model (Section \ref{sec:hierarchical}) are already taken into account, and detailed derivations are given in the Supplemental Material (Sections \ref{sec:supplement_proofs_fcx} to \ref{sec:appendix_par_g}). Our sampler is implemented in the \texttt{R} package \texttt{BayesPIM} which is available from the Supplemental Material, CRAN, and GitHub (\url{https://github.com/thomasklausch2/BayesPIM}).\\

After suitable initialization of the parameters, $g_i$, and $t_i$ (for $i=1,\dots,n$), the Gibbs sampler first performs data augmentation \citep{albert_bayesian_1993} of the latent variables:
\begin{align}
	g_i^{(k+1)} &\sim f_g \big(g_i \mid {\D}_i,\ \betaf^{(k)},\ \sigma^{(k)},\ \thetaf^{(k)}, \ \kappa^{(k)} \big), \quad &\text{if} \ r_i=0 \ \text{or} \ (r_i=1 \ \text{and} \ y_{i1}=0)   \label{eq::fc.g} \\
	t_i^{(k+1)} &\sim f_t \big(t_i \mid {\D}_i,\ g_i^{(k+1)},\ \betaf^{(k)},\ \sigma^{(k)}, \kappa^{(k)} \big), \quad &\text{for all } i \label{eq::fc.x}
\end{align}

These data augmentation steps are discussed in detail in sections \ref{sec:augment_x} and \ref{sec:augment_g}. Latent prevalence indicator $g_i$ is only augmented when it is missing, that is, without baseline test ($r_i=0$) or a negative baseline test ($r_i=1$ and $y_{i1}= 0$). In case of a positive baseline test ($r_i=1$ and $y_{i1}= 1$), $g_i=1$ is known due to the assumption of perfect specificity (see Sections \ref{sec:introduction} and \ref{sec:notation}). Here, we note already that step (\ref{eq::fc.g}) marginalizes over $t_i$, which is also known as collapsing \citep{liu_collapsed_1994}; see Section \ref{sec:augment_g} for a motivation. \\

The subsequent draws of the model parameters in the Gibbs sampler are performed conditional on the augmented (completed) data, 
\begin{align}
	(\betaf,\sigma)^{(k+1)} &\sim f_{\betaf,\sigma} \big(\betaf,\ \sigma \mid \tf^{(k+1)}, \mathbf{X} \big) \label{eq::fc.parx}  \\
	\thetaf^{(k+1)} &\sim f_{\thetaf} \big(\thetaf \mid \g^{(k+1)}, \mathbf{X} \big)	\label{eq::fc.parw} \\
	\kappa^{(k+1)} &\sim f_\kappa \big(\kappa \mid \Df,\ \g^{(k+1)},\ \tf^{(k+1)} \big) \label{eq::fc.kappa}.
\end{align}

Specifically, conditional on the augmented transition times $\tf^{(k+1)}$, the parameters $\betaf, \sigma$ in (\ref{eq::fc.parx}) are sampled from the complete-data posterior distribution, which is proportional to $\mathcal{L}(\betaf,\sigma \mid \tf^{(k+1)}, \mathbf{X}) \ \pi(\betaf,\sigma \mid \tau_{\beta}, \lambda)$, where $\mathcal{L}(\betaf,\sigma \mid \tf^{(k+1)}, \mathbf{X})$ is the complete-data likelihood. For drawing samples we employ a Metropolis step with normal proposal (jumping) distribution centred at the previous parameter draw $(\betaf^{(k)},\sigma^{(k)})$ and user-specified proposal variance. An example for Weibull distributed $t_i$ is given in the Supplemental Material (Section \ref{sec:appendix_par_t}). \\

Furthermore, conditional on the augmented prevalence status $\g^{(k+1)}$, the parameters $\thetaf$ in (\ref{eq::fc.parw}) are also drawn from a complete-data posterior. For this step, \texttt{BayesPIM} internally exploits the latent variable formulation of the probit model  (see Section \ref{sec:AFTmixturemodel}). Specifically, after an additional data augmentation step of the latent probit variable $w_i$ for all $i$, parameters $\thetaf$ can be sampled from a conjugate normal distribution \citep{albert_bayesian_1993}; details are given in the Supplemental Material (Section \ref{sec:appendix_par_g}). Updating the sensitivity parameter $\kappa$ by (\ref{eq::fc.kappa}) is optional; see Section \ref{sec:fc_kappa}.

\subsubsection{Augmenting the transition time} \label{sec:augment_x}
We first consider data augmentation of $t_i$ in step (\ref{eq::fc.x}). The full conditional distribution of $t_i$ in the non-prevalent case ($g_i = 0$) is a finite mixture of non-overlapping truncated distributions of $t_i$,
\begin{align} \label{eq::fc.x.explicit}
	f_t(t_i  \mid  {\D}_i, g_i = 0, \betaf, \sigma, \kappa)= \sum_{j=1}^{c_i-1} \omega_{ij} \  f_t(t_i \mid  v_{ij} < t_i \le v_{ij+1}, \x_i, \betaf, \sigma),
\end{align}

where
\begin{align*}
	\omega_{ij} = \frac{\tilde{\omega}_{ij}}{\sum_{l=1}^{c_i-1}\tilde{\omega}_{il}}.
\end{align*}

and
\begin{align}\label{eq::omega_tilde_weights}
	\tilde{\omega}_{ij} = \kappa^{y_{ic_i}}(1-\kappa)^{(c_i-j-1)} \big[ F_t(v_{ij+1} \mid \x_i, \betaf, \sigma) - F_t(v_{ij} \mid \x_i, \betaf, \sigma) \big] ,
\end{align}

see the Supplemental Material (Section \ref{sec:supplement_proofs_fcx}) for a proof. Specifically, factor $\tilde{\omega}_{ij}$ can be recognized as the weights of an unnormalized mixture distribution of $t_i$, while $\omega_{ij}$ are the normalized mixture weights yielding a normalized mixture distribution for $t_i$. It is interesting to observe that the mixture components of (\ref{eq::fc.x.explicit}) are non-overlapping, because the truncation bounds are equal to the screening times. With perfect test sensitivity, $\kappa=1$, the augmentation simplifies to sampling from the truncated distribution $f_t(t_i  \mid  v_{ic_i-1} < t_i \le v_{ic_i}, \x_i, \betaf, \sigma)$. This means that under perfect sensitivity, $t_i$ is known to lie in the most recent interval, $(v_{ic_i-1}, v_{ic_i}]$ (in case of an event) or $(v_{ic_i-1}, \infty)$ (in case of right censoring). For $\kappa<1$, the relative importance of a mixture component is determined by two main factors of (\ref{eq::omega_tilde_weights}). First, it decreases geometrically in the number of tests conducted since the most recent time $v_{ic_i}$, and, second, it increases with the probability mass of $t_i$ in the truncation interval. In general, more recent intervals are thus given higher weight, which is plausible. \\

Sampling from (\ref{eq::fc.x.explicit}) is straightforward when viewed as a two-stage sampling procedure for mixture distributions. First, sample a mixture component $J$ from $J \sim \text{categorical}(\omega_{i1},\dots,\omega_{i,c_i-1})$, where categorical denotes a draw of a class label $J$ from the set $\{1,\dots, c_i-1\}$ with probabilities $\omega_{i1},\dots,\omega_{i,c_i-1}$. Second, use the associated interval bounds of the draw $J$ as truncation bounds for $t_i$, i.e.\ sample $t_i \sim f_t(t_i \mid v_{iJ} < t_i \le v_{iJ+1}, \x_i, \betaf, \sigma)$.\\

In the prevalent case ($g_i = 1$), the observed screening series contains no information on $t_i$, and hence  
\begin{align} \label{eq::fc.x.explicit.g1}
	f_t(t_i  \mid  {\D}_i, g_i = 1, \betaf, \sigma) = f_t(t_i  \mid  \x_i, \betaf, \sigma),
\end{align}

so that $t_i$ is updated uninformatively.

\subsubsection{Augmenting the prevalence status } \label{sec:augment_g}

As we show in the Supplemental Material (Section \ref{sec:supplement_proofs_fcg}), augmenting the unknown prevalence status $g_i$ from the full conditional distribution requires Bernoulli sampling with probability
\begin{align}\label{eq::fc.g.noncollapsed}
	&f_g(g_i = 1 \mid  {\D}_i, t_i, \thetaf, \kappa) = 
	\begin{cases}
		\frac{ \Phi(\mu_i) (1-\kappa)^{(c_i+r_i-2)} }{ \Phi(\mu_i) (1-\kappa)^{(c_i+r_i-2)}  + (1- \Phi(\mu_i) )  \kappa^{y_{ic_i}}  (1-\kappa)^{m_i} } \quad &\text{if} \  t_i \le v_{ic_i} \\
		1 \quad &\text{if} \ t_i > v_{ic_i}.
	\end{cases}
\end{align}

However, deterministic updating of $g_i$ with probability one if  $t_i > v_{ic_i}$ in (\ref{eq::fc.g.noncollapsed}) can cause a dependency across the Gibbs sampling steps. Suppose we have augmentation $g^{(k)}_i=1$ at any point in the Gibbs sampler. Then $t_i$ is updated uninformatively in $(0,\infty)$ so that $t_i^{(k+1)} > v_{ic_i}$ can occur; see (\ref{eq::fc.x.explicit.g1}). In that case, $g_i^{(k+1)}=1$ with probability one again. This dependency continues unless in a future draw $t_i^{(k')} \le v_{ic_i}$. To avoid this dependency we marginalize (\ref{eq::fc.g.noncollapsed}) over $t_i$, which is also known as collapsing \citep{liu_collapsed_1994}. A collapsed Gibbs sampler preserves the convergence properties of a full-conditional Gibbs sampler but often reduces Markov Chain Monte Carlo (MCMC) autocorrelation when data augmentation is used. As we show in the Supplemental Material (Section \ref{sec:supplement_proofs_fcg}), collapsing $t_i$ yields fully stochastic Bernoulli updates with probability
\begin{align}\label{eq::fc.g.collapsed}
	f_g(g_i = 1  \mid  {\D}_i, \betaf, \sigma, \thetaf, \kappa) = \frac{ \Phi(\mu_i) \kappa^{y_{ic_i}} (1-\kappa)^{(c_i+r_i-2)}}
	{ \Phi(\mu_i) \kappa^{y_{ic_i}} (1-\kappa)^{(c_i+r_i-2)}+ (1- \Phi(\mu_i)) \sum_{l=1}^{c_i-1} \tilde{\omega}_{il} },
\end{align}

where $\tilde{\omega}_{ij}$ is defined in (\ref{eq::omega_tilde_weights}). When testing the Gibbs sampler during development, we found that using (\ref{eq::fc.g.noncollapsed}) instead of (\ref{eq::fc.g.collapsed}) indeed caused apparent bias. All results in sections \ref{sec:simulation_synth} and \ref{sec:application} are, therefore, based on (\ref{eq::fc.g.collapsed}). 

\subsubsection{Updating the test sensitivity parameter} \label{sec:fc_kappa}
Updating the test sensitivity parameter  $\kappa$ is an optional feature of \texttt{BayesPIM}. A common alternative is fixing $\kappa$ at a known value which can be viewed as a point prior at that value. In the simulations (Section \ref{sec:simulation_synth}), we evaluate estimation performance for different prior choices and data settings. For updating, $\kappa$ can be viewed as the parameter of the test outcome likelihood (\ref{eq:prob_bdelta}) and, with a Beta prior on $\kappa$ (see Section \ref{sec:priors}), the full conditional distribution is conjugate Beta,
\begin{align*}
	f_\kappa \big(\kappa \mid \Df, \g^{(k+1)}, \tf^{(k+1)} \big) &\propto
	\bigg[\prod_{i=1}^n \Pr(\yf_i \mid g_i, r_i, t_i, \Vvec_i, \kappa ) \bigg] \pi(\kappa \mid \bm{\alpha})\\
	&\propto
	\bigg[\prod_{i:\ g_i=0} \kappa^{y_{ic_i}} (1-\kappa)^{m_i}\bigg] \bigg[ \prod_{i:\ g_i=1} \kappa^{y_{ic_i}} (1-\kappa)^{(c_i+r_i-2)}\bigg] \mathrm{Beta}(\kappa \mid \bm{\alpha})  \nonumber \\
	&\propto \mathrm{Beta} \big(\kappa  \mid Y + \alpha_1, M + C- 2 G + \alpha_2 \big),
\end{align*}

where $Y=\sum_{i=1}^n y_{ic_i}$, $M=\sum_{i:g_i=0} m_i$, $C = \sum_{i:g_i=1} (c_i+r_i)$, $G = \sum_{i=1}^{n} g_i$. We note that due to the updating order in the Gibbs sampler where $g_i$ is updated first via (\ref{eq::fc.g}) and $t_i$ is subsequently drawn from its full conditional, the test outcome likelihood (\ref{eq:prob_bdelta}) is strictly positive at all sampled states. Hence, the full conditional distribution for $\kappa$ is always well-defined. 

\subsection{Posterior predictive cumulative incidence functions} \label{sec:est_ppd}
As described in Section \ref{sec:AFTmixturemodel}, we draw inference on the conditional CIF $F_t(t \mid g=0, \tilde{\x}, \betaf, \sigma)$ and the mixture CIF $F_{t^*}(t \mid\tilde{\x}, \betaf, \sigma, \thetaf)$ for fixed user-specified covariate values $\tilde{\x}$. In addition, we marginalize $F_t$ and $F_{t^*}$ over the empirical covariate distribution to obtain the marginal (i.e., population-averaged) CIF $F_t(t \mid g=0, \betaf,\sigma)$ and the marginal mixture CIF $F_{t^*}(t \mid \betaf, \sigma, \thetaf)$. For inference, we consider the CIF as functionals of the posterior distribution of the parameters. The posterior (mean) predictive conditional and marginal CIFs are then given by the posterior expectation of $F_t$ and $F_{t^*}$. For estimation and inference, we make use of the push-forward transform of samples $\thetaf^{(k)}, \betaf^{(k)}, \sigma^{(k)}$ via the Gibbs sampler into $F_t$ (or $F_{t^*}$), which yields samples from the posterior predictive CIF for any $t$. The expectation is obtained by the empirical average of the transformed samples. Furthermore, pointwise 95\% credible (posterior) intervals  are obtained by the 2.5\% and 97.5\% quantiles of the samples. For further details, see the Supplemental Material (Section \ref{sec:supplement_sampling_ppdcif}). 

\subsection{Model fit and non-parametric estimator of the marginal CIF} \label{sec:npest}
To choose the best model specification in terms of the distribution of $t_i$ we use the widely applicable information criterion (WAIC) given by $-2 \big[ 2\E_{\betaf, \sigma, \thetaf, \kappa \mid \Df} \big\{ \log \mathcal{L}( \betaf, \sigma, \thetaf, \kappa \mid \Df ) \big\} - \log \E_{\betaf, \sigma, \thetaf, \kappa \mid \Df} \big\{ \mathcal{L}( \betaf, \sigma, \thetaf, \kappa \mid \Df ) \big\}  \big]$, where $\mathcal{L}( \betaf, \sigma, \thetaf, \kappa \mid \Df )$ is given by (\ref{eq::obs_data_ll}) and the expectations are estimated by Monte Carlo integration over the posterior samples of $(\betaf, \sigma, \thetaf, \kappa)$ that are generated by the Gibbs sampler \citep{gelman_understanding_2014}. \\

While the WAIC offers a relative measure of fit, we additionally suggest a non-parametric estimator of the marginal mixture CIF $F_{t^{*}}( t \mid \betaf, \sigma, \thetaf )$. This estimate can then be compared against the parametric \texttt{BayesPIM} estimate as a visual inspection of absolute goodness of fit. Specifically, \citet{witte_em_2017} introduced a non-parametric estimator for $F_t(t \mid g_i=0, \betaf,\sigma)$ for settings without prevalence but interval censoring and misclassification from imperfect screening tests, implemented in function \texttt{em\_mixed}. To include baseline prevalence in  \texttt{em\_mixed}, our strategy is to apply a recoding step, similar to a preprocessing step suggested by \citet{cheung_mixture_2017} to enable use of the Turnbull estimator \citep{turnbull_empirical_1976} for estimating $ F_{t^{*}}$ from interval-censored data when the screening test sensitivity is one.\\

Concretely, we recode $\Vvec_i$ as follows. If $r_i = 1$, then $v_{i1} = 0$ is replaced by
$\min \{ v_{12},\dots,v_{n2} \} \times 0.01$ for all $i$. If $r_i = 0$, we omit $v_{i1}=0$, so that $v_{i2}$ becomes the first screening time. Intuitively, this recoding treats the baseline test as if it occurred immediately after the actual baseline point. Consequently, if $r_i = 0$ for a large proportion of $i$, the estimator may perform poorly because the anchoring to $\min[v_{12},\dots,v_{n2}] \times 0.01$ is not possible. We further investigate the performance of this estimator in the simulation section. We provide the adapted \texttt{em\_mixed} estimator in \texttt{R} package \texttt{EMmixed} available in the Supplemental Material and on GitHub (\url{https://github.com/thomasklausch2/EMmixed}).

\section{Simulation studies} \label{sec:simulation_synth}
We conducted two Monte Carlo simulations for evaluating the performance of \texttt{BayesPIM} compared to \texttt{PIMixture} and the non-parametric \texttt{em\_mixed} estimator. In Simulation 1, we generated synthetic data sets under different conditions repeatedly and evaluated the performance of the estimators. The primary purpose of these experiments was to evaluate the implementation of \texttt{BayesPIM} in a controlled setting with sufficient follow up moments to produce accurate parameter estimates. Simulation 2, to the contrary, was designed to generate data sets that closely resembled the CRC EHR (Section \ref{sec:dataintro}) to evaluate \texttt{BayesPIM} in a realistic screening setting. 

\subsection{Set-up of Simulation 1} \label{sec:simulation_synth_setup}
We generated two covariates that were distributed $x_{1i} \sim N(0,1)$ and $x_{2i} \sim \text{Bernoulli}(0.5)$. The incidence model (\ref{eq::mod_x}) was parametrized as $\betaf = (5,0.2,0.2)^T$ and $\sigma=0.2$, so that
\begin{align*}
	\log t_i = 5 + 0.2 x_{1i} + 0.2 x_{2i} + 0.2 \epsilon_i,
\end{align*}
where the residual $\epsilon_i$ was extreme value distributed, so that $t_i \mid \x_i$ was Weibull distributed. Together with $\sigma=0.2$ this yielded a semi-Markov process (progression hazards increasing over time). Furthermore, the prevalence model (\ref{eq::mod_g}) was parametrized as $\thetaf = (\theta_{0}, 0.2,0.2)^T$, with $\theta_{0}$ the intercept parameter that is varied across simulation conditions (see below), so that
\begin{align*}
	\Pr(g_i = 1 \mid \x_i, \thetaf) = \Phi(\theta_{0} + 0.2 x_{1i} + 0.2 x_{2i}).
\end{align*}

To generate screening times as in (\ref{eq:schedule_mechanism}), we set $v_{i1}=0$ and then iteratively drew $v_{ij}$ for $j\ge2$
\begin{align*}
	v_{ij} \sim \text{uniform}(v_{ij-1}+20, v_{ij-1}+ 30),
\end{align*}

so that at least 20 and at maximum 30 time units elapsed between screening moments. Screening stops if $y_{ij}=1$ generated by (\ref{eq:obsprocess}) with an event (positive test) or due to right censoring, as described in Section~\ref{sec:hierarchical}. The time of right censoring $s_i$ was distributed as
\begin{align*}
	s_{i} = v_{i2} + \tilde{s}_{i}, \quad \tilde{s}_{i} \sim \text{exponential}(80^{-1})
\end{align*}

so that the additional follow-up time after the second screening has mean 80 time units. These choices allowed us to study \texttt{BayesPIM} with substantial follow-up and facilitated that \texttt{BayesPIM} converged in acceptable time. In Simulation 2, Section \ref{sec:simulation_real}, we examine a real-world setting with less frequent screening and stronger right censoring adopted from the CRC EHR application.\\

Our simulation set-up examined estimator performance in a wide range of scenarios concerning sample size, test sensitivity, baseline prevalence, baseline test presence, and prior distribution on the test sensitivity. Specifically, we compared two different sample sizes ($n_{sim} = 1000$ versus $n_{sim} = 2000$) and the true value of the test sensitivity parameter $\kappa$ was set to a high level ($\kappa=0.8$), similar to the sensitivity of a colonoscopy for detection of an adenoma, or to a lower level ($\kappa = 0.4$). Furthermore, the intercept parameter $\theta_{0}$ was employed to control the prevalence rates, where $\theta_{0} = 0.11$ and $\theta_{0} = 0.22$ led to the marginal probabilities of prevalence $\Pr(g_i=1) \approx .13$ and $\Pr(g_i=1) \approx .26$, with the higher prevalence resembling that estimated for the CRC EHR application (Section \ref{sec:application}). Moreover, we compared the conditions when each individual receives a test at baseline ($\Pr(r_i=1) = 1$), again similar to the CRC EHR application (Table \ref{tab:baseline_char}), to the other extreme when nobody receives a baseline test ($\Pr(r_i=1) = 0$). The latter setting was expected to be more challenging in estimation as it led to fully latent prevalence status (i.e., $g_i$ missing for all $i$). Finally, we also expected the estimator performance to depend on the degree of information that is available on the true test sensitivity. In particular, we were interested whether \texttt{BayesPIM} is able to estimate the test sensitivity $\kappa$ when there is no prior information available on the test sensitivity. The latter is an extreme scenario, since, in cancer screening, there is usually some information available on test sensitivity. Hence, we compared a fully uninformative prior $\pi(\kappa \mid \bm{\alpha}) = \text{Beta}(1,1)$ which is equivalent to a uniform$(0,1)$ prior, an informative prior $\pi(\kappa \mid\bm{\alpha}) = \text{Beta}(\kappa \mid\bm{\alpha})$ where $\bm{\alpha}$ was chosen such that the prior distribution was centred at the true sensitivity with a standard deviation of 0.05, and a point-prior fixed at the true sensitivity (i.e., $\kappa$ is treated as known). By a full factorial design we thus obtained $2 \times 2 \times 2 \times 2 \times 3 = 48$ simulation conditions. \\

We generated 200 Monte Carlo data sets for each of the 48 conditions and ran, on each data set, the \texttt{BayesPIM} Gibbs sampler (\ref{eq::fc.g})--(\ref{eq::fc.kappa}) until convergence using 4 randomly initialized MCMC chains. Convergence was assumed to be given when the Gelman-Rubin convergence statistic $\hat{R}$ was below 1.1 for all model parameters and the effective sample size was at least 40 for each parameter (advice by \citet[p. 287]{gelman_bayesian_2013}), while discarding half of all posterior draws as burn-in (warm-up). We evaluated convergence every 20,000 draws, combining draws across chains. The MCMC sampler was interrupted if convergence was not achieved after $5 \times 10^5$ iterations per chain and non-converged runs were replaced by additional converged runs. An analysis of the non-converged runs is presented in the results section.

\subsection{Results from Simulation 1}

Simulation~1 was run on a 2024 state-of-the-art Windows-based computing cloud using 64 CPUs in parallel. Convergence of the Gibbs sampler was achieved, on average over all conditions, in 31.7 minutes (minimum = 5.9 minutes, maximum = 362.8 minutes), requiring in most cases fewer than $2 \times 10^5$ iterations (see Supplemental Figure~\ref{fig:convergence_boxplot}). In total, the simulation workload consumed approximately 211 CPU-days of aggregate compute time, executed in roughly 4 days of wall-clock time by the parallel worker pool. The time until convergence (computing time) varied substantially across conditions (boxplots of all computing times are given in Supplemental Figure~\ref{fig:convergence_time_boxplot}). Fastest convergence, with 9.4 minutes on average (minimum = 8.5 minutes, maximum = 12.9 minutes), was achieved in the condition with smaller sample size ($n_{sim} = 1000$), higher sensitivity ($\kappa = 0.8$), lower prevalence ($\Pr(g_i = 1) = 0.13$), baseline tests, and a point prior on the true test sensitivity. Slowest convergence, with 100.1 minutes on average (minimum = 28.7 minutes, maximum = 362.8 minutes), was achieved in the condition with larger sample size ($n_{sim} = 2000$), lower sensitivity ($\kappa = 0.4$), higher prevalence ($\Pr(g_i = 1) = 0.26$), no baseline tests, and an uninformative prior on $\kappa$. Using a linear regression model with all experimental factors entered as main effects, we examined how these factors influenced average computing time (Supplemental Table~\ref{tab:linregtimes}). The strongest effect was found for sensitivity: high sensitivity ($\kappa = 0.8$) versus low sensitivity ($\kappa = 0.4$) reduced computing time on average by 26.9 minutes. Computing time also increased with sample size, as expected, with $n_{sim} = 2000$ requiring on average 17.0 minutes longer to converge than $n_{sim} = 1000$. Point prior information on $\kappa$ reduced computing time on average by 19.4 minutes compared with an uninformative prior. Higher true prevalence (e.g., $\Pr(g_i = 1) = 0.26$) and the absence of baseline tests ($\Pr(r_i = 1) = 0$) each increased computing time by about 9 minutes on average. We generally note that the absolute computation times reported depend on the specific hardware used and may deviate on other systems. \\

All results are based on converged runs only where non-converged runs were replaced by additional converged runs. Non-convergence until $5 \times 10^5$ Gibbs iterations was rare but occurred in the specific condition with an uninformative prior on $\kappa$, high prevalence ($\Pr(g_i=1) = 0.26$), and no baseline testing ($\Pr(r_i=1) = 0$), which is the setting when model identifiability is weakest (see Supplemental Figure \ref{fig:convergence_ncrates}). Visual inspection of MCMC chains for exemplary datasets revealed that multi-modality of the posterior distribution was the primary cause of non-convergence; for an example see Supplemental Figure \ref{fig:nonconvergence_example}. Crucially, non-convergence did not occur when an informative or point prior on $\kappa$ (known $\kappa$) was used instead of an uninformative prior implying that information on $\kappa$ regularizes an otherwise problematic likelihood surface. \\

\begin{figure}[t]  
	\centering  
	\includegraphics[width=\textwidth, trim=135pt 120pt 140pt 120pt, clip, page = 3]{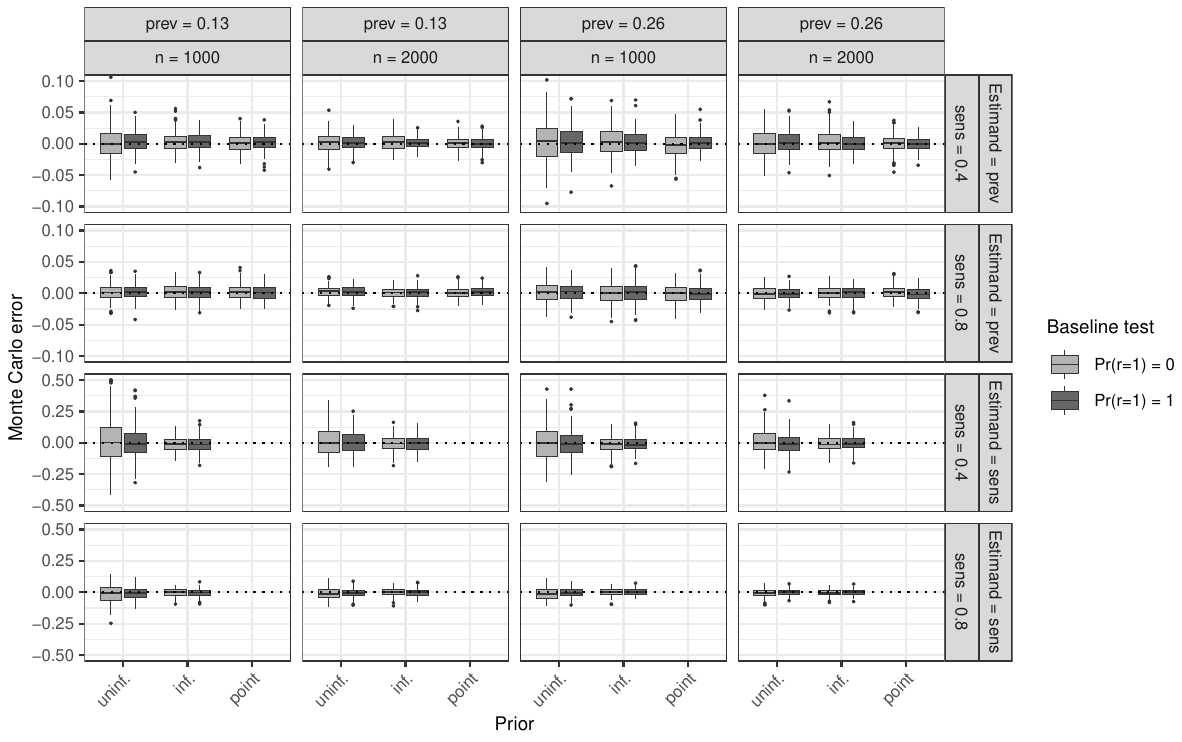}
	\caption{Monte Carlo (estimation) errors of the estimands: marginal prevalence probability ($\Pr(g_i=1)$, denoted "prev") and the test sensitivity ($\kappa$, denoted "sens"). The first two rows give errors for the prevalence and the second two rows for the sensitivity. The priors on the test sensitivity $\kappa$ are either uninformative (uninf.), informative (inf.) or fixed at the true value (point). } 
	\label{fig:kappaprev_boxplot}  
\end{figure}

We evaluated the estimation error (Monte Carlo error) of all parameter estimates, defined as $\hat{\psi} - \psi$, where $\hat{\psi}$ is the posterior median estimate and $\psi$ is the true value (Supplemental Figures \ref{fig:pars_boxplot1} to \ref{fig:pars_boxplot4}). All parameters of the incidence and prevalence models (\ref{eq::mod_x}) and (\ref{eq::mod_g}) were approximately unbiased under all conditions. Parameters in the incidence model were estimated with greater precision than those in the prevalence model. Lower true test sensitivity (0.4 vs. 0.8), absence of a baseline test ($\Pr(r_i=1)=0$ vs. $\Pr(r_i=1)=1$), and higher baseline prevalence probability (0.26 vs. 0.13) increased estimator variance for all parameters. Variance decreased with larger sample sizes ($n_{sim} = 2000$ vs. $n_{sim} = 1000$) and with the use of informative or point priors on $\kappa$. These findings are illustrated by the Monte Carlo error of the marginal baseline prevalence probability $\Pr(g_i=1)$ (Figure \ref{fig:kappaprev_boxplot}). Furthermore, posterior median estimates of the test sensitivity $\kappa$ are also shown and were approximately unbiased. Estimation was reliable for $\kappa = 0.8$, even with uninformative priors, but variance was high for $\kappa = 0.4$ when uninformative priors were used. In this case, informative priors decreased variance. Supplemental Figure \ref{fig:sim1_coverage} shows the approximate frequentist coverage probabilities of the 95\% posterior credible intervals for all parameters. Coverage probabilities closely matched the nominal level across all conditions for all parameters, including test sensitivity.  \\

To further compare model performance across conditions, we utilized posterior predictive mixture CIFs (Section \ref{sec:est_ppd}), which depend jointly on all model parameters and thus serve as summary statistics (Figure \ref{fig:sim1_cdfs_prob_r1} illustrates results for $\Pr(r_i = 1) = 1$; for $\Pr(r_i = 1) = 0$, see Supplemental Figure \ref{fig:sim1_cdfs_prob_r0}). These CIFs visualize the prevalence probability ($\Pr(g_i=1)$) as a point probability at time zero (i.e., a jump of size $\Pr(g_i=1)$ at $t=0$). They enable comparisons between \texttt{BayesPIM}, \texttt{PIMixture}, and the adapted non-parametric CIF estimator \texttt{em\_mixed} (Section \ref{sec:npest}). For \texttt{PIMixture}, we (correctly) specified a Weibull incidence model, included all covariates as predictors of $t_i$ and $g_i$, with logistic regression for the prevalence model. The CIF estimates from \texttt{BayesPIM} were approximately unbiased, consistent with the results for the individual model parameters discussed earlier. Estimator variance was slightly higher when an uninformative prior was used and true $\kappa=0.4$, compared to $\kappa=0.8$. Notably, the considerable uncertainty in $\kappa$ estimation for $\kappa=0.4$ (Figure \ref{fig:kappaprev_boxplot}) did not result in significantly elevated uncertainty in CIF estimation, suggesting that posterior median CIF estimates are robust to posterior uncertainty in test sensitivity. In contrast, \texttt{PIMixture}, which assumes $\kappa=1$, produced biased CIF estimates, with greater bias observed when true $\kappa=0.4$ due to the larger deviation from the assumption of perfect sensitivity. The non-parametric estimator \texttt{em\_mixed} yielded approximately unbiased estimates, provided that sensitivity $\kappa$ was correctly specified. However, \texttt{em\_mixed} performed poorly, as expected (cf. Section \ref{sec:npest}), when $\Pr(r_i=1) = 0$ (Supplemental Figure \ref{fig:sim1_cdfs_prob_r0}). In particular, baseline prevalence estimates were then biased. Additional simulations showed that this bias decreased as more baseline tests became available and was negligible for a moderate proportion of tests done (e.g. $\Pr(r_i=1)=0.5$; Supplemental Figure \ref{fig:additional_results_emmixed}).  \\

\begin{figure}[t]  
	\centering  
	\includegraphics[width=\textwidth, trim=60 120pt 60pt 140pt, clip, page = 3]{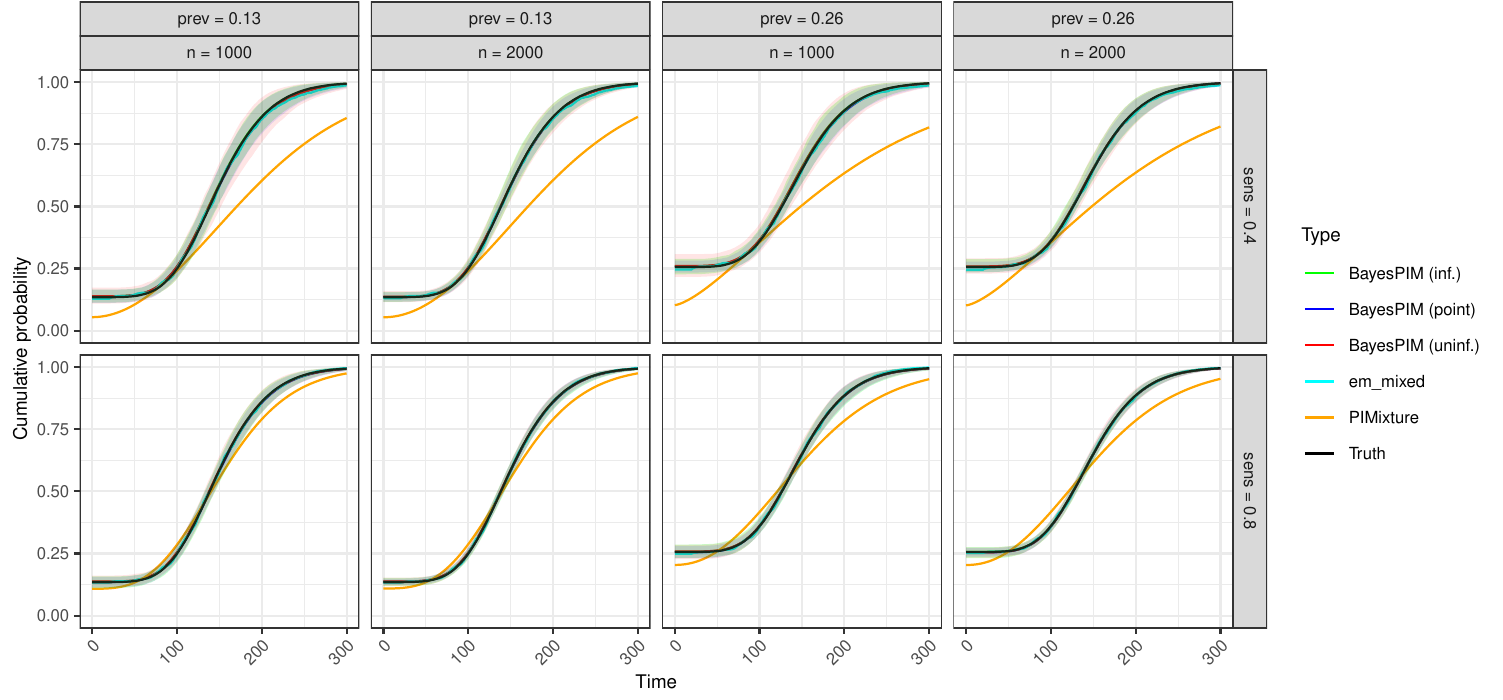}
	\caption{Posterior median estimates of the marginal mixture CIF $F_{t^*}(t \mid \betaf, \sigma, \thetaf)$, point-wise averaged over 200 Monte Carlo simulation runs with 95\% quantiles shown as shaded regions. The condition $\Pr(r_i=1)= 1$ is shown (for $\Pr(r_i=1)= 0$ see Supplemental Figure \ref{fig:sim1_cdfs_prob_r0}). For \texttt{em\_mixed}, $\kappa$ was set to its true value. The lines of all models except \texttt{PIMixture} are overlapping.}
	\label{fig:sim1_cdfs_prob_r1}  
\end{figure}

\subsection{Set-up of Simulation 2} \label{sec:simulation_real}

To evaluate \texttt{BayesPIM} in a realistic setting we conducted a Monte Carlo simulation in which we generated data that strongly resembled the CRC EHR (Simulation 2). The data are described in Section \ref{sec:dataintro} and the results of the analyses with \texttt{BayesPIM} are presented in Section \ref{sec:application}. To obtain covariate and screening times distributed similarly to those observed in the real-world CRC EHR, we devised a resampling procedure, described in detail in the Supplemental Material (Section \ref{sec:supplement:sim2}). \\

For each simulated data set, we generated $n_{sim}$ new individuals with ($g_k, r_k, t_k, \Vvec_k, \x_k$), $k=1,\dots,n_{sim}$. For each $k$, we randomly selected one individual $i'$ with replacement from the CRC EHR and used his/her gender and age ($\x_{i'}$) for the $k$-th newly generated individual (i.e., $\x_k := \x_{i'}$). Subsequently, we generated $t_k$ and $g_k$ according to models (\ref{eq::mod_x})--(\ref{eq::mod_g}), entering both covariates in both models. The true model parameters $\thetaf$, $\betaf$ and $\sigma$ were fixed at the posterior median estimates from the application study presented in Section \ref{sec:application} (Table \ref{tab:coefficients}). The transition time distribution was chosen as Weibull, which had best fit in the application study. The probability of a baseline test was set to its estimate of $\Pr(r_i=1)=0.93$ in the CRC EHR (Table \ref{tab:baseline_char}). \\

To obtain new screening times and run screening tests according to the hierarchical model (Section \ref{sec:hierarchical}), equations (\ref{eq:schedule_mechanism})--(\ref{eq:obsprocess}), we used the observed screening times $\Vvec_{i'}=(v_{i'1},v_{i'2},\dots,v_{i'c_{i'}})$ of the sampled unit $i'$ as donor for new screening times. Specifically, we set $v_{k1}=0$ and stopped screening at baseline with probability $\kappa$, i.e. $y_{k1}=1$ (positive baseline test), only if $g_k=1$ and $r_k=1$. If $r_k=0$, $y_{k1}$ was missing, as described in Section \ref{sec:hierarchical}. Else we set $y_{k1}=0$ and, subsequently, for each $j=2,\dots c_{i'}$ we followed the following steps:
\begin{enumerate}
	\item Set $v_{kj}:=v_{i'j}$
	\item Generate $y_{kj}$ as defined by (\ref{eq:obsprocess})
\end{enumerate}

The procedure was stopped at an occasion $j'$ if $y_{kj'}=1$ such that $\Vvec_k = (v_{k1},\dots,v_{kc_k})$, with $c_k = j'$ the event occasion. Additional steps were taken to simulate a right censoring process that approximated that of the CRC EHR; for details, see the Supplemental Material (Section \ref{sec:supplement:sim2}). As a check, Supplemental Figure \ref{fig:cens_dist_check} compares the sampling distributions of various statistics calculated on repeated samples of the screening times $(\Vvec_1,\dots,\Vvec_{n_{sim}})$ to the observed statistics in the CRC EHR (e.g., mean and standard deviation of the event/censoring time; mean number of test occasions). Results suggested that the generated and observed screening times were similar, which indicated that we were successful in generating screening data similar to the real-world EHR. \\

We simulated 200 data sets per condition and applied \texttt{BayesPIM}, \texttt{PIMixture} and \texttt{em\_mixed}, as in Simulation 1. Specifically, we considered $3 \times 2\times 2 = 12$ simulation conditions comparing different sample sizes ($n_{sim}=$ 810, 1620 or 3240), low and high test sensitivity ($\kappa=$ 0.4 or 0.8), and two different right censoring process (observed as in the CRC EHR or extended by 10 years). The condition with $n_{sim}=810$, $\kappa = 0.8$ was similar to the real CRC EHR and compared to conditions with larger sample sizes ($n_{sim} =1620$ and $n_{sim}=3240$) and lower test sensitivity ($\kappa=0.4$), respectively. In addition, we added a condition with an extended right censoring process where the time of right censoring was simulated to occur ten years later than in the CRC EHR; see the Supplemental Material (Section \ref{sec:sim2_extcens_gen}). This setting was added to assess model performance if follow-up had been longer than that observed in the real CRC EHR.

\subsubsection{Results from Simulation 2}

\begin{figure}[t]  
	\centering  
	\includegraphics[width=\textwidth, trim=100pt 130pt 80pt 130pt, clip]{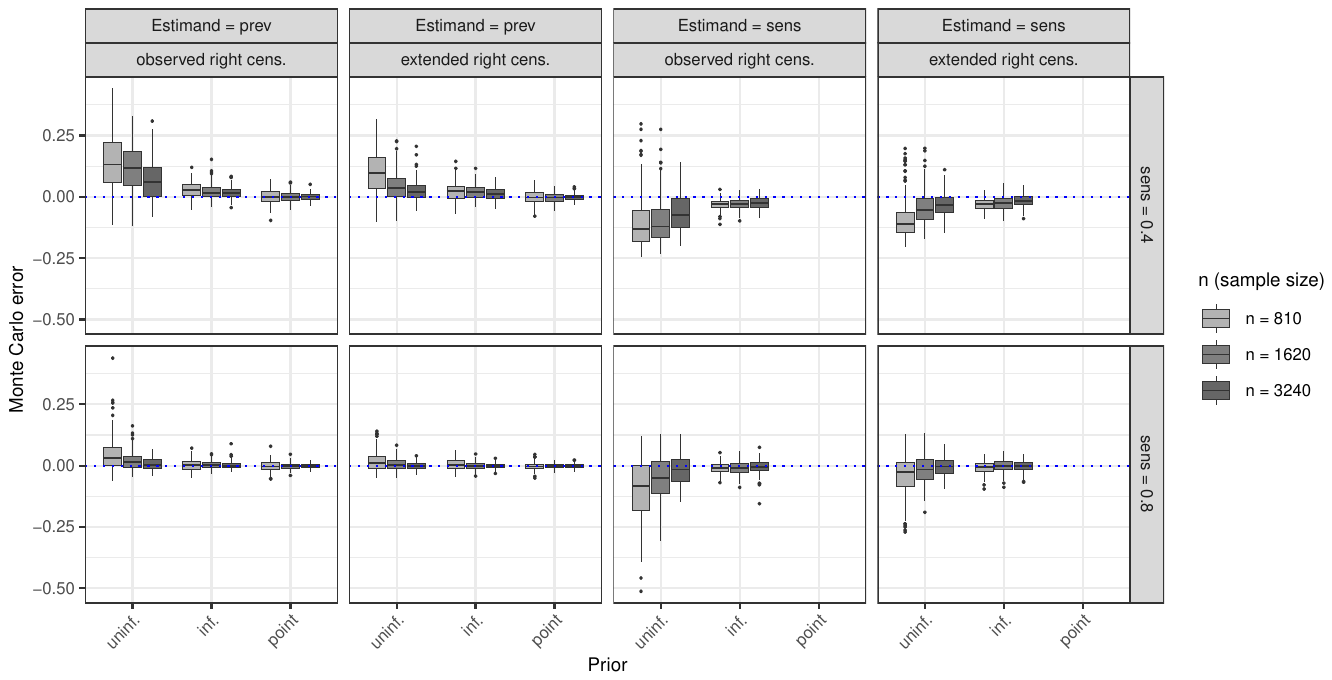}
	\caption{Monte Carlo (estimation) error of the marginal prevalence probability ($\Pr(g_i=1)$, denoted "prev") and the test sensitivity ($\kappa$, denoted "sens") estimands. The first two columns give errors for the prevalence and the second two columns for the sensitivity. The priors on the test sensitivity $\kappa$ are either uninformative (uninf.), informative (inf.) or fixed at the true value (point). } 
	\label{fig:sim2_pars_boxplots_kappaprev}  
\end{figure}

With informative or point (fixed) priors on the test sensitivity, the Gibbs sampler reliably converged within $5\times10^5$ draws (see Supplemental Figures \ref{fig:convergence_ncrates_sim2} and \ref{fig:convergence_boxplot_sim2}). However, with uninformative priors, non-convergence was observed in the condition $n_{sim}=810$, $\kappa=0.4$, and observed right censoring. The impact of uninformative priors was also particularly evident in explaining variations in Monte Carlo (estimation) errors across conditions. Pronounced bias was observed in the intercepts $\beta_{0}$ and $\theta_{0}$ of the incidence (\ref{eq::mod_x}) and prevalence models (\ref{eq::mod_g}) under the $\kappa=0.4$ condition (Supplemental Figures \ref{fig:pars_boxplot1_sim2} and \ref{fig:pars_boxplot2_sim2}). This bias diminished with larger sample sizes, higher test sensitivity ($\kappa=0.8$), or extended follow-up. With an informative prior (centred at the true $\kappa$) this bias was strongly decreased and bias vanished when a point prior (fixed at the true $\kappa$) was employed. In these settings, estimation was also more efficient compared to the uninformative prior case. In the $\kappa=0.8$ condition, bias was approximately zero when informative and point priors were put on the test sensitivity, and it was small when the uninformative prior was used. \\

We also studied the frequentist coverage probability of the 95\% posterior intervals for all parameters (Supplemental Figure \ref{fig:sim2_coverage}), demonstrating that coverage was approximately nominal if the intercept and $\kappa$ parameters did not suffer substantial bias which was the case when informative or point priors were used.\\

\begin{figure}[t]  
	\centering  
	\includegraphics[width=\textwidth, trim=90pt 50pt 100pt 50pt, clip, page = 2]{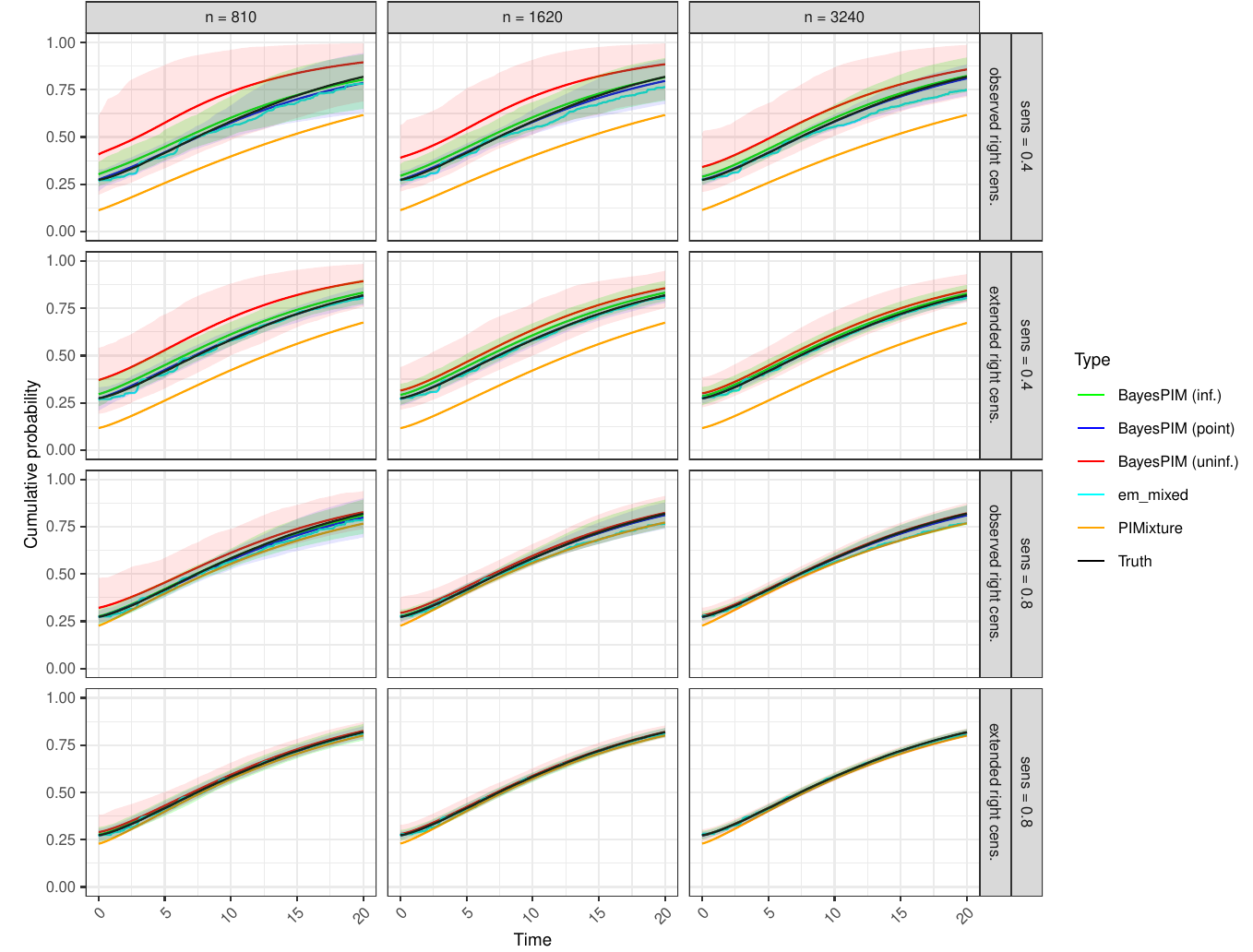}
	\caption{Marginal mixture CIFs, $F_{t^*}(t \mid \betaf, \sigma, \thetaf)$, point-wise averaged over 200 Monte Carlo simulation runs with 95\% quantiles shown as shaded regions (for clarity these bounds have been omitted for \texttt{PIMixture} and \texttt{em\_mixed}). For \texttt{BayesPIM} the posterior median of the posterior predictive marginal mixture CIF is shown. For \texttt{PIMixture} and \texttt{em\_mixed} the corresponding maximum likelihood estimate is shown. Lines of \texttt{BayesPIM} (point) and "Truth" are overlapping in most graphs.}
	\label{fig:sim2_cdfs}  
\end{figure}

These results are summarized in Figure \ref{fig:sim2_pars_boxplots_kappaprev}, which shows the Monte Carlo (estimation) errors, as defined in Simulation 1, for the marginal prevalence probability $\Pr(g_i=1)$ which depends on the prevalence model parameters $\thetaf$, and the test sensitivity $\kappa$. In the uninformative prior setting for $\kappa$, the estimation of $\kappa$ was inaccurate in both the $\kappa=0.4$ and $\kappa=0.8$ conditions, reflecting results on the model parameter estimates described above. Although estimation appeared consistent, meaning that bias decreased with larger sample sizes or extended follow-up, substantially larger samples (beyond $n_{sim}=3240$) or longer follow-up would have been required to eliminate the bias in the $\kappa$ estimates. With informative prior, bias on the $\kappa$ estimates was small although still visible in the $\kappa=0.4$ condition and absent in the $\kappa=0.8$ condition, translating to smaller bias on the parameter estimates and prevalence estimates. This residual bias was fully eliminated with correctly specified point priors (fixed) $\kappa$. In summary, these results suggest that estimating test sensitivity was challenging given the CRC EHR screening time process especially without prior information and the $\kappa=0.4$ setting. \\

A key question was how the Monte Carlo errors affected the estimates of the marginal mixture CIFs, $F_{t^*}(t \mid \betaf, \sigma, \thetaf)$ (Figure \ref{fig:sim2_cdfs}; for marginal CIFs of the non-prevalent population, $F_{t}(t \mid g=0, \betaf,\sigma)$, see Supplemental Figure \ref{fig:sim2_cdfs_x}). As in Simulation 1, these CIFs facilitated comparisons with \texttt{PIMixture} and \texttt{em\_mixed}. When \texttt{BayesPIM} used an uninformative $\kappa$ prior, the resulting bias was evident in the CIFs, though the bias was smaller in the $\kappa=0.8$ conditions. In contrast, \texttt{BayesPIM} with $\kappa$ fixed at its true value (point prior) provided the most precise estimates, even in the most challenging condition ($n_{sim}=810$, $\kappa=0.4$). The non-parametric \texttt{em\_mixed} estimator was generally accurate but exhibited some bias beyond ten years when $\kappa=0.4$. This bias disappeared with longer follow-up (extended right censoring). Given that most cases in the observed right-censoring process were censored after approximately ten years (Supplemental Figure \ref{fig:rightcensoring_bootstrapped}), it is evident that the non-parametric estimator lacked sufficient data to produce accurate incidence estimates beyond this time. This limitation also resulted in larger variance, which is not shown in Figure \ref{fig:sim2_cdfs} but can be seen in Supplemental Figure \ref{fig:sim2_cdfs_emmixed}. In contrast, \texttt{BayesPIM} leveraged model-based extrapolation, providing better estimates in these sparse-data settings. For the comparison model \texttt{PIMixture}, the results were consistent with Simulation 1. Specifically, under $\kappa=0.4$, \texttt{PIMixture} exhibited greater bias compared to $\kappa=0.8$, due to its stronger deviation from the assumption of perfect sensitivity.

\section{Application to the CRC EHR} \label{sec:application} 
We apply \texttt{BayesPIM}, \texttt{PIMixture}, and \texttt{em\_mixed} to the CRC EHR introduced in Section \ref{sec:dataintro}. In doing so, we specify \texttt{BayesPIM} models with Weibull, log-logistical, log-normal, and exponential distributed transition times $t_i$ (model (\ref{eq::mod_x})). To obtain the exponential model, we use a Weibull specification and constrain $\sigma=1$. This model is equivalent to a Markov model due to its constant hazard property and it is primarily included for comparison with the less restrictive models. We included subjects' gender and age as covariates in both the incidence and the prevalence models (models (\ref{eq::mod_x}) and (\ref{eq::mod_g})). We used priors as specified in Section \ref{sec:priors}. For the sensitivity $\kappa$, we compared an uninformative prior that sets $\bm{\alpha} = (1,1)$ to an informative prior that we based on a range of test sensitivities published in the literature (Section \ref{sec:introduction}). Specifically, test sensitivities for colonoscopy detection of adenomas range between 0.65 and 0.92 \citep{van_rijn_polyp_2006}. We therefore centred the Beta prior for $\kappa$ at 0.8 with standard error 0.05; using numerical root-finding we determined $\bm{\alpha} = (50.4,12.6)$, such that $\Pr(\kappa \in (0.696, 0.890))\approx0.95$. In addition, we added the setting of perfect test sensitivity by constraining $\kappa=1$. \\

We ran Gibbs sampler steps (\ref{eq::fc.g})--(\ref{eq::fc.kappa}) until convergence, where step (\ref{eq::fc.kappa}) is omitted in the setting that constrains $\kappa$ to one. Convergence was assumed when the upper confidence bound of the Gelman-Rubin convergence diagnostic $\hat{R}$ calculated over the most recent half of all iterations first came under the value of 1.1 for all model parameters and the effective sample size of all MCMC samples was at least $10^3$ for each parameter. We ran four randomly initialized MCMC chains in parallel and assessed convergence every $5\times10^4$ draws. \\

WAIC slightly favoured a Weibull model over the alternative distributions (Table \ref{tab:WAIC}). Furthermore, an informative prior gave slightly better WAIC than an uninformative prior. The models with perfect sensitivity ($\kappa=1$) had worse fit than the models allowing $\kappa<1$, indicating that relaxing the assumption of perfect sensitivity made by \texttt{PIMixture} improved model fit.\\

\begin{table}[b]
	\centering
	\caption{Model WAIC and number of draws per chain ($\times 10^5$) after convergence. The prior for the test sensitivity $\kappa$ is either chosen informative (Inf.), uninformative (Uninf.) or $\kappa$ is constrained to one ($\kappa=1$).}
	\begin{tabular}{lccc c ccc}
		& \multicolumn{3}{c}{WAIC} & & \multicolumn{3}{c}{Draws per chain ($\times 10^5$)} \\
		\cmidrule(r){2-4} \cmidrule(r){6-8}
		model & Inf. & Uninf. & $\kappa=1$ & & Inf. & Uninf. & $\kappa=1$ \\
		\hline
		exponential & 1598.0 & 1600.4 & 1599.5 & & 1.5 & 2.0 & 1.0 \\
		Weibull & 1595.4 & 1598.4 & 1597.0 & & 4.0 & 6.0 & 2.5 \\
		log-logistic & 1596.1 & 1599.7 & 1599.1 & & 2.0 & 4.0 & 1.0 \\
		log-normal & 1596.8 & 1600.2 & 1601.0 & & 2.5 & 5.5 & 1.5 \\
		\hline
	\end{tabular}
	\label{tab:WAIC}
\end{table}

Estimates for the test sensitivity $\kappa$ differed (.79 vs .70) between informative and uninformative priors but the uninformative prior model had a wide credible interval ([0.44, 0.94]), suggesting that $\kappa$ was only weakly identified in this data set. This conclusion is supported by the fact that the credible interval of $\kappa$ in the model with informative $\kappa$ prior had similar length ([0.69, 0.88]) as the 2.5 to 97.5 quantile range of the Beta prior specified for $\kappa$; see above. However, the added certainty on $\kappa$ encoded in the informative Beta prior decreased the credible interval lengths of all other model parameters as compared to the model without informative $\kappa$ prior. Having said this, the regression coefficient estimates were similar across models, including the exponential model, suggesting robustness of inference on risk factors to the model assumptions on $\kappa$ and the distribution of $t_i$.\\

\begin{table}[t]
	\centering
	\caption{Posterior median estimates with 95\% credible interval for the coefficients from the Weibull and exponential \texttt{BayesPIM} models for different prior settings on the test sensitivity $\kappa$.}
	\begin{tabular}{lrrr}
		& Weibull (inf.) & Weibull (uninf.) & Exponential (Inf.) \\
		\hline
		\addlinespace[0.4em]
		\small\underline{\textit{Model for $t_i$:}} \\
		$\beta_{0}$ & 2.79 [2.50, 3.21] & 2.78 [2.43, 3.31] & 3.07 [2.77, 3.42] \\
		female & -0.11 [-0.44, 0.20] & -0.13 [-0.61, 0.22] & -0.11 [-0.52, 0.27] \\
		age$^*$ & -0.17 [-0.32, -0.03] & -0.16 [-0.33, 0.04] & -0.17 [-0.36, 0.02] \\
		$\sigma$ & 0.74 [0.55, 1.02] & 0.72 [0.48, 1.04] & 1.00 [1.00, 1.00] \\
		\addlinespace[0.4em]
		\small\underline{\textit{Model for $g_i$:}} \\
		$\theta_{0}$ & -0.51 [-0.69, -0.31] & -0.41 [-0.70, 0.09] & -0.52 [-0.70, -0.32] \\
		female & -0.24 [-0.46, -0.02] & -0.25 [-0.49, -0.01] & -0.24 [-0.46, -0.02] \\
		age$^*$ & 0.33 [0.21, 0.45] & 0.34 [0.21, 0.49] & 0.34 [0.22, 0.46] \\
		\addlinespace[0.4em]
		\small\underline{\textit{Sensitivity:}} \\
		$\kappa$ & 0.79 [0.69, 0.88] & 0.70 [0.44, 0.94] & 0.79 [0.69, 0.88] \\
		\hline
		\multicolumn{4}{l}{\footnotesize $^*$age was z-standardized (before standardization: mean = 52.9 years with sd = 7.71 years).}
	\end{tabular}
	\label{tab:coefficients}
\end{table}

The sign and size of the coefficient estimates (Table \ref{tab:coefficients}) suggested that older age increased incidence probability of an adenoma, while gender did not have a clear effect on incidence. Specifically, individuals who had an age of one standard deviation (7.71 years) above the sample mean of 52.9 years at baseline were expected to have transition time to an adenoma reduced by the factor $\exp(-0.175)=0.839$ (16.1\% faster transition). Both gender and age were predictors of prevalence status at baseline, with female gender decreasing and older age increasing the probability of prevalence. These estimates are in line with epidemiological expectations suggesting that older subjects and men are at a higher risk of developing adenomas and CRC. \\

\begin{figure}[t]  
	\centering  
	\includegraphics[width=\textwidth, trim=140pt 160pt 140pt 160pt, clip, page = 3]{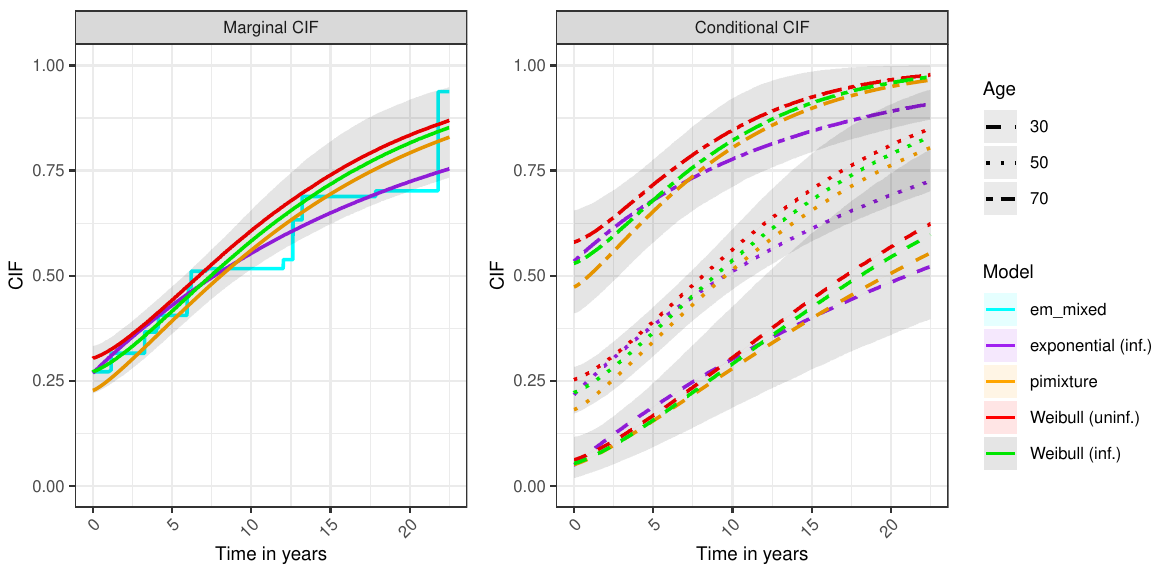}
	\caption{Marginal and conditional mixture cumulative incidence functions (CIF), $F_{t^*}(t \mid \betaf, \sigma, \thetaf)$ and the conditional CIF, $F_{t^*}(t \mid \tilde{\x}, \betaf, \sigma, \thetaf)$, for Weibull \texttt{BayesPIM} with uninformative (uninf.) prior and informative (inf.) priors. For comparison, corresponding CIFs from the Weibull \texttt{PIMixture} model, the exponential \texttt{BayesPIM} model (inf.), and \texttt{em\_mixed} are given. The lines represent posterior median estimates and the shaded regions indicate the 95\% credible interval of the Weibull (inf.) models (for overlayed intervals from the Weibull (uninf.) model see Supplemental Figure \ref{fig:CIF_margcond_complete}). } 
	\label{fig:CIF_margcond}  
\end{figure}

Figure \ref{fig:CIF_margcond} gives the marginal mixture CIF, $F_{t^*}(t \mid \betaf, \sigma, \thetaf)$, and the conditional CIF, $F_{t^*}(t \mid \tilde{\x}, \betaf, \sigma, \thetaf)$, for different age groups (i.e. $\tilde{\x}$ was age set to 30, 50 and 70). The Weibull and the exponential models are shown, respectively. The corresponding CIFs obtained from a fit by \texttt{PIMixture} are added. For this, we used a \texttt{PIMixture} model that assumes the same Weibull incidence model (\ref{eq::mod_x}) as \texttt{BayesPIM}. Furthermore, we added the non-parametric estimator \texttt{em\_mixed} of $F_{t^{*}}( t \mid \betaf, \sigma, \thetaf)$ (see Section \ref{sec:npest}) which allows a visual inspection of goodness of fit of the marginal CIF. We set the sensitivity $\kappa$ to 0.80 in \texttt{em\_mixed}. See Supplemental Figure \ref{fig:CIF_margcond_healthy} for an equivalent plot for the non-prevalent population. \\

The mixture CIFs depict the sum of prevalence and follow-up incidence probability until a given point in time. The prevalence probability can be read at the intercept and is interpreted as the proportion of the population that has an adenoma at the point of inclusion in the surveillance program ($\Pr(g_i=1)$).  Specifically, the models with informative and uninformative prior estimated this proportion at 0.274 [0.222, 0.333] and 0.307 [0.217, 0.475], respectively; the exponential (inf.) model at 0.269 [0.219, 0.330]. Again we note the wider credible interval of the Weibull (uninf.) model as a consequence of the uninformative $\kappa$ prior. In addition, the prevalence estimate was higher, relating to the fact the Weibull (uninf.) model estimated $\kappa$ lower (Table \ref{tab:coefficients}). The observed prevalence was 0.204 (Table \ref{tab:baseline_char}), suggesting that an estimated 34\% (0.07 points) of prevalence was unobserved (latent). The non-parametric estimator \texttt{em\_mixed} estimated the prevalence probability at 0.271, which is very close to the Weibull (inf.) and exponential model estimates and reassured us about model fit of the best model according to WAIC. To the contrary, \texttt{PIMixture} estimated the prevalence lower at 0.228, owing to the assumption of $\kappa=1$. The \texttt{PIMixture} prevalence estimate was similar to that of the \texttt{BayesPIM} model that assumed $\kappa=1$ (.217; not shown in Figure \ref{fig:CIF_margcond}).\\

While there were pronounced differences in prevalence estimates between models, the incidence model estimates were similar (Table \ref{tab:coefficients}). The marginal CIF of Weibull (inf.), Weibull (uninf.), and \texttt{PIMixture} had an almost parallel trajectory (Figure \ref{fig:CIF_margcond}, left side). The marginal CIF of the exponential model had different curvature, however, owing to its constraint of $\sigma=1$, which led in particular in the time after 10 years to different predicted probabilities as compared to the Weibull models. It is important to note that, in this data set, most censoring occurs before the time of 10 years, and the last adenoma event occurred at 22.5 years (Table \ref{tab:baseline_char}). Hence, the uncertainty of the estimates increases after 10 years, as indicated by wider credible interval bounds shown for the Weibull (inf.) model. The non-parametric fit of \texttt{em\_mixed} can be used as a visual inspection of incidence model fit. In this case, the Weibull (inf.) which has lowest WAIC indeed had similar fit to \texttt{em\_mixed}, but the exponential model fits similarly well in this visual inspection. Both models fit notably better to \texttt{em\_mixed} than the \texttt{PIMixture} model. \\

The right panel of Figure \ref{fig:CIF_margcond} shows the conditional mixture CIF, $F_{t^*}(t \mid \tilde{\x}, \betaf, \sigma, \thetaf)$, for selected age groups. The higher risk of prevalence for older individuals was reflected by the increasing ordinate intercepts across groups. Furthermore, in the 30 and 70 year groups which were further from the sample mean of 52.9 years than the 50 year group, some pronounced differences between models were found. This suggests that model selection, in this case, was more important for the estimation of conditional risk than for estimation of marginal risk.

\section{Discussion} \label{sec:discussion}
We introduced \texttt{BayesPIM}, a modelling framework for estimating time- and covariate-dependent incidence probabilities from screening and surveillance data. The model is used when the disease may be prevalent at baseline, tests at baseline are fully or partially missing, and the test for disease has imperfect sensitivity. In this regard, we extend the available prevalence-incidence mixture model \texttt{PIMixture}, which assumes perfect sensitivity \citep{cheung_mixture_2017, hyun_flexible_2017}. We applied the model to data from high familial risk CRC EHR surveillance by colonoscopy, demonstrating that conditioning incidence and prevalence estimates on covariates explains substantial heterogeneity in adenoma risk (cf. Figure \ref{fig:CIF_margcond}). These estimates can serve as valuable input for targeted screening strategies, such as higher intensity for older individuals with familial colorectal cancer risk and recognizing that older individuals have a high probability of having adenomas at baseline. More generally, \texttt{BayesPIM} is readily applicable to other screening data that often have a similar structure (Section \ref{sec:dataintro}). In our study, we assumed perfect specificity, which was a plausible assumption for the CRC EHR because colonoscopy findings are validated by pathology; similar confirmatory tests are common in other screening programs. \\

In estimation, we leveraged a Bayesian Metropolis-within-Gibbs sampler with data augmentation for the latent transition times and the latent prevalence status. Our Bayesian approach facilitates the inclusion of prior information on the model parameters. We have found that weakly informative priors on the incidence and prevalence model parameters (cf. Section \ref{sec:priors}) provide effective regularization without biasing the parameter estimates. Arguably, the prior information on $\betaf$, model (\ref{eq::mod_x}), and $\thetaf$, model (\ref{eq::mod_g}), is scarce in practice, making a weakly informative prior choice an attractive default. Furthermore, we studied the role of the prior on the test sensitivity $\kappa$ in substantial detail. When a Beta prior is used, updating $\kappa$ is conjugate (see (\ref{eq::fc.kappa})) and hence fast. However, through simulations we demonstrated that a fully uninformative prior, chosen uniform$(0,1)$, can lead to estimation problems, such as non-convergence, multi-modality of the posterior, and bias in $\kappa$ and other model parameters. Adding some information on $\kappa$ through a centred prior on the most likely value yielded reliably converging Markov chains and approximately zero bias in the \texttt{BayesPIM} parameter estimates. This included more challenging settings than the one actually observed in the CRC EHR, such as low sensitivity $\kappa = 0.4$ or the absence of a baseline test ($\Pr(r_i=1)=0$). We, therefore, advise the use of informative priors in practice, which is facilitated by the fact that test sensitivity is usually roughly known in cancer screening, as demonstrated for the CRC EHR (Section \ref{sec:application}). Alternatively, several analyses may be run with a range of fixed $\kappa$ values, and the model with the highest WAIC given preference. Clearly, if $\kappa$ is known with high certainty, fixing $\kappa$ at this value is the best option. These measures are particularly important in small samples or when there is only short follow-up, as in the CRC EHR (Table 1, Supplemental Figure \ref{fig:rightcensoring_bootstrapped}).\\

A natural question is how sensitive our modelling framework is to misspecification of the prior for the test sensitivity $\kappa$. In general, the impact on parameter estimates and CIFs will depend both on the discrepancy between the prior assumptions and the true value of $\kappa$ and on how informative the data are about $\kappa$: when the data are highly informative, the likelihood will pull the posterior away from a misspecified prior, whereas when information on $\kappa$ is weak, inferences will inherit more of the prior assumptions. An extreme instance of misspecification occurs for \texttt{PIMixture}, which corresponds to fixing $\kappa = 1$ and thus assuming perfect sensitivity a priori. In Simulations~1 and 2 we observed that, when the true sensitivity was high ($\kappa = 0.8$), the resulting bias in CIF estimates was moderate, whereas for lower sensitivity ($\kappa = 0.4$) the bias became substantial. In practice, we therefore recommend routinely performing sensitivity analyses over a plausible range of fixed $\kappa$ values (or alternative priors on $\kappa$) to assess the impact of potential misspecification of test sensitivity on the substantive conclusions. \\

Model fit evaluation is an important aspect of modelling with \texttt{BayesPIM}, and we proposed the WAIC as a means to select the best transition times ($t_i$) distribution. In the application, WAIC only slightly favored a Weibull model over the alternative models (log-normal, log-logistic, and exponential). In addition, we adapted the non-parametric estimator \texttt{em\_mixed}, initially proposed by \citet{witte_em_2017} for screening data without baseline prevalence but imperfect test sensitivity, to our setting with prevalence. In simulations, \texttt{em\_mixed} had excellent performance, provided baseline tests were available for a moderate to high proportion of the sample. Hence, the estimator serves as an additional fit assessment in these settings. In the application, the visual comparison of the \texttt{em\_mixed} CIF to that of \texttt{BayesPIM} was indeed reassuring on model fit. In future work, \texttt{em\_mixed} might be extended to work well in settings with only a few or no baseline tests. However, while useful for marginal assessment of fit, \texttt{em\_mixed} cannot replace \texttt{BayesPIM} which allows relating incidence and prevalence to covariates to identify risk factors, make personalized predictions, and estimate conditional CIFs (see Figure \ref{fig:CIF_margcond}). Furthermore, as a parametric model \texttt{BayesPIM} can be used to extrapolate CIFs beyond the observed screening times distribution (e.g. maximum censoring time), which is not feasible with non-parametric estimators. \\

A strong aspect of simulation our study of the CRC EHR is that we assessed the performance of \texttt{BayesPIM} under the observed screening times and right censoring process (Section \ref{sec:simulation_real}). Our results indicated that \texttt{BayesPIM} can reliably estimate the model parameters under that condition ($n_{sim}=810$, $\kappa=0.8$, observed right censoring), and also provided good performance in the more challenging setting with $\kappa=0.4$, provided that an informative $\kappa$ prior was used. Here, we found a small bias in the $\kappa$ and prevalence estimates, unless $\kappa$ was set to its true value through a point prior. This finding illustrates the importance of evaluating estimation performance in real-world settings as a supplemental analysis to the data analysis. To this end, Supplemental Material Sections \ref{sec:sim2_obscens_gen} and \ref{sec:sim2_extcens_gen} explain how to generate screening times from observed EHR screening data. This approach should be adopted in future work. \\

Based on our results, we recommend using \texttt{BayesPIM} in settings with imperfect test sensitivity, whereas in settings where perfect sensitivity can be assumed, \texttt{PIMixture} remains an appropriate choice. Although \texttt{BayesPIM} is computationally more demanding than \texttt{PIMixture}, Simulation~1 (Section~\ref{sec:simulation_synth}) showed that convergence was typically achieved within minutes to a few hours, depending on the experimental condition. With contemporary hardware, such runtimes are compatible with routine use, as model fitting is typically performed offline and on single data sets.\\

Limitations of \texttt{BayesPIM} currently include the reliance on assumptions (a) to (e), Section \ref{sec:hierarchical}, as well as our focus on two-parameter survival distributions. The conditional independence of prevalence-incidence components (a) and missing at random baseline test outcomes (b), appear necessary, albeit realistic identifying assumptions. Although also the assumption of uninformative scheduling of screening (c) was plausible for the CRC EHR, because most adenomas are not symptomatic, the assumption may need to be relaxed in future research for settings where symptomatic pre-state diseases or diseases let patients initiate a screening test. A good starting point seems to be a parametric modelling assumption for the association between the scheduling times $v_{ij}$ and $(g_i, t_i)$. Furthermore, the assumption of uninformative censoring (d) is a well-known standard assumption in survival modelling which may be violated, for example, in settings with right censoring due to competing events such as death caused by the disease. In this regard, \texttt{BayesPIM} could be extended to competing event modelling similar to the \citet{hyun_sample-weighted_2020} extension for the \texttt{PIMixture} model. Assumption (e) concerns the stability of test sensitivity across time. This assumption may be violated if adenomas become easier to detect over time (e.g., due to bleeding or increase in volume) which may affect the probability of discovery in screening. Letting $\kappa$ change over time seems to be a viable path for further research. Extending available multi-state semi-Markov models for progressive diseases \citep[e.g.,][]{klausch_bayesian_2023} for prevalence and imperfect sensitivity is also a viable option. In addition, extending \texttt{BayesPIM} to account for recurring adenomas after the first adenoma will be needed for modelling adenoma occurrence in some high risk populations (e.g., patients with the so-called Lynch syndrome develop adenomas frequently). \texttt{BayesPIM}'s focus on two-parameter distributions, estimates may be biased if transition times strongly deviate from the assumed distribution. Semi-parametric extensions (e.g., Cox-type transitions) may be considered in the future to make \texttt{BayesPIM} more flexible, but they may require large samples. \\

In conclusion, \texttt{BayesPIM} brings a flexible, Bayesian approach to handling latent prevalence and imperfect test sensitivity in screening and surveillance data. By incorporating informative priors, assessing model fit through both parametric and non-parametric approaches, and demonstrating robust performance under realistic conditions, \texttt{BayesPIM} enables more accurate, data-driven, and ultimately patient-centred screening strategies.

\section*{Supplemental Material}
The supplemental material contains the following files.
\begin{enumerate}
	\item A pdf with with following sections:
	\begin{itemize}
		\item Section A: Proofs and technical details
		\item Section B: Additional results from Simulation 1
		\item Section C: Additional details on the set-up of Simulation 2
		\item Section D: Additional results from Simulation 2
		\item Section E: Additional results from the CRC application
	\end{itemize}
	\item  A zip-Archive with \texttt{R} code, in particular:
	\begin{itemize}
		\item A tar.gz file with the \texttt{BayesPIM} package (see \texttt{readme\_packages})
		\item A tar.gz file with the \texttt{EMmixed} package (see \texttt{readme\_packages})
		\item A zip-Archive with \texttt{R} code and additional documentation (see \texttt{readme\_simulations}) for running and analyzing the simulation studies
	\end{itemize}
\end{enumerate}

\clearpage
\newpage
\bibliographystyle{apalike}
\bibliography{lit_bayespim}

\newpage
\appendix

\pagenumbering{gobble} 
\thispagestyle{empty}  

\clearpage
\vspace*{0.33\textheight} 
\begin{center}
	{\Large \textbf{"A Bayesian prevalence-incidence mixture model \\[0.5em]
			for screening outcomes with misclassification" \\[2em]
			Supplemental Material}}\\[2cm]
	\large by Thomas Klausch, Birgit Lissenberg-Witte \& Veerle Coup\'e \\
	\vspace{0.5cm}
	\small Amsterdam University Medical Center, Department of Epidemiology and Data Science, \\ Amsterdam, The Netherlands
\end{center}
\clearpage

\pagenumbering{arabic}
\setcounter{page}{1}

\counterwithin{figure}{section}
\counterwithin{table}{section}

\setcounter{equation}{0}

\renewcommand{\theequation}{S-\arabic{equation}}

\section{Proofs} \label{sec:supplement_proofs}
\subsection{Observed-data likelihood} \label{sec:supplement_proofs_obsll}
We derive the observed-data likelihood $\mathcal{L}( \betaf, \sigma, \thetaf, \kappa  \mid  \Df )$. For the following, recall the notation from Section \ref{sec:notation} of the main paper. In particular, screening stops at the first positive test (event) or at right censoring. In the case of an event and baseline test ($r_i=1$) we have $\yf_i = (y_{i1}, y_{i2},\dots, y_{ic_i-1}, y_{ic_i})=(0, 0, \dots, 0, 1)^T$ with screening times $\Vvec_i = (v_{i1},v_{i2},\dots, v_{ic_i})^T$ and $v_{ic_i}<\infty$ (finite event time). In the case of right censoring we have $\yf_i = (0, 0, \dots, 0)^T$ with times $\Vvec_i = (v_{i1},v_{i2},\dots, v_{ic_i-1}, \infty)^T$ so that $v_{ic_i}=\infty$ and $y_{ic_i} = 0$ indicates right censoring.  Hence, the observed vector of test outcomes is complete. Let $\yf_{i,\text{com}}$ denote the complete-data vector of test outcomes, then $\yf_{i,\text{com}} = \yf_i$ if $r_i=1$. \\

Without a baseline test ($r_i=0$), we have $y_{i1}$ is missing, so that the observed test outcomes are $\yf= (y_{i2},\dots, y_{ic_i})$. Hence, the complete-data vector of outcomes is $\yf_{i,\text{com}} = (y_{i1}, \yf)$ with $y_{i1}$ unobserved (missing).  We will discuss below how this missing test is handled under the model assumptions laid out in Section $\ref{sec:hierarchical}$. 

\subsubsection{Likelihood factorization}

The observed-data likelihood consists of two factors, one for the individuals with baseline test ($r_i=1$) and one for the individuals without ($r_i=0$). When there is no baseline test, the missing outcomes are integrated out of the likelihood as follows
\begin{align} \label{eq::likelihood_derivation1}
	\mathcal{L}( \betaf, \sigma, \thetaf, \kappa  \mid  \Df ) &= \bigg( \prod_{i: r_i=1} f_{r,\Vvec,\yf_{\text{com}}}(r_i=1,\Vvec_i,\yf_i  \mid  \x_i, \betaf, \sigma, \thetaf, \kappa) \bigg) \nonumber\\ 
	\quad &\times \bigg( \prod_{i: r_i=0} \sum_{k=0}^{1} f_{r,\Vvec,\yf_{\text{com}}}(r_i=0,\Vvec_i, (y_{i1}=k, \yf_{i}) \mid  \x_i, \betaf, \sigma, \thetaf, \kappa) \bigg)
\end{align}

By the DGM defined in the hierarchical model in Section \ref{sec:hierarchical} and the DAG in Figure \ref{fig:dag} we can derive:
\begin{align} \label{eq::likelihood_derivation2}
	f_{r,\Vvec,\yf_{\text{com}}}(r_i,\Vvec_i,\yf_i  \mid  \x_i, \betaf, \sigma, \thetaf, \kappa)
	&= \sum_{l = 0}^{1} \int_0^\infty f_{g, t, r, \Vvec,\yf_{\text{com}}}(g_i = l,  t_i, r_i, \Vvec_i, \yf_i  \mid  \x_i, \betaf, \sigma, \thetaf, \kappa) dt_i \nonumber \\
	&= \sum_{l = 0}^{1} \int_0^\infty 
	f_g(g_i = l\mid \x_i,\thetaf)
	f_t(t_i \mid \x_i,\betaf,\sigma)
	f_r(r_i\mid \x_i) \nonumber \\
	\quad &\times f_{\Vvec,\yf_{\text{com}}}(\Vvec_i, \yf_i \mid g_i=l, r_i, t_i,\x_i,  \kappa) dt_i \nonumber \\
	&\propto \sum_{l = 0}^{1} f_g(g_i = l\mid \x_i,\thetaf) \int_0^\infty   f_t(t_i \mid \x_i,\betaf,\sigma) \nonumber \\ \quad &\times f_{\Vvec,\yf_{\text{com}}}(\Vvec_i, \yf_i \mid g_i = l, r_i, t_i,\x_i,  \kappa) dt_i \nonumber \\
	&=  (1 - \Phi( \mu_i) )  \int_0^\infty f_t(t_i \mid \x_i,\betaf,\sigma)   f_{\Vvec,\yf_{\text{com}}}(\Vvec_i, \yf_i \mid g_i = 0, r_i, t_i,\x_i, \kappa) dt_i \nonumber \\ &+ \Phi( \mu_i) \int_0^\infty f_t(t_i \mid \x_i,\betaf,\sigma)   f_{\Vvec,\yf_{\text{com}}}(\Vvec_i, \yf_i \mid g_i = 1, r_i, t_i,\x_i, \kappa) dt_i
\end{align}

We note that the term $	f_r(r_i\mid \x_i) $ is dropped in the third line of (\ref{eq::likelihood_derivation2}), because it is does not involve parameters $(\betaf, \sigma, \thetaf, \kappa)$. This is a consequence of the missing at random baseline test outcomes assumption \citep{little_statistical_2002} which will allow ignore missing baseline test outcomes by integrating them out, as discussed below. \\

The density $ f_{\Vvec,\yf_{\text{com}}}(\Vvec_i, \yf_i \mid g_i , r_i, t_i, \x_i, \kappa)$ can be viewed as the likelihood contribution (given $\kappa$) of the test outcomes and screening schedule times conditional on known values for the prevalence status $g_i$ and transition time $t_i$. These values are, however, unknown and, therefore, integrated out in (\ref{eq::likelihood_derivation2}). In the following, we derive likelihood $f_{\Vvec,\yf_{\text{com}}}$ while distinguishing between individuals with ($r_i=1$) and without ($r_i=0$) baseline test and non-prevalent ($g_i=0$) and prevalent individuals ($g_i=1$). 

\subsubsection{Likelihood contribution of the test outcomes with baseline test ($r_i=1$)}

Regardless of prevalence status, we have for individuals with baseline test complete test outcomes ($\yf_{i,\text{com}} = \yf_i$):
\begin{align} \label{eq::likelihood_derivation3}
	 f_{\Vvec,\yf_{\text{com}}}(\Vvec_i, \yf_i \mid g_i, r_i = 1, t_i, \x_i, \kappa) &= \prod_{j=1}^{c_i} 	 f_{v,y}(v_{ij}, y_{ij} \mid g_i , r_i = 1, t_i,\bar{\bm{v}}_{ij}, \x_i, \kappa) \nonumber \\
	&= \prod_{j=1}^{c_i} \sum_{m=0}^{1} f_{v,y,y^*}(v_{ij}, y_{ij}, y_{ij}^* = m \mid g_i , r_i = 1, t_i,\bar{\bm{v}}_{ij},\x_i,  \kappa) \nonumber \\
	&= \prod_{j=1}^{c_i} \sum_{m=0}^{1} \Pr( y_{ij} \mid v_{ij}, y_{ij}^* = m, \kappa )
	\Pr( y_{ij}^* = m \mid g_i , t_i, v_{ij}) \nonumber \\ &\times
	f_v(v_{ij} \mid r_i = 1, \bar{\bm{v}}_{ij}, \x_i) \nonumber \\
\end{align}

where from  (\ref{eq:def_y_star}) it follows that $\Pr( y_{ij}^* = 1 \mid g_i = 1, t_i, v_{ij}) =1 $ and $\Pr( y_{ij}^* = 1 \mid g_i = 0, t_i, v_{ij}) = \mathds{1} _{ \{v_{ij} \ge t_i \}}$, and from (\ref{eq:obsprocess}) we have
\begin{align}\label{eq::likelihood_derivation4}
\Pr( y_{ij} = y \mid v_{ij}, y_{ij}^*, \kappa )
= \bigl(\kappa y_{ij}^* \mathds{1}_{\{v_{ij} < \infty\}}\bigr)^{y}
\bigl(1 - \kappa y_{ij}^* \mathds{1}_{\{v_{ij} < \infty\}}\bigr)^{1-y},
\quad y \in \{0,1\}.
\end{align}

Since the last term in (\ref{eq::likelihood_derivation3}) does not depend on $(\betaf, \sigma, \thetaf, \kappa)$, $t_i$ or $g_i$ it can be ignored in the likelihood. The remaining term is the likelihood of the observed test outcomes and will be denoted 
\begin{align}\label{eq::likelihood_derivation5}
	 \Pr(\yf_i \mid g_i, r_i=1, t_i, \Vvec_i, \kappa) = \prod_{j=1}^{c_i} \sum_{m=0}^{1} \Pr( y_{ij} \mid v_{ij}, y_{ij}^* = m, \kappa )
	\Pr( y_{ij}^* = m \mid g_i , t_i, v_{ij}),
\end{align}

so that (\ref{eq::likelihood_derivation2}) simplifies as follows
\begin{align} \label{eq::likelihood_derivation6}
	f_{r,\Vvec,\yf_{\text{com}}}(r_i = 1,\Vvec_i,\yf_i \mid \x_i, \betaf, \sigma, \thetaf, \kappa)
	&\propto (1 - \Phi( \mu_i) )  \int_0^\infty f_t(t_i \mid \x_i,\betaf,\sigma)   \Pr(\yf_i \mid g_i = 0, r_i=1, t_i, \Vvec_i, \kappa)  dt_i \nonumber \\ &+ \Phi( \mu_i) \int_0^\infty f_t(t_i \mid \x_i,\betaf,\sigma)   \Pr(\yf_i \mid g_i = 1, r_i=1, t_i, \Vvec_i, \kappa)  dt_i.
\end{align}

We derive $ \Pr(\yf_i \mid g_i, r_i=1, t_i, \Vvec_i, \kappa) $ below for prevalent and non-prevalent conditioning.

\subsubsection{Special case: Likelihood contribution from positive baseline tests}

A special case now arises when $r_i=1$ and $y_{i1} = 1$ (positive baseline test). This implies that, in (\ref{eq::likelihood_derivation5}), $\Pr( y_{i1} = 1 \mid v_{i1}, y_{i1}^* = 0, \kappa ) = 0$ where $y_{i1}^* = 0$ with probability one in case of non-prevalence ($g_i=0$). Furthermore, $y_{i1}^* = 1$ with probability one in case of prevalence ($g_i=1$), with $\Pr( y_{i1} = 1 \mid v_{i1}, y_{i1}^* = 1, \kappa ) = \kappa$. Hence,

\begin{align}
	 \Pr(y_{i1} = 1 \mid g_i = 1, r_i=1, t_i, v_{i1}, \kappa) = \kappa
\end{align}

and 

\begin{align}
	\Pr(y_{i1} = 1 \mid g_i = 0, r_i=1, t_i, v_{i1}, \kappa) = 0
\end{align}

Therefore, if $y_{i1} = 1$, (\ref{eq::likelihood_derivation6}) simplifies as

\begin{align} \label{eq::likelihood_derivation7}
	f_{r,\Vvec,\yf_{\text{com}}}(r_i = 1,\Vvec_i = (v_{i1}), \yf_{i}=(1) \mid \x_i, \betaf, \sigma, \thetaf, \kappa)
	&\propto (1 - \Phi( \mu_i) )  \int_0^\infty f_t(t_i \mid \x_i,\betaf,\sigma)   \times 0 \ dt_i \nonumber \\ &+ \Phi( \mu_i) \int_0^\infty f_t(t_i \mid \x_i,\betaf,\sigma) \times \kappa \ dt_i \nonumber \\&= \Phi( \mu_i) \ \kappa.
\end{align}

This effectively collapses the mixture structure of the likelihood over latent $g_i$: the non-prevalent component has zero likelihood, so $g_i=1$ is known whenever the baseline test is positive (assumption of perfect test specificity), while the baseline test still contributes information through the factor $\kappa$.

\subsubsection{Likelihood contribution of the test outcomes when the baseline test is missing ($r_i=0$)}
Without a baseline test ($r_i=0$), we have that $y_{i1}$ in $\yf_{i,\text{com}} = (y_{i1},\yf_i)$ is missing and integrated out of the likelihood. From (\ref{eq::likelihood_derivation1})--(\ref{eq::likelihood_derivation2})we have

\begin{align} 
	&\sum_{k=0}^{1} f_{r,\Vvec,\yf_{\text{com}}}(r_i=0,\Vvec_i, (y_{i1}=k, \yf_i) \mid  \x_i, \betaf, \sigma, \thetaf, \kappa) \nonumber \\
	&\propto \sum_{l = 0}^{1}\Phi(\mu_i)^l (1-\Phi(\mu_i))^{(1-l)} 
	\int_0^\infty   f_t(t_i \mid \x_i,\betaf,\sigma) \sum_{k=0}^{1} f_{\Vvec,\yf_{\text{com}}}(\Vvec_i, (y_{i1}=k, \yf_i) \mid g_i = l, r_i =0, t_i,\x_i,  \kappa) dt_i
\end{align}

where 

\begin{align}\label{eq::likelihood_derivation8}
	\sum_{k=0}^{1} f_{\Vvec,\yf_{\text{com}}}(\Vvec_i,(y_{i1}=k,\yf_i)\mid g_i=l,r_i=0,t_i,\x_i,\kappa)
	&= \Bigg[\sum_{k=0}^1 f_{v,y}(v_{i1},y_{i1}=k\mid g_i=l,r_i=0,t_i,\x_i,\kappa)\Bigg] \nonumber \\ &\times
	\prod_{j=2}^{c_i} f_{v,y}(v_{ij},y_{ij}\mid g_i=l,r_i=0,t_i,\bar{\bm v}_{ij},\x_i,\kappa) \nonumber\\
	&= f_v(v_{i1}\mid r_i=0,\x_i) \nonumber \\ &\times
	\prod_{j=2}^{c_i} f_{v,y}(v_{ij},y_{ij}\mid g_i=l,r_i=0,t_i,\bar{\bm v}_{ij},\x_i,\kappa) \nonumber\\
	&\propto \prod_{j=2}^{c_i} f_{v,y}(v_{ij},y_{ij}\mid g_i=l,r_i=0,t_i,\bar{\bm v}_{ij},\x_i,\kappa),
\end{align}

where the first equality and the proportionality hold because
\begin{align}
	\sum_{k = 0}^1 f_{v,y}(v_{i1}, y_{i1} = k \mid g_i , r_i = 0, t_i,\x_i, \kappa)&= 
	\sum_{k = 0}^1 \sum_{m=0}^{1} \Pr( y_{i1} = k \mid v_{i1}, y_{i1}^* = m, \kappa )  \nonumber \\
	&\times	\Pr( y_{i1}^* = m \mid g_i , t_i, v_{i1})
	f_v(v_{i1} \mid r_i = 0, \x_i) \nonumber \\
	&= f_v(v_{i1} \mid r_i = 0, \x_i) \propto 1.
\end{align}

Therefore, using the same results as in (\ref{eq::likelihood_derivation3}) and (\ref{eq::likelihood_derivation5}) we have that regardless of prevalence status the likelihood of the observed test outcomes (i.e. all test outcomes except the baseline test) is

\begin{align}\label{eq::likelihood_derivation9}
	\Pr(\yf_i \mid g_i, r_i=0, t_i, \Vvec_i, \kappa ) = \prod_{j=2}^{c_i} \sum_{m=0}^{1} \Pr( y_{ij} \mid v_{ij}, y_{ij}^* = m, \kappa )
	\Pr( y_{ij}^* = m \mid g_i , t_i, v_{ij}).
\end{align}

Hence, when $r_i=0$, the observed test outcomes correspond to occasions indexed $j=2,\dots,c_i$, and the likelihood factorizes as a product over these observed occasions. Result (\ref{eq::likelihood_derivation9}) is equivalent to result (\ref{eq::likelihood_derivation5}) except that the factorization starts at the first moment after baseline ($j=2$) instead of at baseline which "ignores" baseline outcomes in individuals for whom the outcome is missing \citep{little_statistical_2002}. \\

As a result we have that  (\ref{eq::likelihood_derivation2}) simplifies as follows

\begin{align} \label{eq::likelihood_derivation10}
	f_{r,\Vvec,\yf_{\text{com}}}(r_i = 0,\Vvec_i,\yf_i \mid \x_i, \betaf, \sigma, \thetaf, \kappa)
	&\propto (1 - \Phi( \mu_i) )  \int_0^\infty f_t(t_i \mid \x_i,\betaf,\sigma)   \Pr(\yf_i \mid g_i = 0, r_i=0, t_i, \Vvec_i, \kappa)  dt_i \nonumber \\ &+ \Phi( \mu_i) \int_0^\infty f_t(t_i \mid \x_i,\betaf,\sigma)   \Pr(\yf_i \mid g_i = 1, r_i=0, t_i, \Vvec_i, \kappa)  dt_i.
\end{align}

We derive $ \Pr(\yf_i \mid g_i, r_i=0, t_i, \Vvec_i, \kappa) $ below for prevalent and non-prevalent conditioning.

\subsubsection{Likelihood contribution of the test outcomes from non-prevalent individuals}

We now derive $\Pr(\yf_i \mid g_i = 0, r_i = 1, t_i, \Vvec_i, \kappa)$ and $\Pr(\yf_i \mid g_i = 0, r_i = 0, t_i, \Vvec_i, \kappa)$ in (\ref{eq::likelihood_derivation6})  and (\ref{eq::likelihood_derivation10}) by plugging probability (\ref{eq::likelihood_derivation4}) together with  $\Pr( y_{ij}^* = 1 \mid g_i = 0, t_i, v_{ij}) = \mathds{1} _{ \{v_{ij} \ge t_i \}}$  into (\ref{eq::likelihood_derivation5})  and (\ref{eq::likelihood_derivation9}). \\ 

We first note that with $g_i=0$ (non-prevalence) all baseline tests that are done ($r_i=1$) are necessarily negative (no false positives). Formally, $\Pr( y_{i1} = 0 \mid v_{i1}=0, y_{i1}^* = 0, \kappa ) = 1$ with $\Pr( y_{i1}^* = 0 \mid g_i=0 , t_i, v_{i1} = 0) = 1$. Therefore:
\begin{align}\label{eq::likelihood_derivation11}
	\Pr(\yf_i \mid g_i=0, r_i=1, t_i, \Vvec_i, \kappa ) &= \Pr(\yf_i \mid g_i=0, r_i=0, t_i, \Vvec_i, \kappa ) \nonumber \\ 
	&=	\prod_{j=2}^{c_i}  \bigg[ \mathds{1} _{ \{v_{ij} \ge t_i \}} (1-\kappa)^{(1-y_{ij})\mathds{1} _{\{v_{ij} < \infty\}} } \kappa^{y_{ij}} + \mathds{1} _{ \{v_{ij} < t_i \}} (1-y_{ij})\bigg] \nonumber \\
	&= \bigg[ \prod_{j : v_{ij} \ge t_i } (1-\kappa)^{(1-y_{ij})\mathds{1} _{\{v_{ij} < \infty\}} } \kappa^{y_{ij}} \bigg] \times \bigg[ \prod_{j : v_{ij} < t_i } (1-y_{ij}) \bigg] \nonumber \\
	&= \begin{cases}
		\prod_{j : v_{ij} \ge t_i } (1-\kappa)^{(1-y_{ij})\mathds{1} _{\{v_{ij} < \infty\}} } \kappa^{y_{ij}} \quad &\text{if} \ v_{ic_i} \ge t_i   \\
		0\quad &\text{if} \ v_{ic_i} < t_i.
	\end{cases}
\end{align}

The restriction $\Pr(\yf_i \mid g_i = 0, t_i, \Vvec_i, \kappa ) =0$ if $v_{ic_i} < t_i $ emerges, because
\begin{align*}
	\prod_{j : v_{ij} < t_i } (1-y_{ij}) = (1 - 0) \times (1-0) \times \dots \times (1-y_{ic_i}) = 0 \quad \text{if}\ v_{ic_i} < t_i,
\end{align*} 

since whenever $ v_{ic_i} < t_i$ we have that $y_{ic_i}=1$ (note that $v_{ic_i} < t_i $ cannot occur in case of right censoring because then $v_{ic_i} = \infty$). In other words, the situation that $v_{ic_i} < t_i$ while a test was positive (event) has zero likelihood and hence cannot occur in the data (no false positive tests allowed). Technically, this restriction is relevant, because (\ref{eq::likelihood_derivation2}) integrates $t_i$ in $\Pr(\yf_i \mid g_i = 0, t_i, \Vvec_i, \kappa )$ unrestricted over its full support $(0,\infty)$. \\

For $v_{ic_i} \ge t_i $, we have
\begin{align} \label{eq::likelihood_derivation12}
	\Pr(\yf_i \mid g_i = 0, t_i, \Vvec_i, \kappa ) &= (1-\kappa)^{\sum_{j : v_{ij} \ge t_i }(1-y_{ij})\mathds{1} _{\{v_{ij} < \infty\}} } \kappa^{\sum_{j : v_{ij} \ge t_i} y_{ij}} \nonumber \\
	&= (1-\kappa)^{ [\sum_{j : v_{ij} \ge t_i }1 ] - 1 } \kappa^{\sum_{j : v_{ij} \ge t_i} y_{ij}} \nonumber \\
	&= (1-\kappa)^{\sum_{j =1 }^ {c_i} [ \mathds{1} _{\{v_{ij} \ge t_i\}} ]- 1 } \kappa^{y_{ic_i}}
	\nonumber \\
	&= (1-\kappa)^{ m_i } \kappa^{y_{ic_i}},
\end{align}

with $m_i = \sum_{j =1 }^ {c_i} [ \mathds{1} _{\{v_{ij} \ge t_i\}} ]- 1 = \sum_{j=1}^{c_i-1} y^*_{ij}$ the number of falsely negative screening tests until $v_{ic_i}$. For the second equation, we have used from (\ref{eq::likelihood_derivation4}) the fact that $y_{ij} = 0$ whenever $y^ *_{ij} = 0$ (i.e., $v_{ij} < t_i$). For the third equation,  we have used the fact that $y_{ij} = 0$ for all $j$  in $\yf_i$ except for the last element $y_{ic_i}$. If $y_{ic_i} = 1$ (event at $v_{ic_i} < \infty$), $1-y_{ic_i} = 0$. If  $y_{ic_i} = 0$ we have right censoring with $v_{ic_i} = \infty$, so that $(1-y_{ic_i})\mathds{1} _{\{v_{ic_i} < \infty\}} = 0$. Hence
\begin{align} \label{eq::likelihood_derivation13}
	\Pr(\yf_i \mid g_i = 0, t_i, \Vvec_i, \kappa ) =
	\begin{cases}
		(1-\kappa)^{ m_i } \kappa^{y_{ic_i}}  &\text{if} \ v_{ic_i} \ge t_i   \\
		0\quad &\text{if} \ v_{ic_i} < t_i .
	\end{cases}
\end{align}

A useful result is now that (\ref{eq::likelihood_derivation13}) can be written as a sum
\begin{align} \label{eq::likelihood_derivation14}
	\Pr(\yf_i \mid g_i = 0, t_i, \Vvec_i, \kappa ) = \kappa^{y_{ic_i}} \sum_{j=1}^{c_i-1} (1-\kappa)^{(c_i-j-1)} \mathds{1} _{ \{v_{ij} < t_i \le v_{ij+1}\} },
\end{align}

where we emphasize again that we consider either individuals without baseline test ($r_i=0$) or negative baseline test $y_{i1}=0$, so that $c_i>1$. For the special case when $y_{i1}=1$, see (\ref{eq::likelihood_derivation7}). \\

After plugging back into (\ref{eq::likelihood_derivation6}) we obtain closed form integrals over $t_i$
\begin{align} \label{eq::likelihood_derivation15}
	&\int_0^\infty f_t(t_i \mid \x_i,\betaf,\sigma)  \Pr(\yf_i \mid g_i = 0, r_i=1, t_i, \Vvec_i, \kappa ) dt_i \nonumber \\
	&= \int_0^\infty f_t(t_i \mid \x_i,\betaf,\sigma) \kappa^{y_{ic_i}} \sum_{j=1}^{c_i-1} (1-\kappa)^{(c_i-j-1)} \mathds{1} _{ \{v_{ij} < t_i \le v_{ij+1}\} }  dt_i \nonumber \\
	&=  \kappa^{y_{ic_i}} \sum_{j=1}^{c_i-1} (1-\kappa)^{(c_i-j-1)} \big[ F_t(v_{ij+1} \mid \x_i,\betaf,\sigma) - F_t(v_{ij} \mid \x_i,\betaf,\sigma) \big].
\end{align}

The same result follows for $r_i=0$ in (\ref{eq::likelihood_derivation10}).

\subsubsection{Likelihood contribution of the test outcomes from prevalent individuals}

We now derive $\Pr(\yf_i \mid g_i = 1, r_i = 1, t_i, \Vvec_i, \kappa)$ and $\Pr(\yf_i \mid g_i = 1, r_i = 0, t_i, \Vvec_i, \kappa)$ in (\ref{eq::likelihood_derivation6})  and (\ref{eq::likelihood_derivation10}) for prevalent individuals by plugging in probability (\ref{eq::likelihood_derivation4}) together with $\Pr( y_{ij}^* = 1 \mid g_i = 1, t_i, v_{ij}) =1 $ in (\ref{eq::likelihood_derivation5})  and (\ref{eq::likelihood_derivation9}). \\

In case of prevalence ($g_i=1$) and baseline tests ($r_i=1$) we have
\begin{align}\label{eq::likelihood_derivation16}
	\Pr(\yf_i \mid g_i = 1, r_i=1, t_i, \Vvec_i, \kappa ) &=
	\prod_{j=1}^{c_i}  \kappa^{y_{ij}} (1-\kappa)^{(1-y_{ij})\mathds{1} _{\{v_{ij} < \infty\}} } = \kappa^{y_{ic_i}}(1-\kappa)^{(c_i-1)}.
\end{align}

In case of prevalence ($g_i=1$) and no baseline tests ($r_i=0$), we have
\begin{align}\label{eq::likelihood_derivation17}
	\Pr(\yf_i \mid g_i = 1, r_i=0, t_i, \Vvec_i, \kappa ) &=
	\prod_{j=2}^{c_i}  \kappa^{y_{ij}} (1-\kappa)^{(1-y_{ij})\mathds{1} _{\{v_{ij} < \infty\}} } = \kappa^{y_{ic_i}}(1-\kappa)^{(c_i-2)}.
\end{align}

Combining these results we obtain the general result
\begin{align}\label{eq::likelihood_derivation18}
	\Pr(\yf_i \mid g_i = 1, r_i, t_i, \Vvec_i, \kappa ) &= \kappa^{y_{ic_i}}(1-\kappa)^{(c_i+r_i-2)}.
\end{align}

After plugging back into (\ref{eq::likelihood_derivation6}) and (\ref{eq::likelihood_derivation10}), we may integrate $t_i$ out, obtaining
\begin{align} \label{eq::likelihood_derivation19}
	&\int_0^\infty f_t(t_i \mid \x_i,\betaf,\sigma)  \Pr(\yf_i \mid g_i = 1, r_i, t_i, \Vvec_i, \kappa ) dt_i \nonumber \\
	&= \kappa^{y_{ic_i}}(1-\kappa)^{(c_i+r_i-2)},
\end{align}

where we give a general expression for both $r_i=1$ and $r_i=0$. \\

Note that in the special case of a positive baseline test (i.e. $r_i=1$, $c_i=1$, $y_{i1}=1$ so $g_i=1$), equation (\ref{eq::likelihood_derivation18}) reduces to
$\Pr(\yf_i \mid g_i=1,r_i=1,t_i,\Vvec_i,\kappa) = \kappa$, which is consistent with result (\ref{eq::likelihood_derivation7}). In this situation the non-prevalent component of (\ref{eq::likelihood_derivation6}) has zero likelihood, so the mixture over $g_i$ collapses.

\newpage
\subsubsection{Full expression for the observed-data likelihood}

We are now ready to write down a complete expression for the observed-data likelihood (\ref{eq::likelihood_derivation1})--(\ref{eq::likelihood_derivation2}). Substituting (\ref{eq::likelihood_derivation6}) and (\ref{eq::likelihood_derivation10}) into (\ref{eq::likelihood_derivation2}) and factoring in result (\ref{eq::likelihood_derivation7}), we obtain
\begin{align} 
	\mathcal{L}( \betaf, \sigma, \thetaf, \kappa  \mid  \Df ) &\propto \prod_{i:r_i=1, y_{i1}=0} \bigg[ (1 - \Phi( \mu_i) )  \int_0^\infty f_t(t_i \mid \x_i,\betaf,\sigma)   \Pr(\yf_i \mid g_i = 0, r_i=1, t_i, \Vvec_i, \kappa)  dt_i \nonumber \\ &+ \Phi( \mu_i) \int_0^\infty f_t(t_i \mid \x_i,\betaf,\sigma)   \Pr(\yf_i \mid g_i = 1, r_i=1, t_i, \Vvec_i, \kappa)  dt_i \bigg] \nonumber \\
	&\times \prod_{i:r_i=0}  \bigg[(1 - \Phi( \mu_i) )  \int_0^\infty f_t(t_i \mid \x_i,\betaf,\sigma)   \Pr(\yf_i \mid g_i = 0, r_i=0, t_i, \Vvec_i, \kappa)  dt_i \nonumber \\ &+ \Phi( \mu_i) \int_0^\infty f_t(t_i \mid \x_i,\betaf,\sigma)   \Pr(\yf_i \mid g_i = 1, r_i=0, t_i, \Vvec_i, \kappa)  dt_i \bigg] 
	\nonumber \\
	&\times  \prod_{i:r_i=1, y_{i1}=1} \Phi( \mu_i) \ \kappa
\end{align}

Now substituting results (\ref{eq::likelihood_derivation15}) and (\ref{eq::likelihood_derivation19}) we obtain 	that $\mathcal{L}( \betaf, \sigma, \thetaf, \kappa  \mid  \Df )$ is proportional to
\begin{align}
	\prod_{i:r_i=1, y_{i1}=0} &\bigg[ (1 - \Phi( \mu_i) )  \kappa^{y_{ic_i}} \sum_{j=1}^{c_i-1} (1-\kappa)^{(c_i-j-1)} \big[ F_t(v_{ij+1} \mid \x_i,\betaf,\sigma) - F_t(v_{ij} \mid \x_i,\betaf,\sigma) \big] \nonumber \\ &+ \Phi( \mu_i)  \kappa^{y_{ic_i}}(1-\kappa)^{(c_i-1)} \bigg] \nonumber \\
	&\times \prod_{i:r_i=0} \bigg[ (1 - \Phi( \mu_i) )  \kappa^{y_{ic_i}} \sum_{j=1}^{c_i-1} (1-\kappa)^{(c_i-j-1)} \big[ F_t(v_{ij+1} \mid \x_i,\betaf,\sigma) - F_t(v_{ij} \mid \x_i,\betaf,\sigma) \big] \nonumber \\ &+ \Phi( \mu_i) \kappa^{y_{ic_i}}(1-\kappa)^{(c_i-2)} \bigg] 
	\nonumber \\
	&\times  \prod_{i:r_i=1, y_{i1}=1} \Phi( \mu_i)  \ \kappa
\end{align}

which is  more compactly written as
\begin{align}
	 \prod_{i \in \mathcal{I}_0} &\bigg[ (1 - \Phi( \mu_i) )  \kappa^{y_{ic_i}} \sum_{j=1}^{c_i-1} (1-\kappa)^{(c_i-j-1)} \big[ F_t(v_{ij+1} \mid \x_i,\betaf,\sigma) - F_t(v_{ij} \mid \x_i,\betaf,\sigma) \big] \nonumber \\ &+ \Phi( \mu_i)  \kappa^{y_{ic_i}}(1-\kappa)^{(c_i+r_i-2)} \bigg] \times  \prod_{i \in \mathcal{I}_1} \Phi( \mu_i)  \ \kappa.
\end{align}

where $\mathcal{I}_0$ denotes the set of all individuals $i$ with a negative baseline test ($r_i=1$ and $y_{i1}=0$) or a missing baseline test ($r_i=0$) and $\mathcal{I}_1$ denotes the set of all individuals $i$ with a positive baseline test ($r_i=1$ and $y_{i1}=1$).

\newpage

\subsection{Gibbs sampler}

\subsubsection{Notation}

To derive the full conditional distributions used in the Gibbs sampler, we will apply a full factorization of joint density of the random variables and the data. In the factorization, the joint likelihood of the test outcomes and screening times (\ref{eq::likelihood_derivation2}) will appear, but depending on whether a baseline test outcome is observed, the likelihood is either integrated over missing outcomes or not integrated (i.e., all outcomes observed); see Supplemental Material Section~\ref{sec:supplement_proofs_obsll}. To facilitate the subsequent derivations, we introduce a unifying notation for the joint likelihood of screening times and the observed test outcomes as follows:
\begin{align}\label{eq:mcmc_notation1}
	\tilde{f}_{\Vvec,\yf}(\Vvec_i, \yf_i \mid g_i , r_i, t_i,\x_i,  \kappa) =
	\begin{cases}
		f_{\Vvec,\yf_{\text{com}}}(\Vvec_i, \yf_i \mid g_i , r_i=1, t_i,\x_i,  \kappa), & \text{if} \ r_i=1 \\
		\sum_{k=0}^{1}f_{\Vvec,\yf_{\text{com}}}(\Vvec_i, (y_{i1},\yf_i) \mid g_i , r_i=0, t_i,\x_i,  \kappa), & \text{if} \ r_i=0
	\end{cases}
\end{align}

The two likelihoods for $r_i=1$ and $r_i=0$ are derived in equation (\ref{eq::likelihood_derivation3}) and (\ref{eq::likelihood_derivation9}), respectively. The special case of a positive baseline test is given in (\ref{eq::likelihood_derivation7}). Specifically,
\begin{align}\label{eq:mcmc_notation2}
	\tilde{f}_{\Vvec,\yf}(\Vvec_i, \yf_i \mid g_i , r_i, t_i,\x_i,  \kappa) \propto
	\begin{cases}
		\Pr(\yf_i \mid g_i, r_i=1, t_i, \Vvec_i, \kappa), & \text{if} \ r_i=1 \\
		\Pr(\yf_i \mid g_i, r_i=0, t_i, \Vvec_i, \kappa), & \text{if} \ r_i=0
	\end{cases}
\end{align}

where $	\Pr(\yf_i \mid g_i=0, r_i=1, t_i, \Vvec_i, \kappa) = 	\Pr(\yf_i \mid g_i=0, r_i=0, t_i, \Vvec_i, \kappa)$ is given in equation (\ref{eq::likelihood_derivation13}) with its rewritten sum representation in (\ref{eq::likelihood_derivation14}) and $	\Pr(\yf_i \mid g_i=1, r_i, t_i, \Vvec_i, \kappa) $ is given in (\ref{eq::likelihood_derivation18}). \\

\newpage
\subsubsection{Full conditional distribution of $t_i$} \label{sec:supplement_proofs_fcx}
We derive the full conditional distribution of $t_i$. We first consider the non-prevalent case ($g_i =0$), i.e.
\begin{align}
	f_t(t_i  \mid  {\D}_i, g_i = 0, \betaf, \sigma, \thetaf, \kappa ) 
	&\propto f_{g, r, t,\Vvec, \yf}(g_i = 0, r_i, t_i,\Vvec_i,  \yf_i  \mid \x_i, \betaf, \sigma, \thetaf, \kappa )  \nonumber \\
	&=	f_g(g_i = 0\mid \x_i, \thetaf) f_r(r_i \mid \x_i) f_t(t_i \mid \x_i, \betaf, \sigma) \tilde{f}_{\Vvec, \yf}( \Vvec_i, \yf_i \mid g_i = 0, r_i, t_i, \x_i, \kappa)  \nonumber  \\
	&\propto \ f_t(t_i  \mid  \x_i, \betaf, \sigma) \Pr(\yf_i  \mid  g_i=0, r_i, t_i, \Vvec_i, \kappa) \nonumber \\
	&= \sum_{j=1}^{c_i-1} \ f_t(t_i \mid  \x_i, \betaf, \sigma) \ \big( \kappa^{y_{ic_i}} (1-\kappa)^{(c_i-j-1)} \mathds{1} _{ \{v_{ij} < t_i \le v_{ij+1}\} } \big) \nonumber \\
	&= \sum_{j=1}^{c_i-1} f_t(t_i \mid  v_{ij} < t_i \le v_{ij+1}, \x_i, \betaf, \sigma) \nonumber \\ &\times \big[ F_t(v_{ij+1}  \mid \x_i, \betaf, \sigma) - F_t(v_{ij}  \mid   \x_i, \betaf, \sigma) \big] \ \big( \kappa^{y_{ic_i}} (1-\kappa)^{(c_i-j-1)} \big)\nonumber \\
	&= \sum_{j=1}^{c_i-1} \tilde{\omega}_{ij}  f_t(t_i \mid  v_{ij} < t_i \le v_{ij+1}, \x_i, \betaf, \sigma), 
\end{align}

where
\begin{align} \label{eq::tilde_omega}
	\tilde{\omega}_{ij} = \kappa^{y_{ic_i}}(1-\kappa)^{(c_i-j-1)} \big[ F_t(v_{ij+1} \mid \x_i, \betaf, \sigma) - F_t(v_{ij} \mid \x_i, \betaf, \sigma) \big] .
\end{align}

The second equation follows from the factorization into the joint data generating densities through the hierarchical model described in Section \ref{sec:hierarchical} and the DAG (Figure \ref{fig:dag}) together with notation (\ref{eq:mcmc_notation1}). The third equation (proportionality) follows due to independence of various terms from $t_i$ and after applying the result (\ref{eq:mcmc_notation2}). The fourth equation substitutes result (\ref{eq::likelihood_derivation14}), the sum representation of the likelihood of the test outcomes under $g_i=0$. The fifth equation uses the fact that $\mathds{1} _{ \{v_{ij} < t_i \le v_{ij+1}\} } f_t(t_i \mid  \x_i, \betaf, \sigma)$ is an unnormalized truncated distribution. The last two factors in the fifth equation can then be recognized as the weights $\tilde{\omega}_{ij} $ of an unnormalized mixture of truncated distributions of $t_i$; see the sixth equation. After normalizing the weights through
\begin{align}
	\omega_{ij} = \frac{\tilde{\omega}_{ij}}{\sum_{l=1}^{c_i-1}\tilde{\omega}_{il}}
\end{align}

the result in (\ref{eq::fc.x.explicit}) is obtained, i.e. 
\begin{align}
	f_t(t_i  \mid  {\D}_i, g_i = 0, \betaf, \sigma, \kappa ) = \sum_{j=1}^{c_i-1} \omega_{ij}  f_t(t_i \mid  v_{ij} < t_i \le v_{ij+1}, \x_i, \betaf, \sigma)
\end{align}

Finally, in the prevalent case, when $g_i = 1$, we have after substituting (\ref{eq::likelihood_derivation18})
\begin{align}
	f_t(t_i  \mid  {\D}_i, g_i = 1, \betaf, \sigma, \thetaf, \kappa ) &\propto  f_t(t_i  \mid  \x_i, \betaf, \sigma) \big(\kappa^{y_{ic_i}} (1-\kappa)^{(c_i+r_i-2)} \big) \propto f_t(t_i  \mid \x_i, \betaf, \sigma),
\end{align}

irrespective of $r_i=0$ or $r_i=1$. Hence, $t_i$ is updated uninformatively when $g_i=1$.

\newpage

\subsubsection{Full conditional distribution and collapsed distribution of $g_i$} \label{sec:supplement_proofs_fcg}

We begin by noting again, as in Section \ref{sec:notation}, that $g_i=1$ is known if $r_i=1$ and $y_{i1} =1 $ (positive baseline test). In all other cases $g_i$ is missing (latent) and augmented as part of the Gibbs sampler.  We now derive the full conditional distribution of (missing) $g_i$. We have

\begin{align}
	f_g(g_i \mid {\D}_i , t_i, \betaf, \sigma, \thetaf, \kappa ) &\propto f_{g, r, t, \Vvec, \yf}(g_i, r_i, t_i,\Vvec_i,  \yf_i  \mid \x_i, \betaf, \sigma, \thetaf, \kappa )  \nonumber \\
	&=	f_g(g_i \mid \x_i, \thetaf) f_r(r_i \mid \x_i) f_t(t_i \mid \x_i, \betaf, \sigma) \tilde{f}_{\Vvec, \yf}( \Vvec_i, \yf_i \mid g_i, r_i, t_i, \x_i, \kappa) \nonumber\\
	&\propto	f_g(g_i \mid \x_i, \thetaf) \Pr(\yf_i  \mid  g_i, r_i, t_i, \Vvec_i, \kappa) \nonumber\\
	&= \begin{cases}
		0  \quad &\text{if} \ g_i = 0 \ \text{and} \ t_i > v_{ic_i} \\
		\big(1-\Phi(\mu_i)\big) \kappa^{y_{ic_i}} (1-\kappa)^{m_i} \quad  &\text{if} \ g_i = 0 \ \text{and} \ t_i \le v_{ic_i} \\
		\Phi(\mu_i) \kappa^{y_{ic_i}} (1-\kappa)^{(c_i+r_i-2)} \quad &\text{if} \ g_i = 1.
	\end{cases} 
\end{align}

The second equation follows from the factorization into the joint data generating densities through the hierarchical model described in Section \ref{sec:hierarchical} and the DAG (Figure \ref{fig:dag}) together with notation (\ref{eq:mcmc_notation1}). The third equation (proportionality) follows due to independence of various terms from $g_i$ and after applying the result (\ref{eq:mcmc_notation1}). The fourth equation substitutes results (\ref{eq::likelihood_derivation13}) and (\ref{eq::likelihood_derivation18}). Hence, after normalizing,
\begin{align}\label{eq::app_fc_g_on_t}
	&f_g(g_i = 1 \mid  {\D}_i, t_i, \thetaf, \kappa ) = 
	\begin{cases}
		\frac{ \Phi(\mu_i) \kappa^{y_{ic_i}}  (1-\kappa)^{(c_i+r_i-2)} }{ \Phi(\mu_i) \kappa^{y_{ic_i}}  (1-\kappa)^{(c_i+r_i-2)}  + \big(1- \Phi(\mu_i) \big)  \kappa^{y_{ic_i}}  (1-\kappa)^{m_i} } \quad &\text{if} \  t_i \le v_{ic_i} \\
		1 \quad &\text{if} \ t_i > v_{ic_i}.
	\end{cases}
\end{align}

As can be seen from (\ref{eq::app_fc_g_on_t}), $g_i$ is updated to one with probability one (deterministic update), except when $t_i \le v_{ic_i}$ (stochastic update). The determinacy can cause inefficiencies in the Gibbs sampler due to the following feedback. Suppose $g^{(k)}_i=1$ at any point in the Gibbs sampler, then $t_i$ is updated uninformatively in $(0,\infty)$ so that $t_i^{(k+1)} > v_{ic_i}$ has positive probability. In that case, $g_i^{(k+1)}=1$ with probability one in the next step. This dependency continues unless in a future draw $t_i^{(k')} \le v_{ic_i}$, in which case $g_i^{(k')}=0$ has positive probability again. \\

To avoid this problem, we integrate additionally over the latent times $t_i$ and obtain 
\begin{align}
	f_g(g_i \mid {\D}_i ,  \betaf, \sigma, \thetaf, \kappa)  &\propto \int_{0}^{\infty} f_{g, r, t, \Vvec, \yf}(g_i, r_i, t_i,\Vvec_i,  \yf_i  \mid \x_i,  \betaf, \sigma, \thetaf, \kappa) dt_i \nonumber \\
	&= \int_{0}^{\infty} f_g(g_i \mid \x_i, \thetaf) f_r(r_i \mid \x_i) f_t(t_i \mid \x_i, \betaf, \sigma) \tilde{f}_{\Vvec, \yf}( \Vvec_i, \yf_i \mid g_i, r_i, t_i, \x_i, \kappa) dt_i \ \nonumber \\
	&\propto \int_{0}^{\infty} f_g(g_i \mid \x_i, \thetaf) f_t(t_i \mid \x_i, \betaf, \sigma) 
	\Pr(\yf_i  \mid  g_i, r_i, t_i, \Vvec_i, \kappa)dt_i
\end{align}

Hence, $f_g(g_i = 0 \mid {\D}_i ,  \betaf, \sigma, \thetaf, \kappa)$ is proportional to
\begin{align}
	&\big (1- \Phi(\mu_i)) \int_0^{\infty} \sum_{j=1}^{c_i-1} \kappa^{y_{ic_i}} (1-\kappa)^{(c_i-j-1)} \mathds{1} _{ \{v_{ij} < t_i \le v_{ij+1}\} } f_t(t_i \mid  \x_i, \betaf, \sigma) dt_i \nonumber \\
	&= (1- \Phi(\mu_i)) \sum_{j=1}^{c_i-1} \kappa^{y_{ic_i}} (1-\kappa)^{(c_i-j-1)} \big[ F_t(v_{ij+1}  \mid \x_i, \betaf, \sigma) - F_t(v_{ij}  \mid  \x_i, \betaf, \sigma) \big] \nonumber \\
	&= (1- \Phi(\mu_i)) \sum_{j=1}^{c_i-1} \tilde{\omega}_{ij} 
\end{align}

where in the first equation (proportionality) we used result (\ref{eq::likelihood_derivation14}) and in the last equation definition (\ref{eq::tilde_omega}). Furthermore, $f_g(g_i = 1 \mid {\D}_i ,  \betaf, \sigma, \thetaf, \kappa)$ is proportional to
\begin{align}
	\Phi(\mu_i) \int_0^{\infty}  \kappa^{y_{ic_i}} (1-\kappa)^{(c_i+r_i-2)} f_t(t_i \mid   \x_i, \betaf, \sigma) dt_i  = \Phi(\mu_i) \kappa^{y_{ic_i}} (1-\kappa)^{(c_i+r_i-2)}
\end{align}

so that after normalization
\begin{align}
	f_g(g_i = 1  \mid  {\D}_i,  \betaf, \sigma, \thetaf, \kappa) = \frac{ \Phi(\mu_i) \kappa^{y_{ic_i}} (1-\kappa)^{(c_i+r_i-2)}}
	{ \Phi(\mu_i) \kappa^{y_{ic_i}} (1-\kappa)^{(c_i+r_i-2)}+ (1- \Phi(\mu_i)) \sum_{l=1}^{c_i-1} \tilde{\omega}_{il} }.
\end{align}

\newpage
\subsubsection{Updating  the parameters of $t_i$} \label{sec:appendix_par_t}

We begin by noticing that the full conditional distribution of ($\betaf, \sigma$) is proportional to a distribution of ($\betaf, \sigma$) that, conditional on ($\tf, \mathbf{X}$), is independent of the other model variables, specifically:
\begin{align}\label{eq::compldat.pst.x}
	f_{\betaf,\sigma} \big(\betaf, \sigma \mid {\Df}, \g, \tf, \thetaf, \kappa \big) 
	&\propto \prod_{i = 1}^n f_{g, r , t, \Vvec, \yf, \betaf, \sigma}(  g_i, r_i , t_i, \Vvec_i, \yf_i, \betaf, \sigma \mid \x_i, \thetaf, \kappa ) \nonumber \\
	&= \prod_{i = 1}^n \bigg[ f_g(g_i \mid \x_i, \thetaf) f_r(r_i \mid \x_i) f_t(t_i \mid \x_i, \betaf, \sigma) \tilde{f}_{\Vvec,\yf}( \Vvec_i, \yf_i \mid g_i, r_i, t_i, \x_i, \kappa) \bigg] \nonumber \\
	& \quad \times \pi(\betaf, \sigma \mid \tau_{\beta}, \lambda) \nonumber \\
	&\propto \prod_{i = 1}^n \bigg[  f_t(t_i \mid \x_i, \betaf, \sigma) \bigg]  \pi(\betaf \mid \tau_{\beta}) \ \pi_{\sigma}(\sigma \mid \lambda) \nonumber\\
	&= \mathcal{L}( \betaf, \sigma \mid \tf, \mathbf{X}) \ \pi(\betaf \mid \tau_{\beta}) \ \pi(\sigma \mid \lambda) \nonumber \\
	&\propto f_{\betaf,\sigma} (\betaf, \sigma \mid \tf, \mathbf{X} ) 
\end{align}

where $ \mathcal{L}( \betaf, \sigma \mid \tf, \mathbf{X}) $ is the complete-data likelihood, so that $ f_{\betaf,\sigma} (\betaf, \sigma \mid \tf, \mathbf{X} ) $ is the complete-data posterior of ($\betaf,\sigma$). The specific form of this posterior follows from the distributional assumption on the transition times $t_i$, which is implied by the distribution of the AFT residuals $\epsilon_i$ in model (\ref{eq::mod_x}). As an example, we consider a Weibull model. Then the residuals are extreme value distributed with $f_{\epsilon}(\epsilon_i) = \exp(\epsilon_i - \exp(\epsilon_i))$ so that, by a change of variable, it follows that $t_i$ has a Weibull density, 
\begin{align}
	f_t(t_i\mid\betaf, \sigma, \x_i ) = \frac{\eta}{\gamma_i} \bigg(\frac{t_i}{\gamma_i} \bigg)^{(\eta-1)}\exp\bigg(-\bigg(\frac{t_i}{\gamma_i}\bigg)^\eta \ \bigg),
\end{align}

where $\eta = \sigma^{-1}>0$ and $\gamma_i = \exp(\x_i^T\betaf)$. The log of the posterior (\ref{eq::compldat.pst.x}) with priors as defined in Section \ref{sec:priors} is then proportional to
\begin{align}
	n\log(\eta) + \sum_{i=1}^{n}\bigg[ (\eta-1) \log(t_i) - \eta \log(\gamma_i) - \bigg(\frac{t_i}{\gamma_i} \bigg)^\eta \ \bigg]- \frac{1}{2} \bigg[ \sum_{j=1}^{p}\bigg( \frac{\beta_{j}^2}{\tau_{\beta}}\bigg) + \frac{\eta^{-2}}{\lambda} \ \bigg] .
\end{align} 

In general, (\ref{eq::compldat.pst.x}) does not follow a known distribution and, therefore, we use a Metropolis sampler to generate Markov Chain Monte Carlo (MCMC) samples. The Metropolis sampler applies a multivariate normal proposal (jumping) distribution that is centred at the previous draw $(\betaf^{(k)},\sigma^{(k)})$ and has a diagonal variance-covariance matrix $\Sigma$ chosen by the user. The proposal variance is a tuning parameter that needs to be calibrated such that the proposed jumps have an acceptance probability of approximately 23\% \citep{gelman_bayesian_2013}.

\newpage
\subsubsection{Updating the parameters of $g_i$} \label{sec:appendix_par_g}

Gibbs sampler step (\ref{eq::fc.parw}) asks to obtain repeated draws from the full conditional distribution of $\thetaf$ as follows:
\begin{align}\label{eq::fc_betag_on_g}
	f_{\thetaf} (\thetaf \mid  {\Df}, \g, \tf, \betaf, \sigma, \kappa  ) &\propto \prod_{i = 1}^n f_{ g, r , t, \Vvec, \yf, \thetaf}( g_i, r_i , t_i, \Vvec_i, \yf_i, \thetaf \mid \x_i, \betaf, \sigma, \kappa ) \nonumber \\
	&= \prod_{i = 1}^n \bigg[ f_g(g_i \mid \x_i, \thetaf) f_r(r_i \mid \x_i) f_t(t_i \mid \x_i, \betaf, \sigma) \tilde{f}_{\Vvec,\yf}( \Vvec_i, \yf_i \mid g_i, r_i, t_i, \x_i, \kappa) \bigg] \nonumber \\
	& \quad \times \pi(\thetaf \mid \tau_{\theta}) \nonumber \\
	&=  \prod_{i = 1}^n \bigg[ f_g(g_i \mid \x_i, \thetaf) \bigg] \ \pi(\thetaf \mid \tau_{\theta}) \nonumber \\
	&=  \prod_{i = 1}^n \bigg[ \Phi(\mu_i)^{g_i} (1-\Phi(\mu_i))^{(1-g_i)}\bigg] \ \pi(\thetaf \mid \tau_{\theta}) \nonumber \\
	&= f_{\thetaf} (\thetaf \mid \g, \mathbf{X}),
\end{align}

which is the complete-data posterior of $\thetaf$ given augmented $\g$. Sampling from this posterior could be achieved by a Metropolis step in similar manner as described for the parameters $(\betaf, \sigma)$ in Section \ref{sec:appendix_par_t} of the Supplemental Material. However, \texttt{BayesPIM} achieves more efficient, conjugate normal updating through exploiting the latent variable formulation of the probit model. As described in Section \ref{sec:AFTmixturemodel}, we define a new latent variable
\begin{align}
	w_i = \mu_i + \phi_i \quad, \phi_i \ \sim \ N(0,1)
\end{align}

and
\begin{align}\label{eq::gw_constr}
	g_i = \mathds{1}_{\{w_i>0\}},
\end{align}

so that $\Pr(g_i=1\mid  \x_i, \thetaf)=\Pr(w_i>0\mid \x_i, \thetaf)=\Phi(\mu_i)$. Under this model, equation (\ref{eq::fc_betag_on_g}) can be viewed as integral over latent $w_i$, i.e.
\begin{align}
	f_{\thetaf} (\thetaf \mid  {\Df}, \g, \tf, \betaf, \sigma, \kappa  ) &\propto \prod_{i = 1}^n \int_{-\infty}^{\infty} f_{ g, r , t, \Vvec, w, \yf, \thetaf} (  g_i, r_i , t_i, \Vvec_i, w_i, \yf_i, \thetaf \mid \x_i, \betaf, \sigma, \kappa ) dw_i \nonumber \\
	&\propto  \prod_{i = 1}^n \bigg[ \int_{-\infty}^{\infty}  f_w(w_i \mid \x_i, \thetaf) (\mathds{1}_{\{w_i>0\}})^{g_i} (\mathds{1}_{\{w_i\le0\}})^{(1-g_i)} dw_i\bigg] \pi(\thetaf \mid \tau_{\theta}), 
\end{align}

which directly equates to (\ref{eq::fc_betag_on_g}) because $f_w(w_i \mid \x_i, \thetaf)$ is a  normal density with mean $\mu_i$ and variance 1. Now, instead of integrating over $w_i$ analytically, an alternative is to use data augmentation of $w_i$ in the Gibbs sampler, and subsequently sample $\thetaf$ fully conditional (including $w_i$). This approach mirrors that of data augmentation for the Bayesian probit model \citep{albert_bayesian_1993}, with the difference that in \texttt{BayesPIM} also $g_i$ is partly latent and hence augmented through (\ref{eq::fc.g}), while in the standard probit model the outcome is usually fully observed. Specifically, after updating $g_i$ in Gibbs sampling step (\ref{eq::fc.g}), we update the latent propensity $w_i$ by drawing  from truncated normal distributions
\begin{align} \label{eq::aug_w}
	w_i \mid g_i, \x_i, \thetaf \sim  
	\begin{cases}
		N^{+}(w_i \mid \mu_i, 1 ) \quad &\text{if} \ g_i = 1 \\
		N^{-}(w_i \mid \mu_i, 1 ) \quad &\text{if} \ g_i = 0
	\end{cases}
\end{align}
where $N^{+}(w_i \mid \mu_i, 1)$ denotes the normal density with mean $\mu_i$ and variance 1 truncated $(0,\infty)$ and $N^{-}$ denotes the same density truncated $(-\infty,0)$. This result follows after observing that $g_i$ acts as constraint on $w_i$ due to (\ref{eq::gw_constr}).\\

Subsequently, we use the fact that the parameters $\thetaf$ only depend on the complete-data through $\w = (w_1,\dots,w_n)^T$ and $\mathbf{X}$, i.e.
\begin{align}\label{eq::fc_betag_on_w}
	f_{\thetaf}(\thetaf \mid \Df,  \g, \tf, \w, \betaf, \sigma) &\propto  f_{\thetaf}(\thetaf \mid \w, \mathbf{X}) \nonumber \\
	&= N\big(\thetaf \mid \hat{\betaf}_g(\tau_{\theta}), (\mathbf{X} ^T\mathbf{X}  + \tau_{\theta}^{-1} I_{p})^{-1} \big)
\end{align}

where $\hat{\betaf}_g(\tau_{\theta}) = (\mathbf{X}^T \mathbf{X} + \tau_{\theta}^{-1} I_{p})^{-1} \mathbf{X} ^T \w$. Here, $f_{\thetaf}(\thetaf  \mid  \w, \mathbf{X})$ is the complete-data posterior of $\thetaf$ conditional on augmented $\w$. Internally, \texttt{BayesPIM} augments $w_i$ using (\ref{eq::aug_w}) and then, instead of executing step  (\ref{eq::fc.parw}), updates $\thetaf$ from 
\begin{align}
	\thetaf \mid \w, \mathbf{X} \sim 
	N\big( \hat{\betaf}_g(\tau_{\theta}), (\mathbf{X} ^T\mathbf{X}  + \tau_{\theta}^{-1} I_{p})^{-1} \big).
\end{align}

\newpage
\subsection{Inference on the posterior predictive CIFs} 
\label{sec:supplement_sampling_ppdcif}

\subsubsection{Conditional posterior predictive CIF}
\label{sec:supplement_condCIF}

We draw inference on the conditional CIF of the event time $t$ for a non-prevalent individual with a fixed covariate value $\x_i  = \tilde{\x}$
\begin{align}
	F_t(t \mid g = 0, \tilde{\x}, \betaf, \sigma).
\end{align}

To do so, we view $F_t(t \mid g = 0, \tilde{\x}, \betaf, \sigma)$ as a functional of the parameters $(\betaf, \sigma)$ with posterior distribution
$f_{\betaf, \sigma}(\betaf, \sigma \mid \Df)$. The corresponding posterior (mean) predictive CIF at time $t$ is
defined as the posterior expectation
\begin{align} \label{eq:CIF_cond_apdx}
	F_t(t \mid g = 0, \tilde{\x}, \Df)
	&= \E_{\betaf, \sigma \mid \Df}
	\big[ F_t(t \mid g = 0, \tilde{\x}, \betaf, \sigma) \big] \\
	&= \int_0^\infty \int_{\Theta_t}
	F_t(t \mid g = 0, \tilde{\x}, \betaf, \sigma)
	\, f(\betaf, \sigma \mid \Df)
	\, d\betaf \, d\sigma,
\end{align}

where $\Theta_t = \{\betaf \in \mathbb{R}^{p}\}$. To approximate this expectation and perform inference, we use the retained post-burnin MCMC samples $(\betaf^{(k)}, \sigma^{(k)})$, $k = 1,\dots,K$, from the marginal posterior $f_{\betaf, \sigma}(\betaf, \sigma \mid \Df)$ generated by the Gibbs sampler (\ref{eq::fc.g})--(\ref{eq::fc.kappa}). We apply the push-forward transform
\begin{align}
\nu_k(t, \tilde{\x}) = F_t(t \mid g = 0, \tilde{\x}, \betaf^{(k)}, \sigma^{(k)}),
\end{align}

which yields Monte Carlo samples from the posterior distribution of the CIF at time $t$ for a
non-prevalent individual with covariates $\tilde{\x}$. The posterior mean (Monte Carlo
estimate of the posterior predictive CIF) is then
\begin{align}
	\frac{1}{K} \sum_{k=1}^{K} \nu_k(t, \tilde{\x}).
\end{align}

Furthermore, pointwise 95\% credible intervals are obtained as the empirical 2.5\% and
97.5\% quantiles of the distribution $\{ \nu_k(t, \tilde{\x}) \}_{k=1}^K$.

\subsubsection{Marginal posterior predictive CIF}
\label{sec:supplement_margCIF}

Instead of conditioning on a specific combination of covariate values, we can also marginalize over the distribution of the covariates. This yields the marginal (or population-averaged) CIF at time $t$, defined by
\begin{align}
F_t(t \mid g = 0, \betaf, \sigma)
= \int_{\Theta_{\x}} F_t(t \mid g = 0, \x_{t} , \betaf, \sigma)\,
f_{\x}(\x_{t})\, d\x_{t},
\end{align}

where $f_{\x}$ denotes the distribution of the covariates and $\Theta_{\x}$ its support. Taking $f_{\x}$ to be the empirical distribution of the observed covariates, this reduces to
\begin{align}
F_t(t \mid g = 0, \betaf, \sigma)
= \frac{1}{\sum_{i=1}^{n}(1-g_i)} \sum_{i:g_i=0}
F_t(t \mid g_i = 0, \x_i, \betaf, \sigma).
\end{align}

As in Section~\ref{sec:supplement_condCIF}, we treat $F_t(t \mid g = 0, \betaf, \sigma)$ as a functional of the parameters $(\betaf, \sigma)$ with marginal posterior distribution $f_{\betaf, \sigma}(\betaf, \sigma \mid \Df)$. The posterior (mean) predictive marginal CIF is then
\begin{align}
F_t(t \mid g = 0, \Df)
= \E_{\betaf, \sigma \mid \Df}
\big[ F_t(t \mid g = 0, \betaf, \sigma) \big].
\end{align}

In practice, with MCMC samples $(\betaf^{(k)}, \sigma^{(k)})$, $k = 1,\dots,K$, we compute
\begin{align}
\hat{\nu}_k(t)
=  \frac{1}{\sum_{i=1}^{n}(1-g_i)} \sum_{i:g_i=0}
F_t\big(t \mid g_i = 0, \x_i, \betaf^{(k)}, \sigma^{(k)}\big).
\end{align}

The posterior (mean) predictive marginal CIF at time $t$ is then estimated by
\begin{align}
\widehat{F}_t(t \mid g = 0, \Df)
= \frac{1}{K} \sum_{k=1}^{K} \hat{\nu}_k(t),
\end{align}

and pointwise 95\% credible intervals are given by the empirical 2.5\% and 97.5\% quantiles of
$\{  \hat{\nu}_k(t) \}_{k=1}^{K}$. 

\subsubsection{Posterior predictive mixture CIF}

Besides the CIF of $t$ in the non-prevalent population, we draw inference on the mixture CIF. This is the CIF of $t$ where we include prevalence $g=1$ through a point probability mass at zero. This step requires defining a new variable (cf. Section \ref{sec:AFTmixturemodel})
\begin{align}
	t^* = (1-g) t
\end{align}

so that $t^*=t$ in the non-prevalent case, and $t^*=0$ in the prevalence case. The resulting conditional mixture CIF is 
\begin{align}
		F_{t^*}(t \mid \tilde{\x}, \betaf, \sigma, \thetaf) = \Phi(  \tilde{\x}^T \thetaf)
	+ \bigl(1-\Phi(\tilde{\x}^T \thetaf)\bigr)
	F_t(t \mid g = 0, \tilde{\x}, \betaf, \sigma).
\end{align}

The marginal mixture CIF over the covariate distribution $f_{\x}$ is
\begin{align}
	F_{t^*}(t \mid \betaf, \sigma, \thetaf) = \int_{\Theta_{\x}} 
	F_{t^*}(t \mid \x, \betaf, \sigma, \thetaf)  f_{\x}(\x)\, d \x.
\end{align}

As in the previous sections, we consider $F_{t^*}(t \mid \x_i, \betaf,\sigma,\thetaf,)$ and 	$F_{t^*}(t \mid \betaf,\sigma,\thetaf,)$ as functionals of the posterior distribution $f_{\betaf, \sigma, \thetaf}(\betaf, \sigma, \thetaf \mid \Df)$. To estimate the posterior predictive conditional mixture CIF, we use the push-forward transform for the posterior samples $\betaf^{(k)}, \thetaf^{(k)}, \sigma^{(k)}$, $k=1,\dots,K$, to obtain
\begin{align}
	\nu_k(t, \tilde{\x}) = F_{t^*}(t \mid \tilde{\x}, \betaf^{(k)},\sigma^{(k)}, \thetaf^{(k)})
\end{align}

for a fixed $\tilde{\x}$. Subsequently, the posterior predictive conditional mixture CIF is obtained by the empirical mean of $\{\nu_k(t, \tilde{\x}) \}_{k=1}^K$ and pointwise 95\% credible intervals are obtained by the 2.5\% and 97.5\% quantiles of the empirical distribution $\{ \nu_k(t, \tilde{\x})  \}_{k=1}^K$. For the marginal mixture CIF, we first perform marginalization, as in Section \ref{sec:supplement_margCIF}
\begin{align}
	\hat{\nu}_k(t) = \frac{1}{n} \sum_{i=1}^{n} F_{t^*}(t \mid \x_{i}, \betaf^{(k)},\sigma^{(k)}, \thetaf^{(k)})
\end{align}

and do the inference as described above.

\clearpage
\newpage
\section{Additional results from Simulation 1} \label{sec:supplement_simulation1}

\begin{figure}[h]  
	\centering  
	\includegraphics[width=\textwidth, trim=135pt 180pt 140pt 190pt, clip]{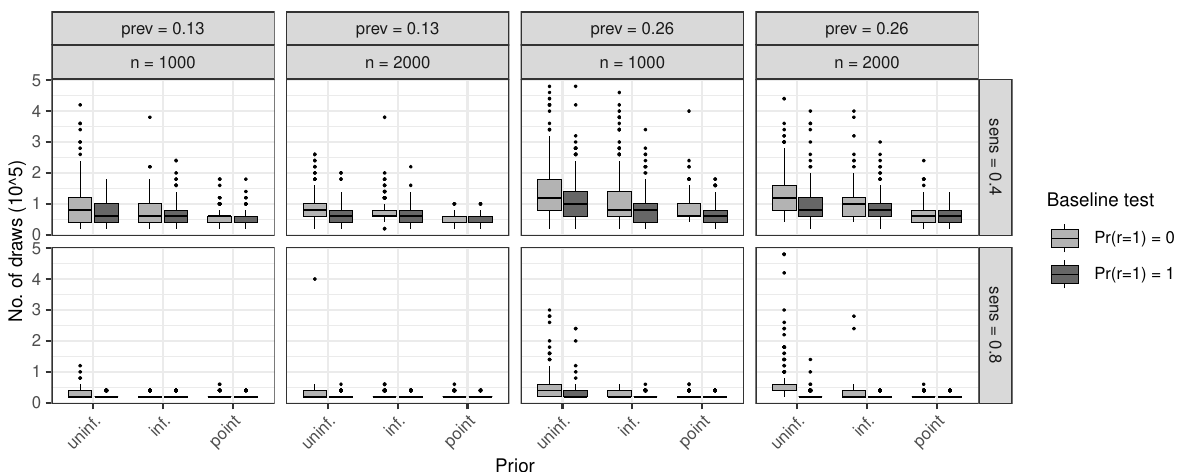}
	\caption{Number of posterior draws until convergence including burn-in (scaled by $10^5$) by simulation conditions. Convergence was evaluated every $2 \times 10^4$ draws. Abbreviations prev and sens denote, respectively, the prevalence probability $\Pr(g_i=1)$ and the test sensitivity $\kappa$. The priors on the test sensitivity $\kappa$ are either uninformative (uninf.), informative (inf.) or fixed at the true value (point).} 
	\label{fig:convergence_boxplot}  
\end{figure}

\begin{figure}[h]  
	\centering  
	\includegraphics[width=\textwidth, trim=100pt 150pt 100pt 175pt, clip]{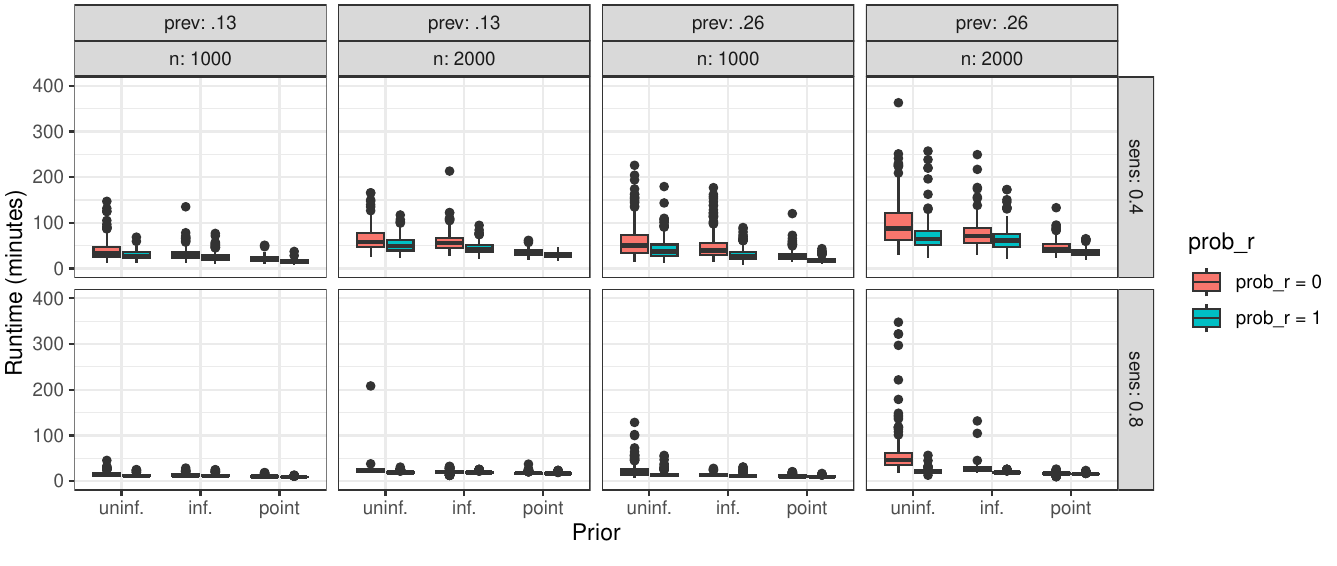}
	\caption{Runtime of \texttt{BayesPIM} until convergence across simulation conditions (in minutes). Abbreviations prev and sens denote, respectively, the prevalence probability $\Pr(g_i=1)$ and the test sensitivity $\kappa$. The priors on the test sensitivity $\kappa$ are either uninformative (uninf.), informative (inf.) or fixed at the true value (point).} 
	\label{fig:convergence_time_boxplot}  
\end{figure}

\begin{table}[ht] \label{tab:linregtimes}
	\centering
	\caption{Linear regression of convergence time on design factors. Reference categories are given in the table footer.}
	\label{tab:linregtimes}
	\centering
	\begin{tabular}{lrrrr}
		\hline
		Predictor & Coefficient & $SE$ & $t$ & $p$ \\
		\hline
		Intercept                      & 45.53  & 0.52 &  86.84 & $< .001$ \\
		$n = 2000$                     & 17.01  & 0.40 &  42.93 & $<.001$ \\
		$\kappa = 0.8$ (sens)                 & -26.85 & 0.40 & -67.77 & $<.001$ \\
		Informative prior              & -8.13  & 0.49 & -16.75 & $<.001$ \\
		Point prior on true value      & -19.36 & 0.49 & -39.88 & $< .001$ \\
		$\Pr(r_i=1) = 1$          & -8.91  & 0.40 & -22.47 & $< .001$ \\
		$\Pr(g_i=1) = 0.26$ (prev)        & 9.53   & 0.40 &  24.06 & $<.001$ \\
		\hline
	\end{tabular}
	\begin{flushleft}
		Notes: Unstandardized coefficients ($B$) from a linear regression predicting convergence time in minutes. 
		Reference categories are $n = 1000$, test sensitivity (sens) $\kappa = 0.4$, uninformative prior, baseline test probability $\Pr(r_i=1) = 0$, and prevalence probability (prev) $\Pr(g_i=1) = 0.13$. 
		Model fit: $R^2 = .49$, adjusted $R^2 = .49$, $F(6, 9587) = 1521$, $p < .001$, residual $SD = 19.41$.
	\end{flushleft}
\end{table}

\begin{figure}[h]  
	\centering  
	\includegraphics[scale = 0.8, trim=130pt 180pt 140pt 190pt, clip, page = 1]{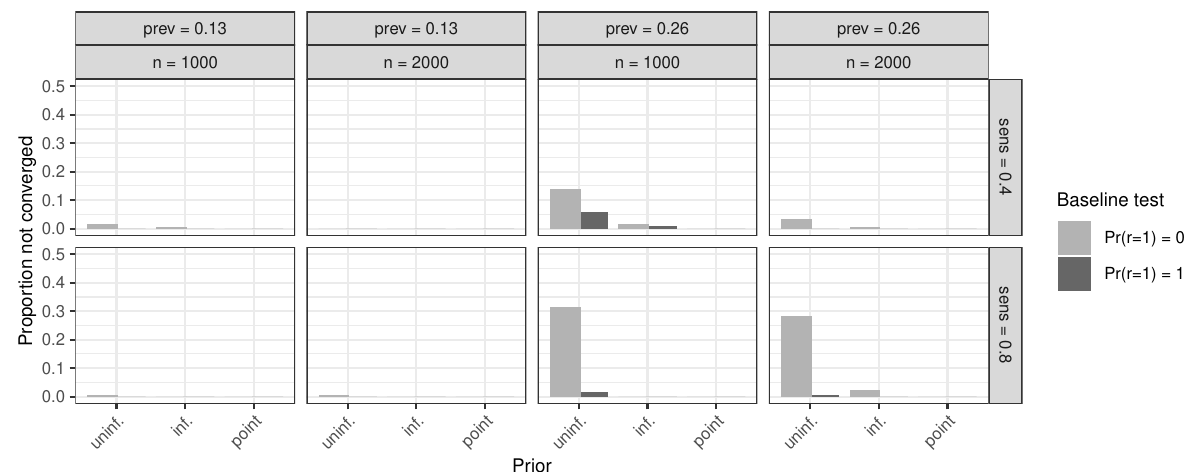}
	\caption{Proportion of runs that did not converge until $5 \times 10^5$ Gibbs iterations, arranged by simulation conditions (200 runs per condition). Convergence was evaluated every $2 \times 10^4$ draws. Abbreviations prev and sens denote, respectively, the prevalence probability $\Pr(g_i=1)$ and the test sensitivity $\kappa$. The priors on the test sensitivity $\kappa$ are either uninformative (uninf.), informative (inf.) or fixed at the true value (point).} 
	\label{fig:convergence_ncrates}  
\end{figure}

\clearpage

\begin{sidewaysfigure}
		\centering
		\includegraphics[scale=0.8]{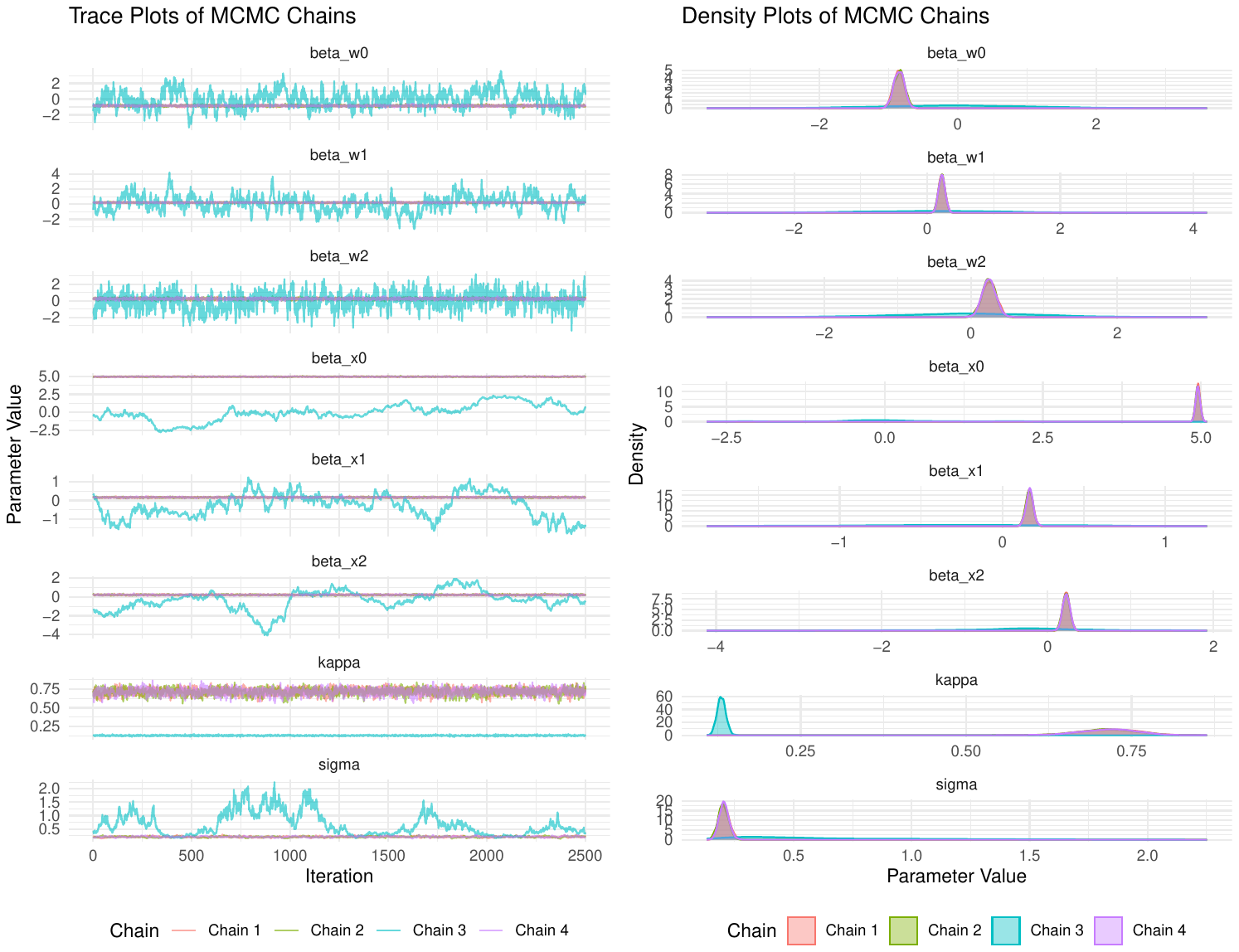}
		\caption{Example of the MCMC chains resulting from a non-convergent run on a data set from Simulation 1 ($n_{sim} = 1000$, $\Pr(r_i=1)=1$, prev.=.26, $\kappa = .80$, uninformative prior on $\kappa$). The Gibbs sampler finds a local posterior mode in chain 2 centred arround an incorrect, very low $\kappa$ estimate. Chains are thinned with an interlace rate of 200 between draws ($5 \times 10^5$ draws in total, of which half are discarded for burn-in).}
		\label{fig:nonconvergence_example}
\end{sidewaysfigure}
\clearpage

\begin{figure}[h]  
	\centering  
	\includegraphics[scale = 0.8, trim=135pt 50pt 135pt 65pt, clip, page = 1]{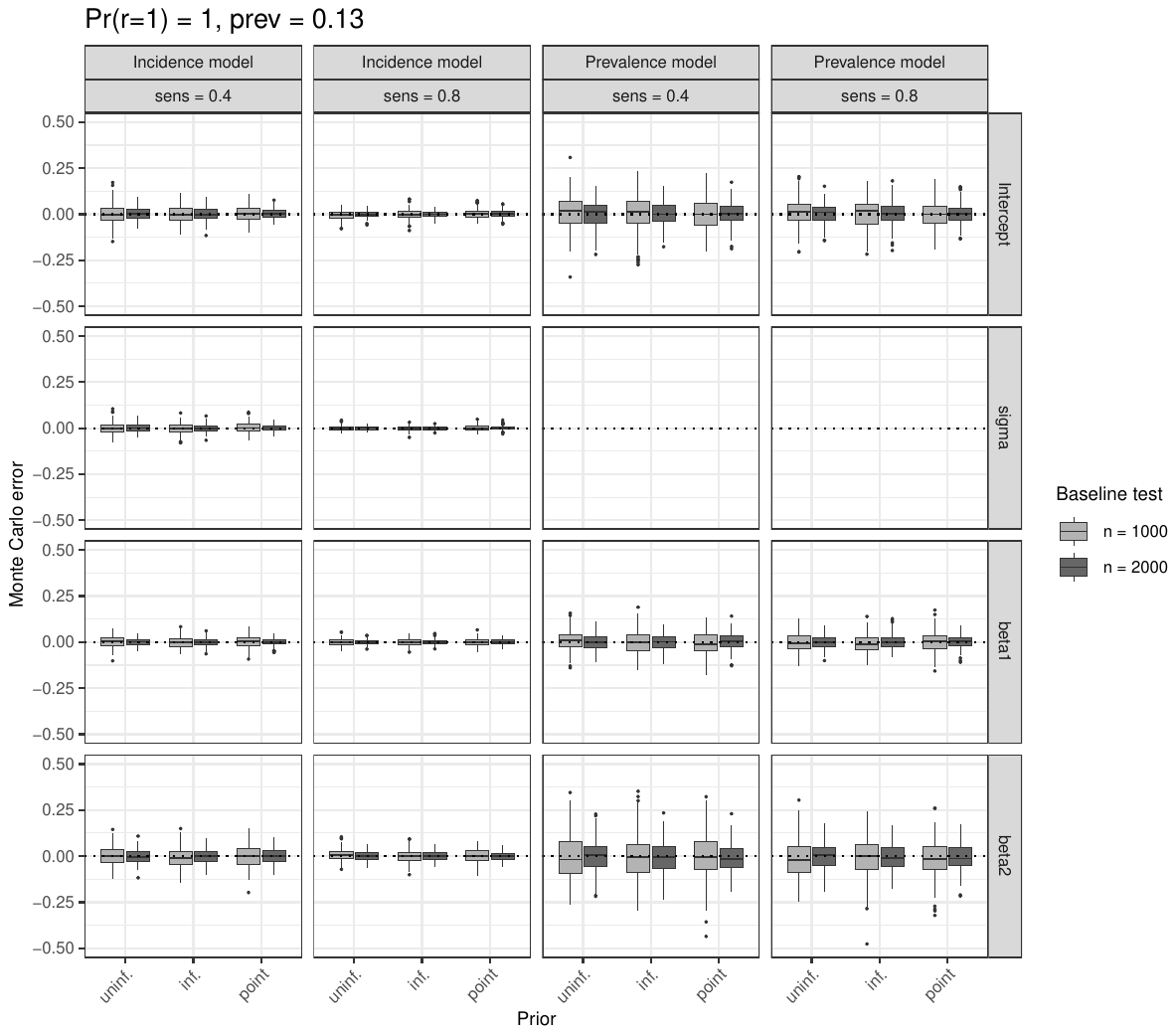}
	\caption{Monte Carlo error of the model parameters, as indicated on the right, for both the incidence model (\ref{eq::mod_x}) and the prevalence model (\ref{eq::mod_g}) in the simulation condition: $\bm{\Pr(g_i=1)=0.13, \ \Pr(r_i=1)= 1}$. Note that there is no $\sigma$ parameter in the prevalence model and hence the corresponding panels are left blank. Row labels beta1 and beta2 refer to $\beta_1$ and $\beta_2$ in the incidence model and $\theta_1$ and $\theta_2$ in the prevalence model. Abbreviations prev and sens denote, respectively, the prevalence probability $\Pr(g_i=1)$ and the test sensitivity $\kappa$. The priors on the test sensitivity $\kappa$ are either uninformative (uninf.), informative (inf.) or fixed at the true value (point).} 
	\label{fig:pars_boxplot1}  
\end{figure}
\clearpage

\begin{figure}[h]  
	\centering  
	\includegraphics[scale = 0.8, trim=135pt 50pt 135pt 65pt, clip, page = 2]{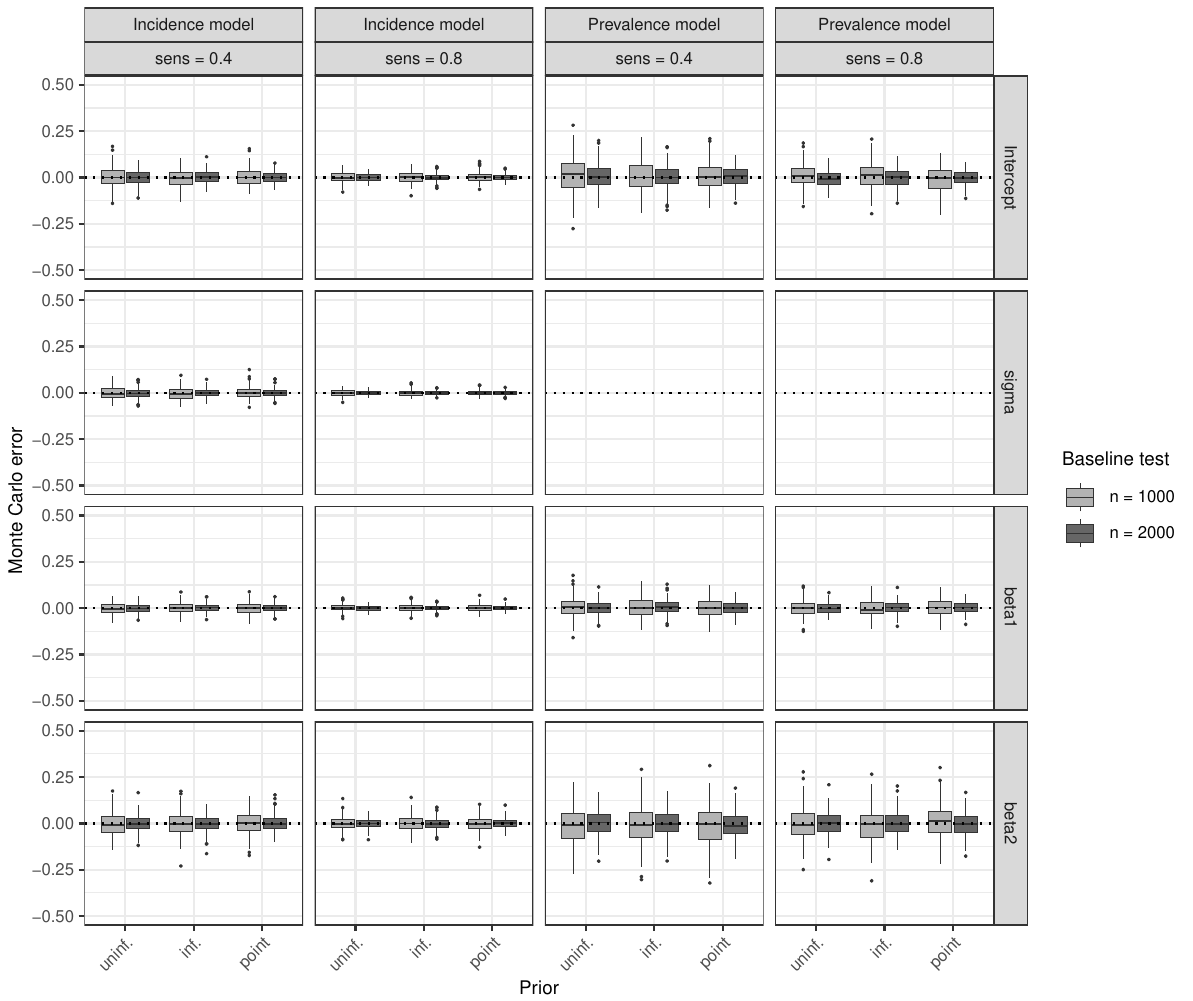}
	\caption{Monte Carlo error of the model parameters, as indicated on the right, for both the incidence model (\ref{eq::mod_x}) and the prevalence model (\ref{eq::mod_g}) in the simulation condition: $\bm{\Pr(g_i=1)=0.26, \ \Pr(r_i=1)= 1}$. Note that there is no $\sigma$ parameter in the prevalence model and hence the corresponding panels are left blank. Row labels beta1 and beta2 refer to $\beta_1$ and $\beta_2$ in the incidence model and $\theta_1$ and $\theta_2$ in the prevalence model. Abbreviations prev and sens denote, respectively, the prevalence probability $\Pr(g_i=1)$ and the test sensitivity $\kappa$. The priors on the test sensitivity $\kappa$ are either uninformative (uninf.), informative (inf.) or fixed at the true value (point).} 
	\label{fig:pars_boxplot2}  
\end{figure}
\clearpage

\begin{figure}[h]  
	\centering  
	\includegraphics[scale = 0.8, trim=135pt 50pt 135pt 65pt, clip, page = 3]{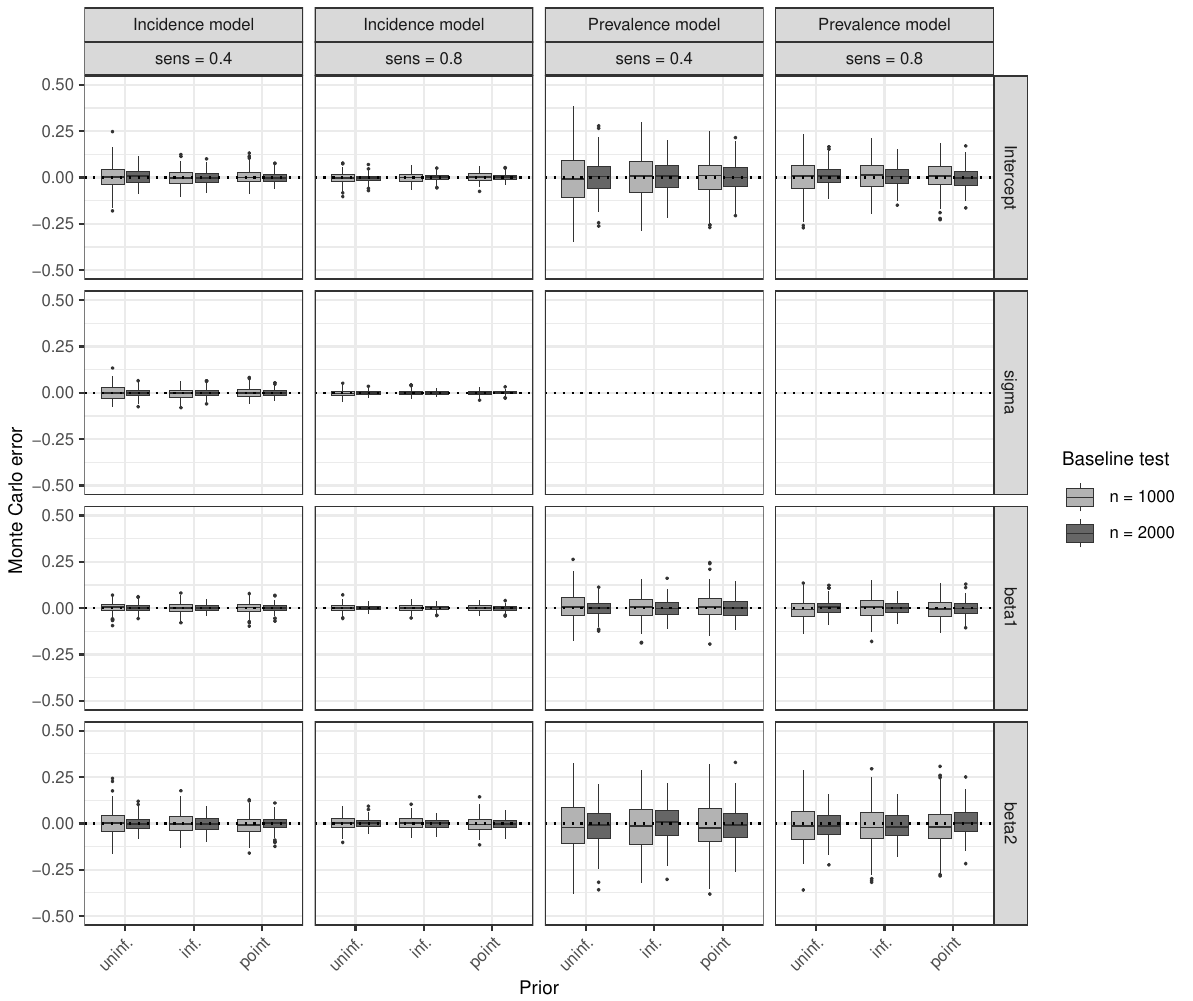}
	\caption{Monte Carlo error of the model parameters, as indicated on the right, for both the incidence model (\ref{eq::mod_x}) and the prevalence model (\ref{eq::mod_g}) in the simulation condition: $\bm{\Pr(g_i=1)=0.13, \ \Pr(r_i=1)= 0}$. Note that there is no $\sigma$ parameter in the prevalence model and hence the corresponding panels are left blank. Row labels beta1 and beta2 refer to $\beta_1$ and $\beta_2$ in the incidence model and $\theta_1$ and $\theta_2$ in the prevalence model. Abbreviations prev and sens denote, respectively, the prevalence probability $\Pr(g_i=1)$ and the test sensitivity $\kappa$. The priors on the test sensitivity $\kappa$ are either uninformative (uninf.), informative (inf.) or fixed at the true value (point).} 
	\label{fig:pars_boxplot3}  
\end{figure}
\clearpage

\begin{figure}[h]  
	\centering  
	\includegraphics[scale = 0.8, trim=135pt 50pt 135pt 65pt, clip, page = 4]{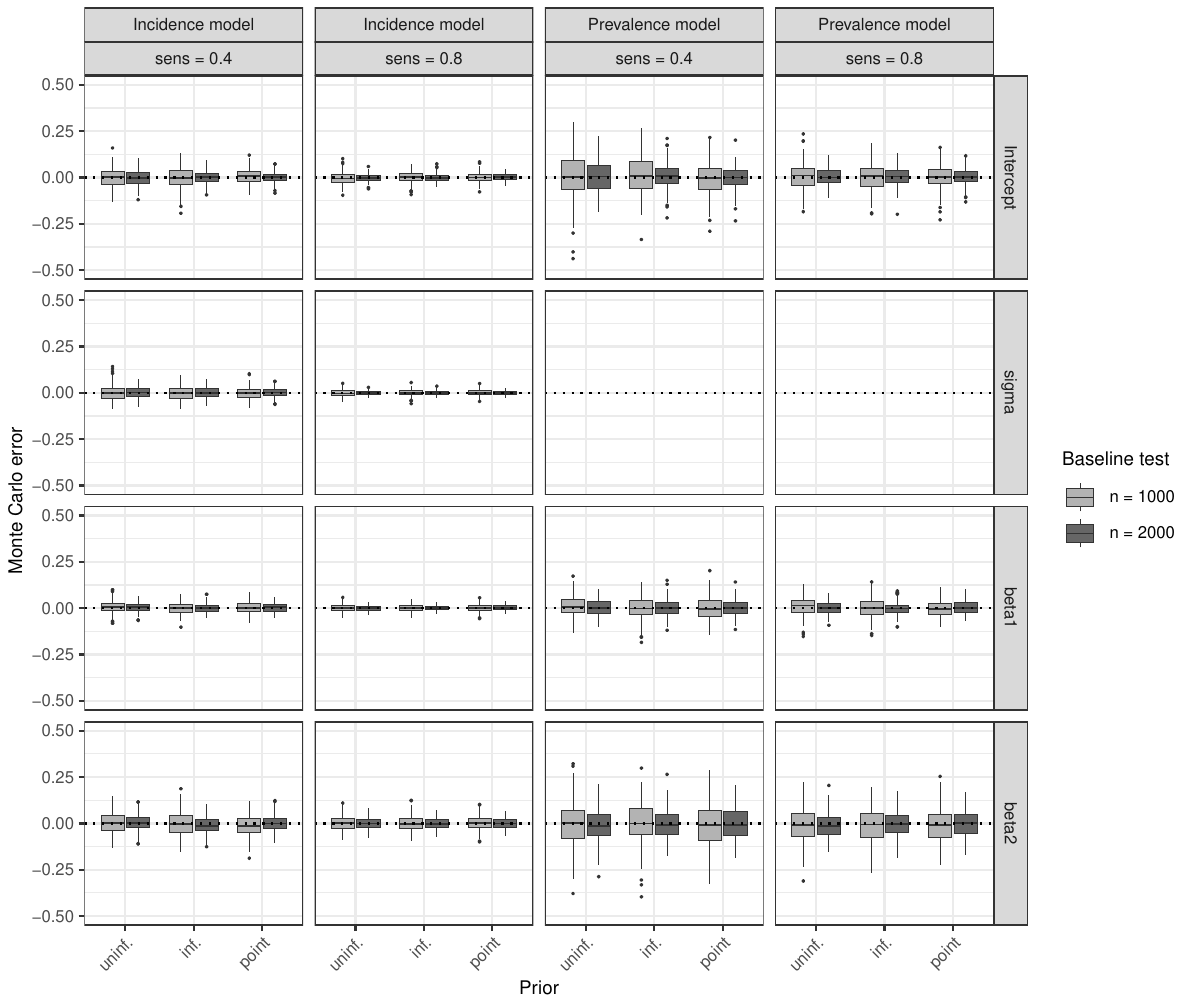}
	\caption{Monte Carlo error of the model parameters, as indicated on the right, for both the incidence model (\ref{eq::mod_x}) and the prevalence model (\ref{eq::mod_g}) in the simulation condition: $\bm{\Pr(g_i=1)=0.26, \ \Pr(r_i=1)= 0}$. Note that there is no $\sigma$ parameter in the prevalence model and hence the corresponding panels are left blank. Row labels beta1 and beta2 refer to $\beta_1$ and $\beta_2$ in the incidence model and $\theta_1$ and $\theta_2$ in the prevalence model. Abbreviations prev and sens denote, respectively, the prevalence probability $\Pr(g_i=1)$ and the test sensitivity $\kappa$. The priors on the test sensitivity $\kappa$ are either uninformative (uninf.), informative (inf.) or fixed at the true value (point).} 
	\label{fig:pars_boxplot4}  
\end{figure}
\clearpage

\begin{figure}[h]  
	\centering  
	\includegraphics[scale = 0.7, trim=60 120pt 60pt 140pt, clip, page = 4]{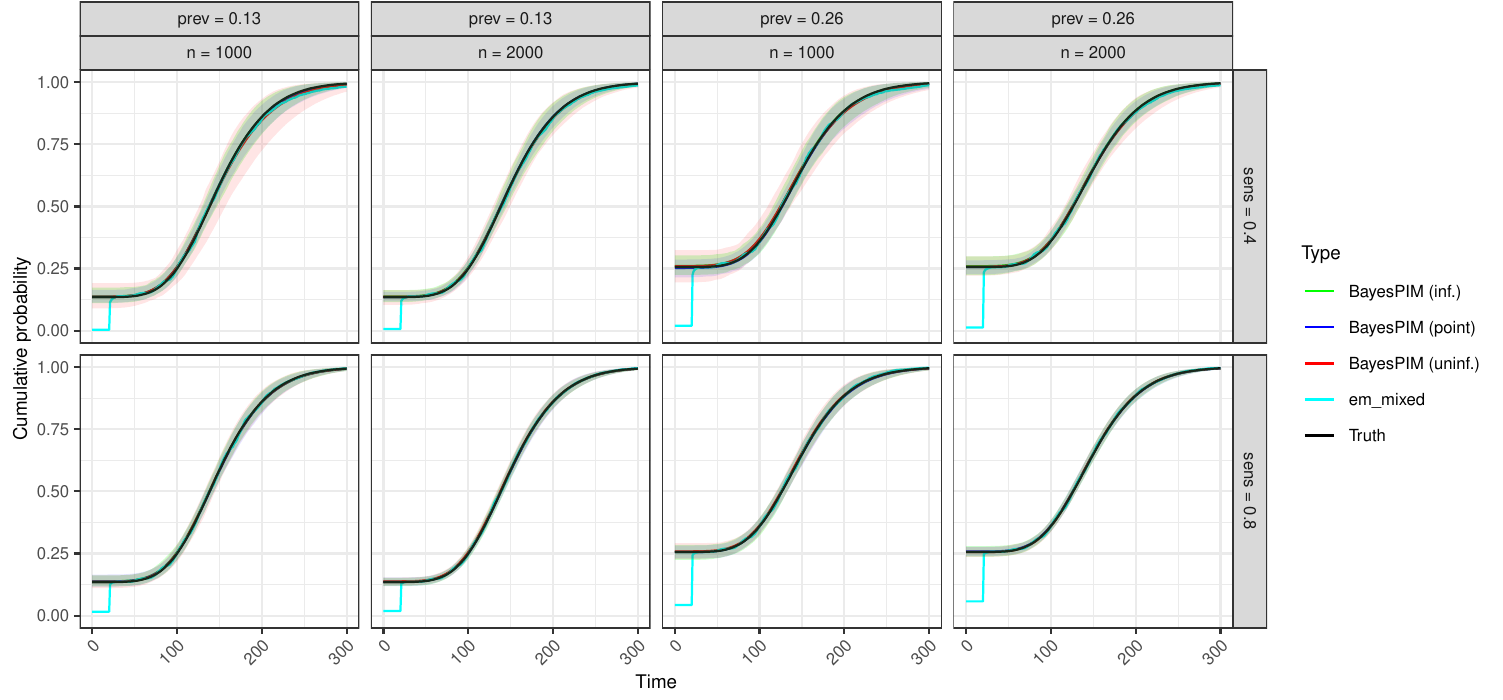}
	\caption{Posterior median marginal mixture CIFs $F_{t^*}(t\mid \thetaf,\betaf, \sigma)$, point-wise averaged over 200 Monte Carlo simulation runs with 95\% quantiles shown as shaded regions. The condition $\bm{\Pr(r_i=1)=  0}$ is shown.}
	\label{fig:sim1_cdfs_prob_r0}  
\end{figure}

\begin{figure}[h]  
	\centering  
	\includegraphics[scale = 0.65, trim=60 120pt 60pt 140pt, clip, page = 1]{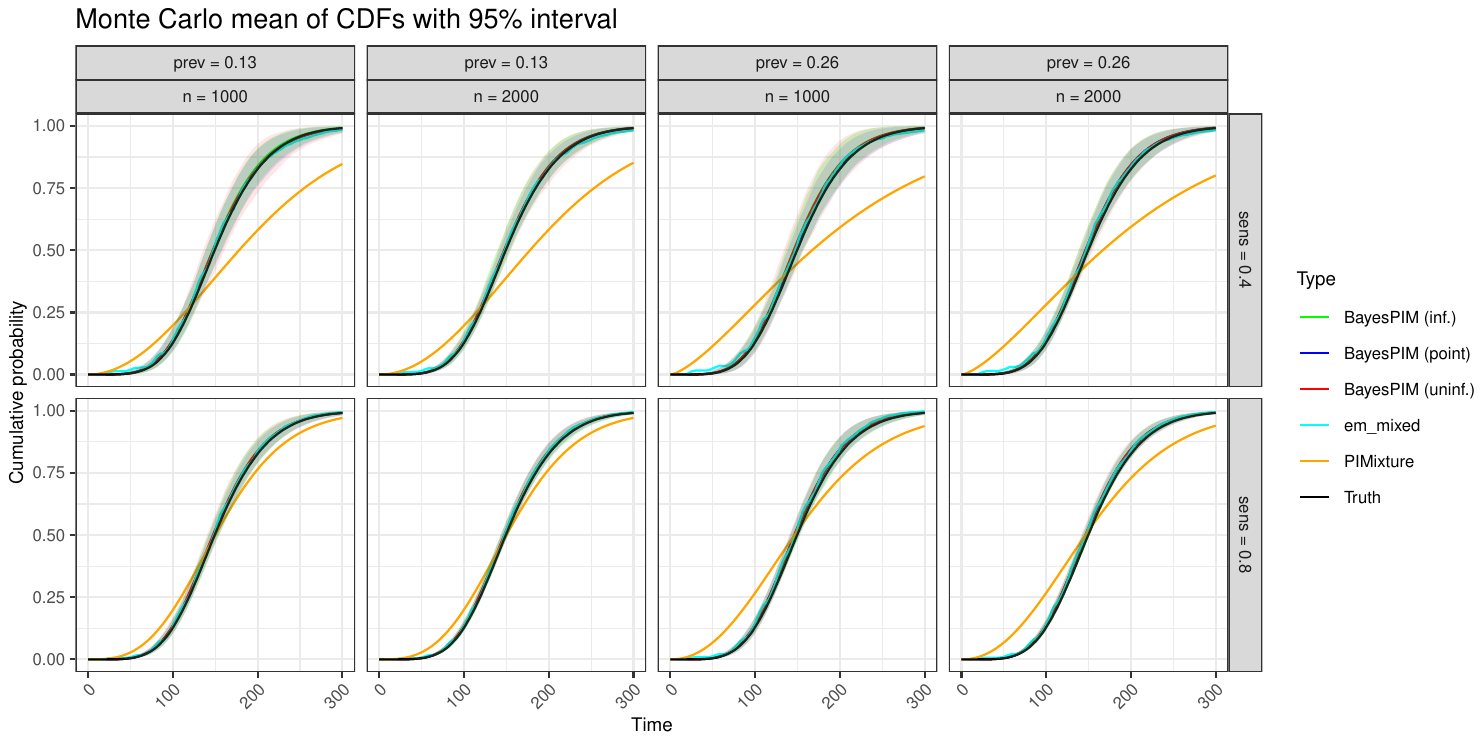}
	\caption{Posterior median marginal CIFs for non-prevalent cases $F_{t}(t\mid g=0, \betaf, \sigma)$, point-wise averaged over 200 Monte Carlo simulation runs with 95\% quantiles shown as shaded regions. The condition $\bm{\Pr(r_i=1)= 1}$ is shown. Lines of all models except \texttt{PIMixture} are overlapping. }
	\label{fig:sim1_cdfs_prob_r1_healthy}  
\end{figure}

\begin{figure}[h]  
	\centering  
	\includegraphics[scale = 0.65, trim=60 120pt 60pt 140pt, clip, page = 2]{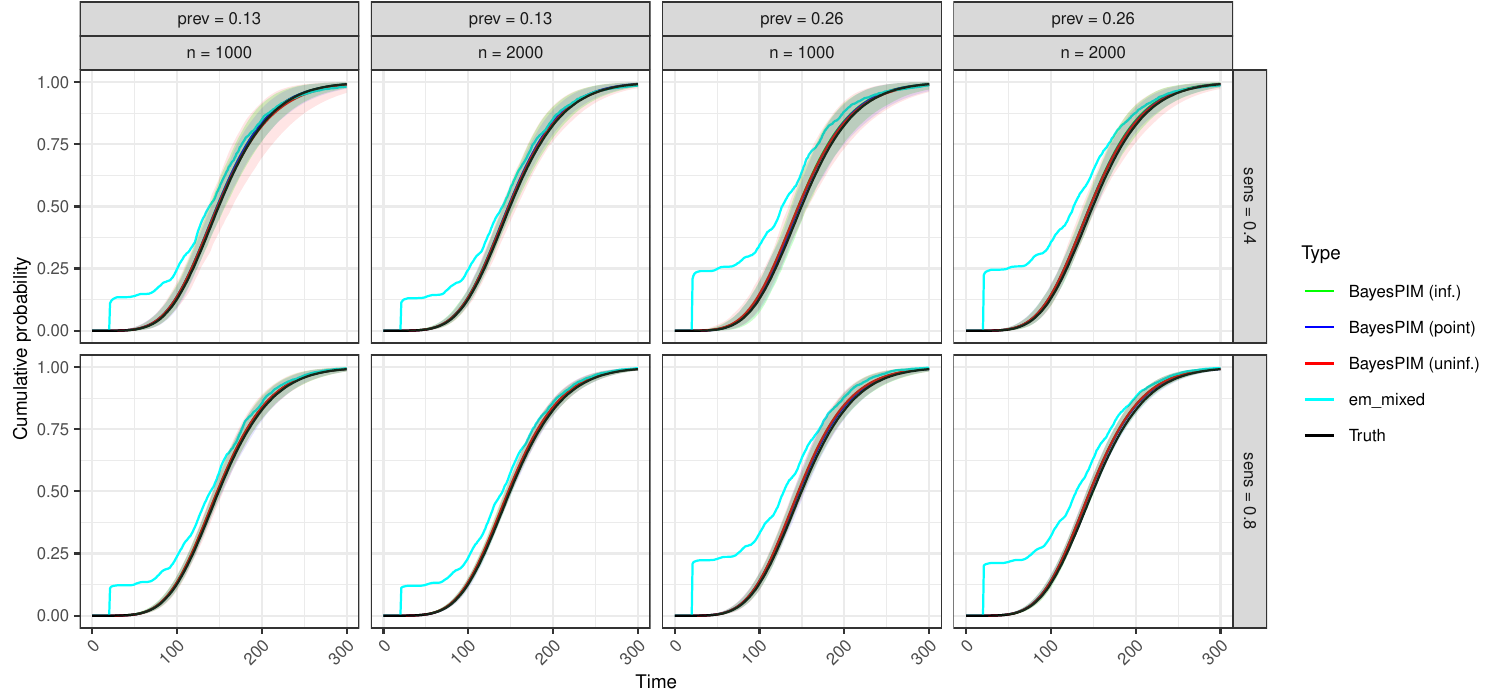}
	\caption{Posterior median marginal CIFs for non-prevalent cases $F_{t}(t\mid g=0, \betaf, \sigma)$, point-wise averaged over 200 Monte Carlo simulation runs with 95\% quantiles shown as shaded regions. The condition $\bm{\Pr(r_i=1)= 0}$ is shown. Lines of all models except \texttt{em\_mixed} are overlapping. }
	\label{fig:sim1_cdfs_prob_r0_healthy}  
\end{figure}

\begin{figure}[h]  
	\centering  
	\includegraphics[scale = 0.8, trim=120 120pt 120pt 120pt, clip, page = 1]{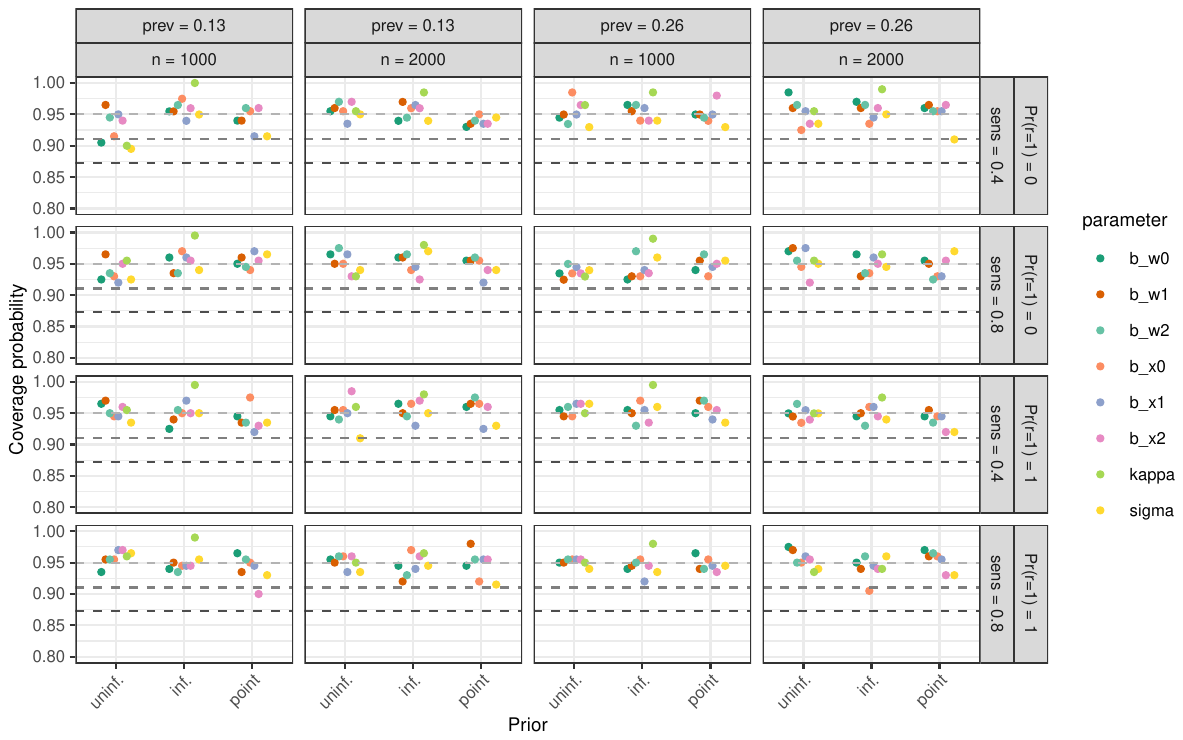}
	\caption{Frequentist coverage probability of the Bayesian 95\% posterior credible intervals for the 48 simulation conditions (estimated from 200 Monte Carlo data sets per condition by the proportion of intervals covering the true parameter value). The gray dotted lines in each panel denote, from top to bottom: (a) the nominal 95\% level, (b) the value of a point estimate whose 95\% confidence upper bound is equal to 95\%, (c) the value of a point estimate whose 95\% confidence upper bound is equal to 95\% with a Bonferroni adjustment for 48 repeated tests.}
	\label{fig:sim1_coverage}  
\end{figure}

\clearpage
\begin{sidewaysfigure}
	\centering  
	\includegraphics[scale = 0.8, trim=0 0 0 0, clip, page = 1]{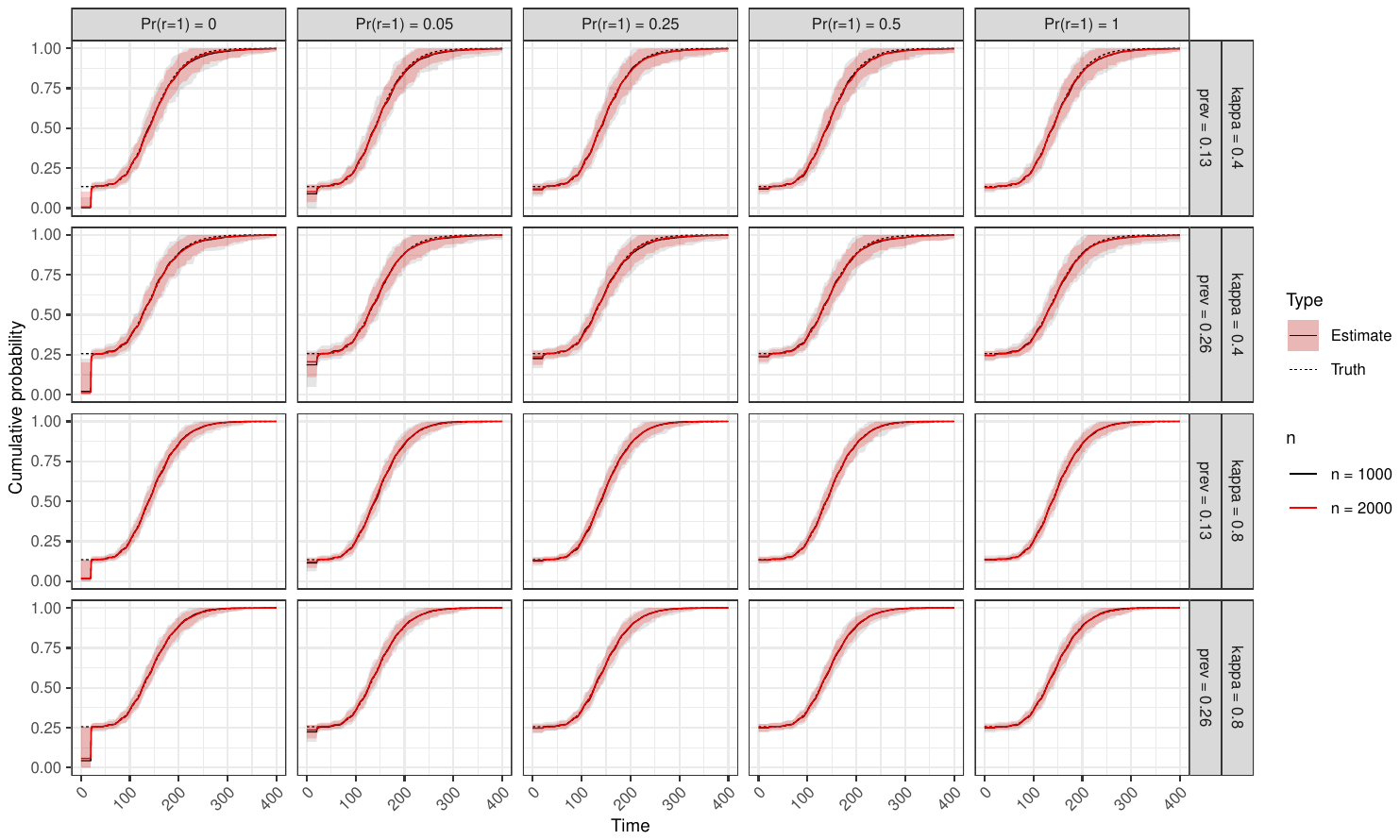}
	\caption{Posterior median marginal mixture CIFs $F_{t^*}(t\mid \thetaf,\betaf, \sigma)$ pointwise averaged over the estimates from 200 data sets. These are additional results on the \texttt{em\_mixed} estimator. The same simulation conditions as in Simulation 1 are used, but a wider range of baseline test probabilities $\Pr(r_i = 1) \in \{0, 0.05, 0.25, 0.5, 1\}$ is reported. The results demonstrate that \texttt{em\_mixed} was only approximately unbiased in the $\Pr(r_i = 1) = 1$ but the bias was negligible when a moderate number of tests was available (e.g. $\Pr(r_i = 1) =0.5$). }
	\label{fig:additional_results_emmixed}  
\end{sidewaysfigure}
	\clearpage

\clearpage
\newpage

\newpage
\section{Additional details on the set-up of Simulation 2}
\label{sec:supplement:sim2}

\subsection{Data generation by resampling the CRC EHR data} \label{sec:sim2_obscens_gen}

\begin{figure}[h]  
	\centering  
	\includegraphics[scale=0.65, trim=130pt 100pt 160pt 130pt, clip, page=2]{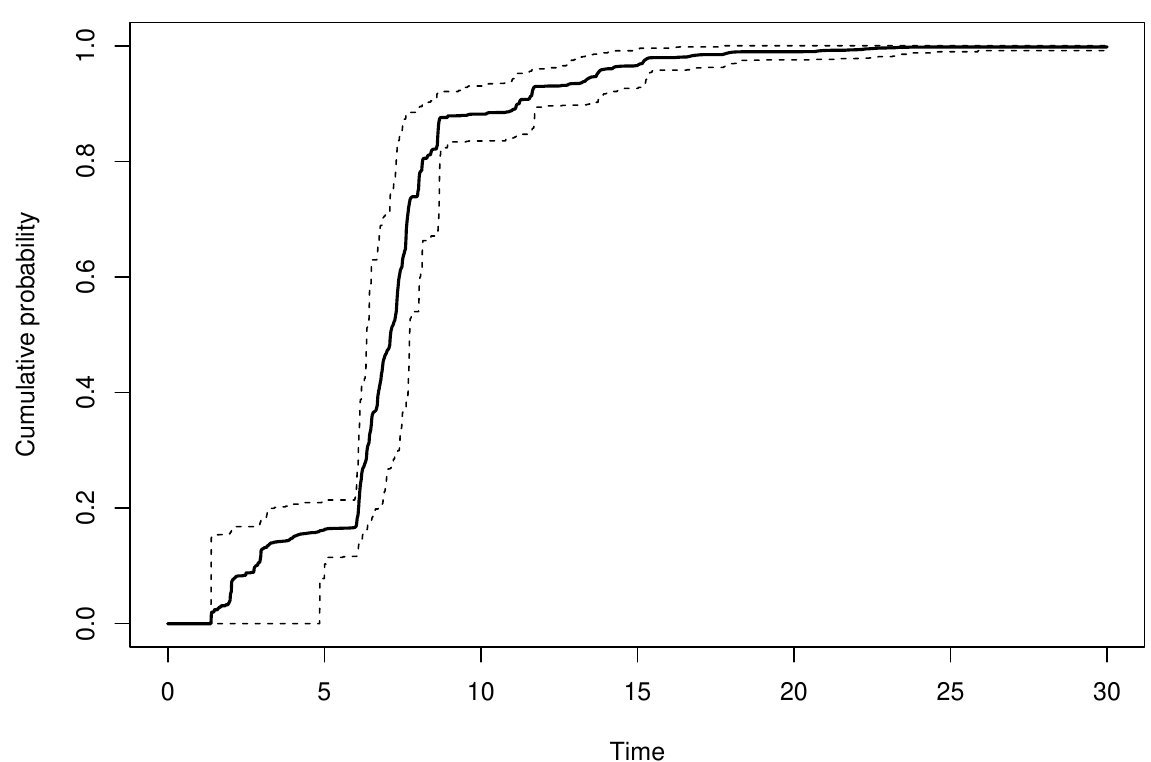}
	\caption{Bootstrapped approximated empirical CDF of $s_{i}$ in the CRC EHR, the time of right censoring in the data. Solid line represents point-wise averages across bootstrapped samples with the interval representing point-wise 95\% confidence intervals.}  
	\label{fig:rightcensoring_bootstrapped}  
\end{figure}

For Simulation 2, we here give details on the data generating process which resampled the CRC EHR to obtain realistic screening times and covariate distributions. Specifically, we explain how we resampled the CRC EHR to obtain new screening times from the observed screening times distribution and determined the event and right censoring times for each newly generated data point $k$. The re-sampled screening times distribution was compared with the observed screening times distribution on a number of benchmark statistics to assure that the  simulated data resembled the CRC EHR (see Figure \ref{fig:rightcensoring_bootstrapped}).  \\

Specifically, we followed these steps:

\begin{enumerate}
	\item Set the true parameters of $(\betaf, \sigma, \thetaf)$ to the posterior median estimates from the application (Table \ref{tab:coefficients}, Section \ref{sec:application}).
\end{enumerate}

Then for each newly generated individual $k=1,...,n_{sim}$ with $f_t$ the Weibull density, we generated the latent incidence time ($t_k$), prevalence status ($g_k$) and baseline test status ($r_k$ with Bernoulli probability 0.93 set to that observed in the CRC EHR data). We proceeded as follows:
\begin{enumerate}[start=2]
	\item Randomly sample integer $i'$ discretely uniform from $\{1,\dots,n\}$
	\item Set $\x_k = \x_{i'} $ (in words: set the sampled covariates to those of the $i'$-th individual in the CRC EHR data; covariates for the incidence and prevalence model are the same, hence $\x_k$ without index g or t)
	\item Generate: $t_k \mid \x_{k}, \betaf, \sigma \sim f_t(t_k \mid \x_{k}, \betaf, \sigma)$
	\item Generate: $g_k \mid \x_{k}, \thetaf \sim f_g(g_k \mid \x_{k}, \thetaf)$
	\item Generate: $r_k \sim \text{Bernoulli}(0.93)$	
\end{enumerate}

This procedure followed 
the hierarchical model described in Section \ref{sec:hierarchical}. To obtain new screening times and run screening tests according to the hierarchical model, equations (\ref{eq:schedule_mechanism})--(\ref{eq:obsprocess}), we used the observed screening times $\Vvec_{i'}=(v_{i'1},v_{i'2},\dots,v_{i'c_{i'}})$ of the sampled unit $i'$ as donor for new screening times. Specifically, we set $v_{k1}=0$ and stopped screening, i.e. $y_{k1}=1$ (positive baseline test), if $g_k=1$ and $r_k=1$ with probability $\kappa$. If $r_k=0$, $y_{k1}$ is missing, as described in Section \ref{sec:hierarchical}. Else we set $y_{k1}=0$ and, subsequently, for each $j=2,\dots c_{i'}$ we followed the following steps:
\begin{enumerate}
	\item Set $v_{kj}:=v_{i'j}$
	\item Generate $y_{kj}$ as defined by (\ref{eq:obsprocess})
\end{enumerate}

The procedure was stopped if $y_{kj}=1$ such that $\Vvec_k = (v_{k1},\dots,v_{kj})$. If the process was not stopped until $v_{k,c_{i'}-1}$, its continuation depended on $y_{i'c_{i'}}$ indicating whether individual $i'$ was not right censored at $v_{i'c_{i'}-1}$. If $y_{i'c_{i'}}=0$, $i'$ was right censored, we set $v_{kc_{i'}}:=v_{i'c_{i'}}=\infty$, to indicate that also $\Vvec_k$ was right censored. However,  when $y_{i'c_{i'}}=1$, $\Vvec_{i'}$ was not right censored because an event was observed and screening stopped at $v_{i'c_{i'}}<\infty$ due to (\ref{eq:obsprocess}). As a consequence, no further screening times beyond $v_{i'c_{i'}}$ were available as donors but for unit $k$ screening should continue if $y_{k,c_{i'}-1}=0$. To generate new times after $v_{i'c_{i'}}$ (if $y_{i'c_{i'}}=1$) we, therefore, approximated the unobserved distribution of screening times that would have occurred if, counter to the fact, after the first $\Vvec_{i'}$ times the series had not been stopped due to an observed event for individual $i'$. To do so, we resampled time difference scores $d_{ij} = v_{ij}-v_{ij-1}$ from the empirical distribution of all observed difference scores in the group that was not right censored (all individuals with $y_{ic_i}=1$). Samples $ \tilde{d}$ from this distribution were obtained by calculating $d_{ij}$ for all $i$ and $j$ in the data for which $y_{ic_i}=1$ and then randomly selecting, with replacement, one $\tilde{d}$ from the pool of all $d_{ij}$. Call the set of all difference scores $D$. Then, the procedure for generating screening times was continued as follows (if $y_{i'c_{i'}}=1$):

\begin{itemize}
	\item Set $v_{kc_{i'}} := v_{i'c_{i'}}$ (the final known screening time at which $i'$ had a positive test can still be used as donor for $k$)
	\item Generate $y_{kc_{i'}}$ as defined by (\ref{eq:obsprocess})
	\item Set $v_{k,c_{i'}+1} :=v_{kc_{i'}} + \tilde{d}$, where $\tilde{d}$ is drawn uniformly from set D
	\item Generate $y_{k,c_{i'}+1}$ as defined by (\ref{eq:obsprocess})
		\item Set $v_{k,c_{i'}+2} :=v_{k,c_{i'}+1} + \tilde{d}$, where $\tilde{d}$ is newly drawn uniformly from set D
	\item Generate $y_{k,c_{i'}+2}$ as defined by (\ref{eq:obsprocess})
	\item \dots
\end{itemize}

This process is continued until stopping due to $y_{kj}=1$  or right censoring which happens as defined by the hierarchical model (Section \ref{sec:hierarchical}) if $v_{kc_{i'}+j} > s_k$, where $s_k$ is the latent time of right censoring. We explain below how we obtained right censoring time $s_k$ from its approximated empirical distribution. \\

The imputation procedure used to generate screening times $v_{kc_{i'}+j}$, $j\ge1$, was based on assumptions that we, here, make explicit. We assumed (a) that the observed time differences $d_{ij} = v_{ij} - v_{ij-1}$ between the observed screening visits $v_{ij}$ were exchangeable with those between the unobserved screening times (i.e. the screening process would have continued as it was observed into the future), (b) that the difference scores $d_{ij}$ were exchangeable between units $i$ and (c) the ordering implied by index $j$ did not matter for future screening moments.  Assumption (b) was needed to avoid very similar distances between successive screening moments (e.g., if $\Vvec$ was short such as $\Vvec = (0,3)$ using only the observed difference $d_i = 3$ to impute subsequent screening times would not have appropriately represented the empirically observed variation in timely distance between screening moments on other units). \\

We now return to the question of how to sample the time of right censoring $s_k$ from the empirical right censoring distribution. Our goal is to use $s_{i'}$ as a donor but we note that $s_{i'}$ is unobserved in all cases and should not be confused with $v_{i'c_{i'}}$. Time $v_{i'c_{i'}}$ is the finite time point at which the event occurred if $y_{i'c_{i'}}=1$ and $v_{i'c_{i'}} = \infty$ in case of right censoring ($y_{i'c_{i'}}= 0$). Then $v_{i',c_{i'}-1}$ with $v_{i',c_{i'}-1} < s_{i'}$ denotes the last follow-up moment (without event), but not the time of right censoring. Hence, the last screening moment is a lower bound for $s_{i'}$. Now we note that right censoring occurs if the time of right censoring $s_{i'}$ is sooner than the time of the next screening moment. Therefore, if $y_{i'c_{i'}}=0$, right censoring occurs when $v_{i',c_{i'}-1} < s_{i'} \le v_{i',c_{i'}-1} + \tilde{v}_{i'} $, where $\tilde{v}_{i'}$ denotes the unknown time until the next screening moment. Hence, an upper bound for $s_{i'}$ is found, namely $ v_{i',c_{i'}-1} + \tilde{v}_{i'}$. Now, returning to the full CRC EHR sample $i=1,\dots,n$ with $y_{i,c_{i}}=0$, if $\tilde{v}_{i}$ was known for all $i$ in the CRC EHR data, which it is not, the distribution of $s_{i}$ could be estimated by treating the time as interval-censored between the last observed times $v_{i,c_{i}-1}$ and the missing $v_{i,c_{i}-1}+\tilde{v}_{i}$. Our strategy for estimating the distribution of $s_{i}$ therefore was defined as follows. For all $i$ for which $y_{ic_i}=0$, set

\begin{enumerate}
	\item $l_i := v_{i,c_i-1}$
	\item $r_i := l_i +  \tilde{v}_{i}$, where $ \tilde{v}_{i}$ is sampled uniformly from set $D$
\end{enumerate}

Subsequently, we estimated $\hat{F}_s(s)$, the empirical CDF of $s_i$, through interval-censored nonparametric maximum likelihood using the Turnbull estimator \citep{turnbull_empirical_1976}, where the intervals were given by $(l_i,r_i]$. This procedure was repeated $b = 1,\dots,1000$ times on bootstrapped samples from the CRC EHR data to reflect the uncertainty in the Turnbull estimator as well as the repeated imputations $\tilde{v}_{i}$ (Figure \ref{fig:rightcensoring_bootstrapped}). For each bootstrapped data set we thusly obtained one $\hat{F}^{(b)}_s(s)$. To draw one right censoring time $s_k$ from this approximated distribution, we first randomly selected one $\hat{F}^{(b)}_s(s)$ and subsequently used the inverse sampling method to obtain  $s_k$ under the constraint that  $s_k > v_{kc_{i'}}$. This is achieved by drawing from the truncated distribution $\hat{F}^{(b)}_s$ subject to truncation rule $s_k > v_{kc_{i'}}$. Technically, we reject draws violating this rule and repeat the draw.

\clearpage

\begin{figure}[h]  
	\centering  
	\includegraphics[scale=0.85, trim=20pt 70pt 30pt 70pt, clip]{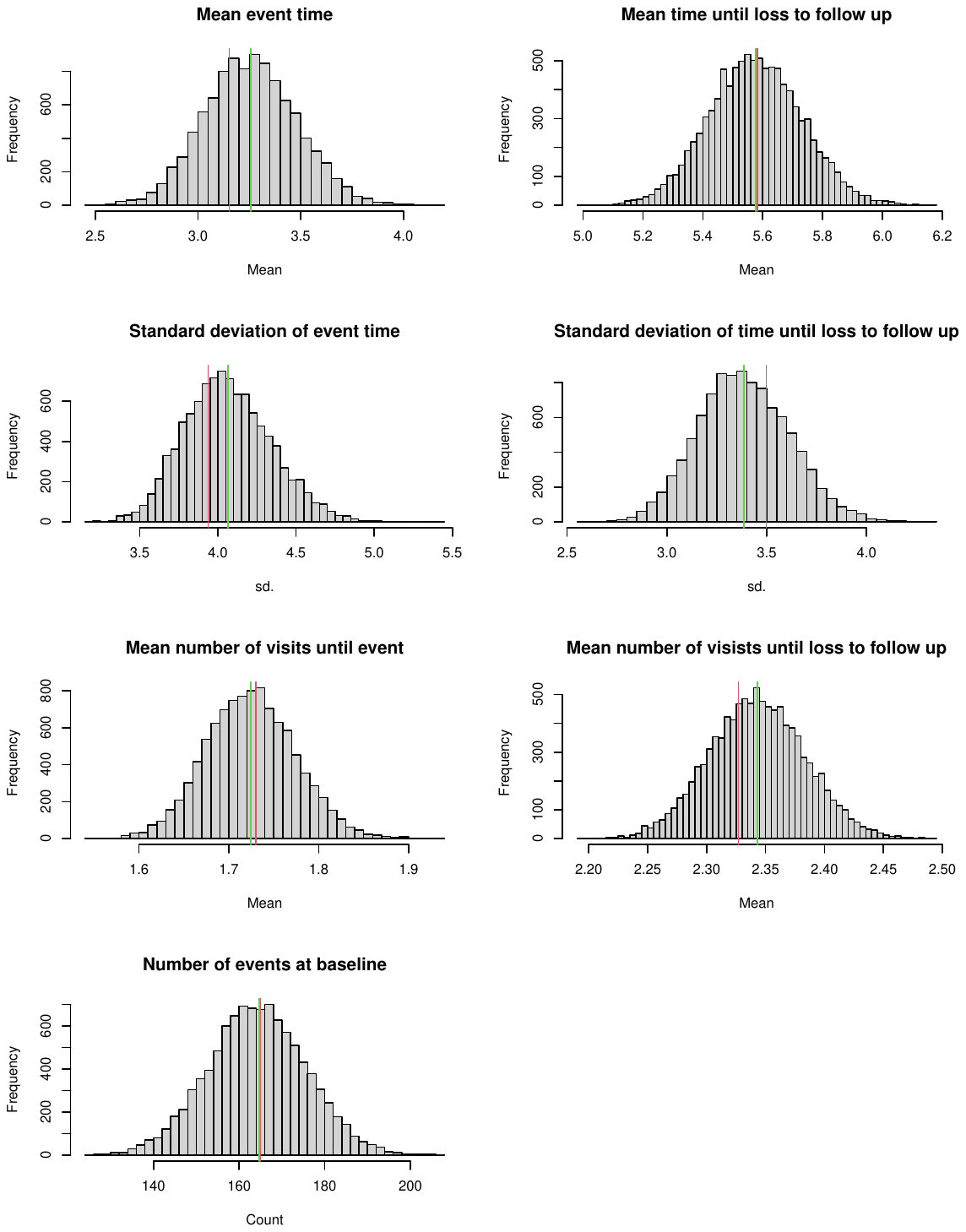}
	\caption{Sampling distributions (Simulation 2) of statistics calculated on 10,000 samples of simulated censoring times $\{\Vvec_k\}_{k=1}^n$ where $n_{sim} = 810$ (sample size of the CRC EHR data). Green shows the mean of the  sampling distribution and red shows the observed value calculated on the CRC EHR data. These comparisons demonstrate that the simulated screening times distribution was similar to the observed screening times distribution in the CRC EHR.}
	\label{fig:cens_dist_check}  
\end{figure}
\clearpage

\subsection{Generation of screening times under extended right censoring} \label{sec:sim2_extcens_gen}

In Simulation 2, we considered an extended right censoring distribution as an addition to the empirical approximation $\hat{F}_s(s)$ to the right censoring distribution in the CRC EHR data described in Section \ref{sec:sim2_obscens_gen}. The extended right censoring distribution was obtained through adding an offset to $s_i$, i.e.
\begin{align}
	s_i^{ext}:=s_i + \Delta,
\end{align}
where $\Delta =10$ years which shifted $\hat{F}_s(s_i)$ to the right. This procedure generally led to longer potential follow up and more screening moments for every $k$.\\ 

The process of generating screening times was similar to that described in Section \ref{sec:sim2_obscens_gen} with two differences. First, instead of sampling right censoring times from $\hat{F}_s(s)$ under constraint $s_k > v_{kc_{i'}}$, we now sample from the shifted (extended) distribution $\hat{F}^{ext}_s(s_i)$ under the constraint $s^{ext}_k > v_{kc_{i'}}$. This distribution was obtained with the same procedure as described in Section \ref{sec:sim2_obscens_gen} with the difference that the $(l_i,r_i]$ interval bounding $s_i$ was shifted $(l_i+\Delta,r_i+\Delta]$ before applying the \citet{turnbull_empirical_1976} estimator. Second, individuals $i$ experiencing an event ($y_{ic_i}=1$) acted as donors for $\Vvec_i$ with times and (extended) right censoring time augmented as described in Section \ref{sec:sim2_obscens_gen}. However, right censored individuals $i$ ($y_{ic_i}=0$) lacked visit times after their time of right censoring until the extended right censoring time $\Delta=10$ years later. Therefore, also their visit times were augmented until extended right censoring in the same way as described in  Section \ref{sec:sim2_obscens_gen}. Note that this was not necessary under standard right censoring when their last moment of follow up could naturally be used as the last moment before right censoring.

\clearpage
\newpage

\section{Additional results from Simulation 2 (CRC EHR)}

\begin{figure}[h]  
	\centering  
	\includegraphics[scale = 0.8, trim=130pt 180pt 140pt 190pt, clip, page = 1]{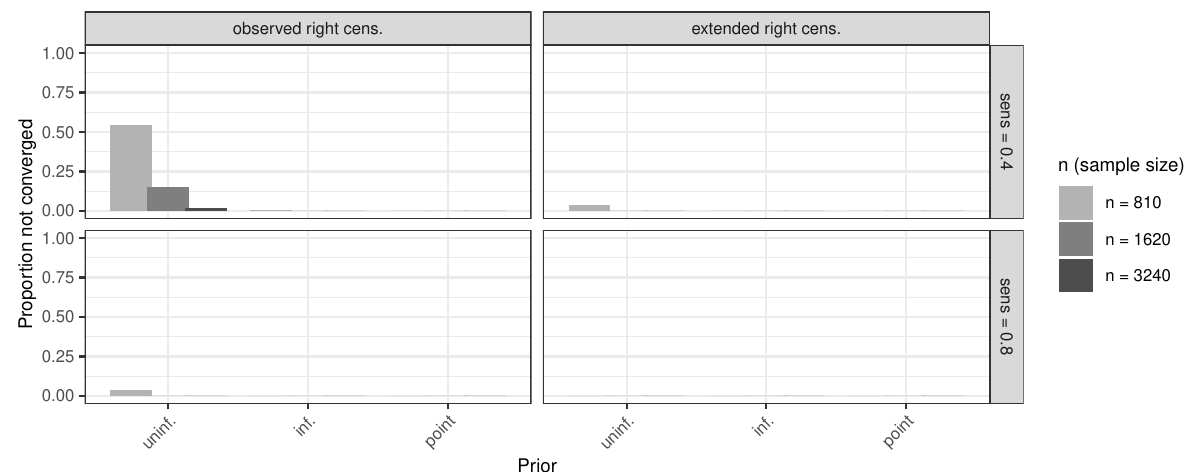}
	\caption{Proportion of runs that did not converge per simulation condition. Convergence was evaluated every $2 \times 10^4$ draws. Abbreviations prev and sens denote, respectively, the prevalence probability $\Pr(g_i=1)$ and the test sensitivity $\kappa$. The priors on the test sensitivity $\kappa$ are either uninformative (uninf.), informative (inf.) or fixed at the true value (point).} 
	\label{fig:convergence_ncrates_sim2}  
\end{figure}

\begin{figure}[h]  
	\centering  
	\includegraphics[scale = 0.8, trim=130pt 180pt 140pt 190pt, clip, page = 1]{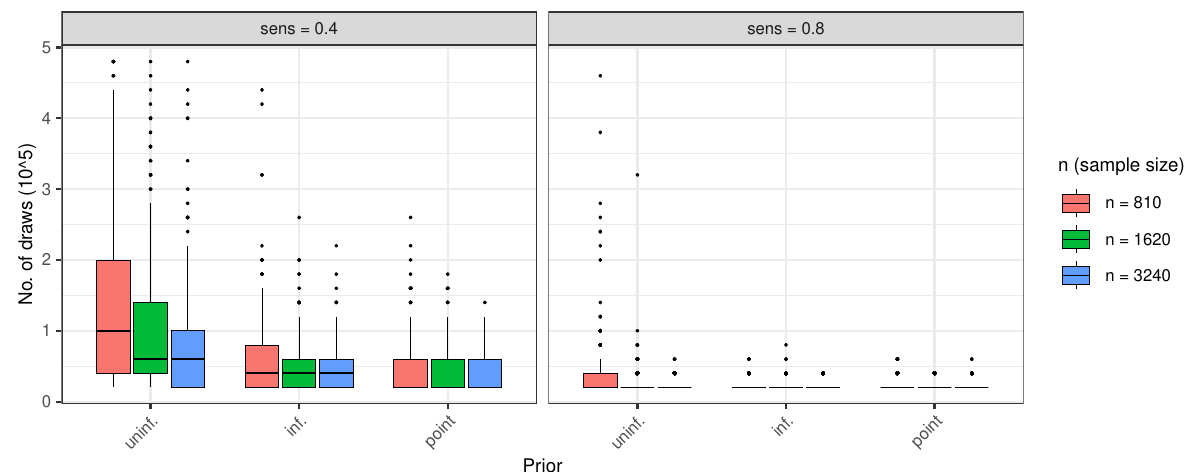}
	\caption{Number of posterior draws until convergence including burn-in (scaled by $10^5$) by simulation conditions. Convergence was evaluated every $2 \times 10^4$ draws. Abbreviation sens denotes the test sensitivity $\kappa$. The priors on the test sensitivity $\kappa$ are either uninformative (uninf.), informative (inf.) or fixed at the true value (point).} 
	\label{fig:convergence_boxplot_sim2}  
\end{figure}

\clearpage

\begin{figure}[h]  
	\centering  
	\includegraphics[scale = 0.6, trim=50pt 10pt 70pt 20pt, clip, page = 1]{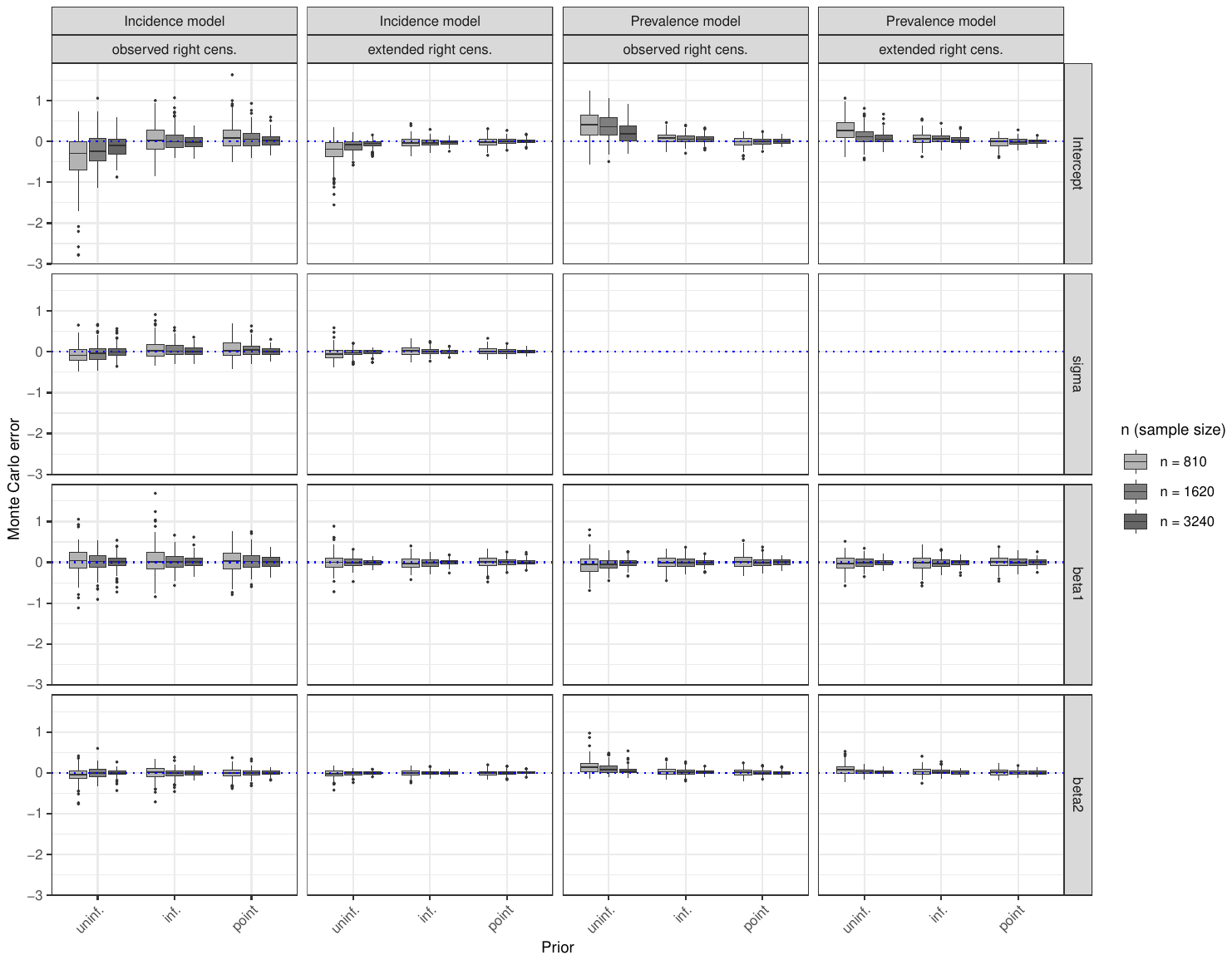}
	\caption{Monte Carlo error of the model parameters (Simulation 2), as indicated on the right, for both the incidence model (\ref{eq::mod_x}) and the prevalence model (\ref{eq::mod_g}) in the simulation condition: $\bm{\kappa=0.4}$. Note that there is no $\sigma$ parameter in the prevalence model and hence the corresponding panels are left blank. Row labels beta1 and beta2 refer to $\beta_1$ and $\beta_2$ in the incidence model and $\theta_1$ and $\theta_2$ in the prevalence model. Abbreviation sens denotes the test sensitivity $\kappa$. The priors on the test sensitivity $\kappa$ are either uninformative (uninf.), informative (inf.) or fixed at the true value (point).} 
	\label{fig:pars_boxplot1_sim2}  
\end{figure}
\clearpage

\begin{figure}[h]  
	\centering  
	\includegraphics[scale = 0.6, trim=50pt 10pt 70pt 20pt, clip, page = 2]{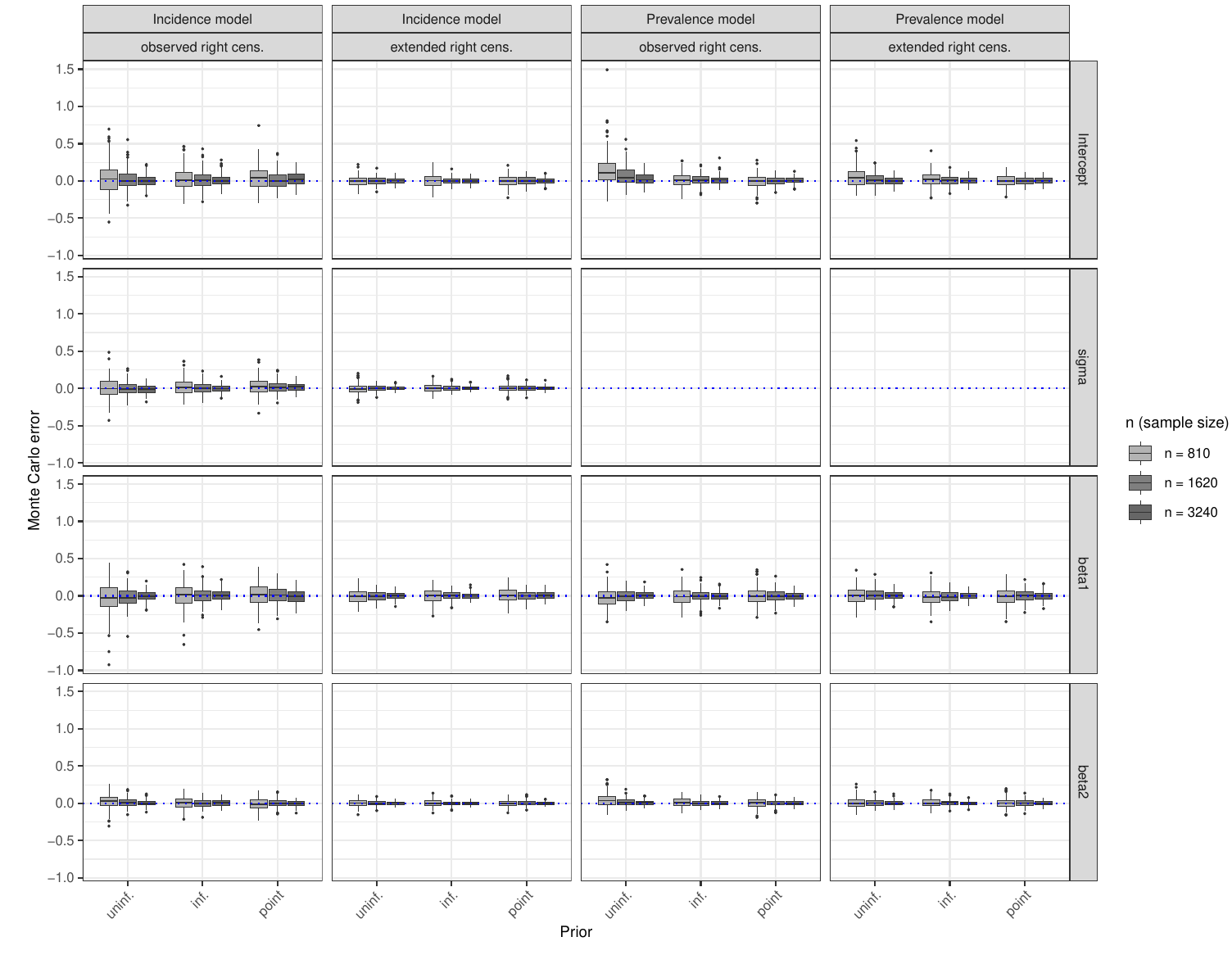}
	\caption{Monte Carlo error of the model parameters  (Simulation 2), as indicated on the right, for both the incidence model (\ref{eq::mod_x}) and the prevalence model (\ref{eq::mod_g}) in the simulation condition: $\bm{\kappa=0.8}$. Note that there is no $\sigma$ parameter in the prevalence model and hence the corresponding panels are left blank. Row labels beta1 and beta2 refer to $\beta_1$ and $\beta_2$ in the incidence model and $\theta_1$ and $\theta_2$ in the prevalence model. Abbreviation sens denotes the test sensitivity $\kappa$. The priors on the test sensitivity $\kappa$ are either uninformative (uninf.), informative (inf.) or fixed at the true value (point).} 
	\label{fig:pars_boxplot2_sim2}  
\end{figure}
\clearpage

\begin{figure}[h]  
	\centering  
	\includegraphics[width=\textwidth, trim=90pt 50pt 100pt 50pt, clip, page = 1]{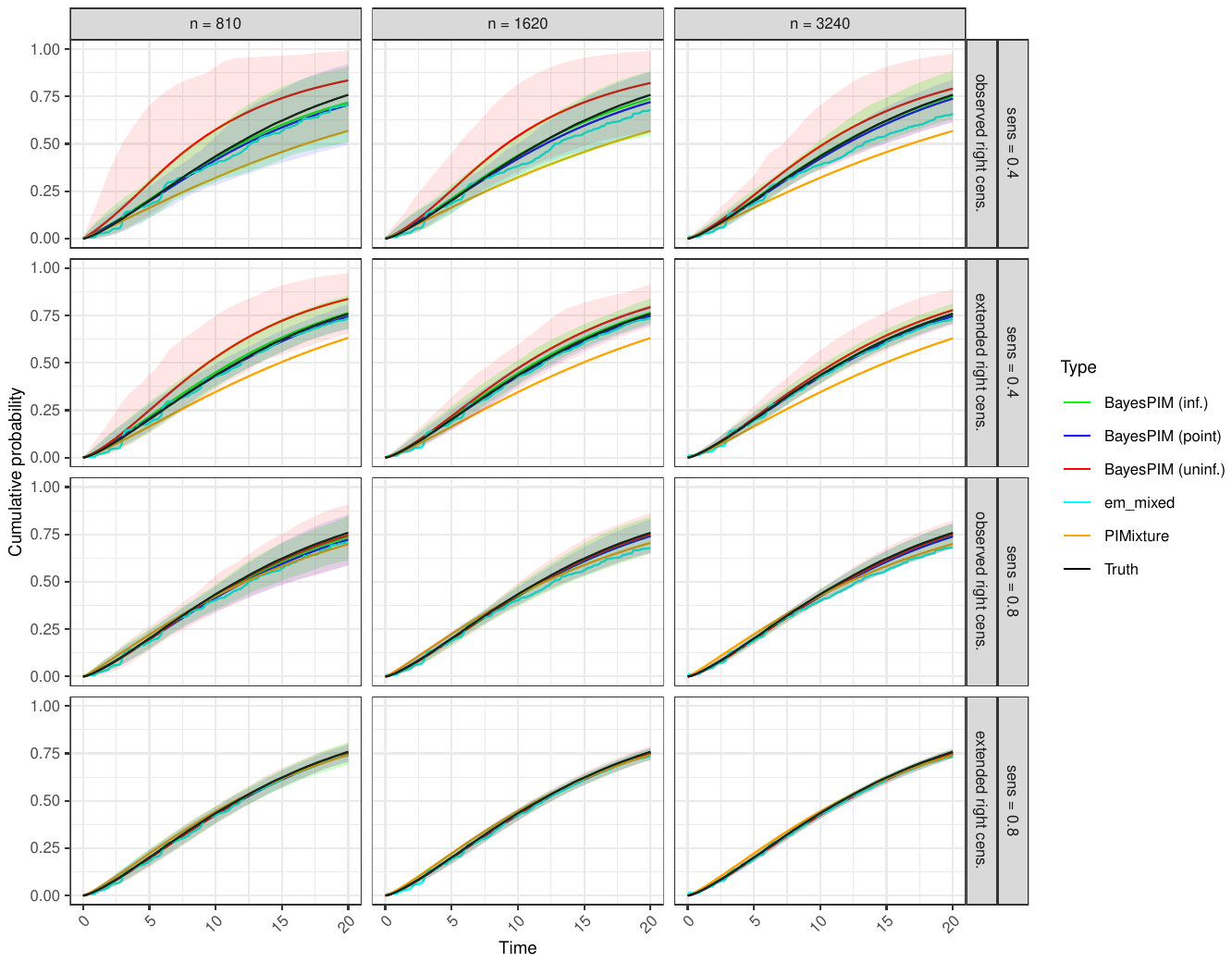}
	\caption{Marginal CIFs for the non-prevalent population, $F_t(t \mid g=0, \betaf, \sigma)$, point-wise averaged over 200 Monte Carlo simulation runs with 95\% quantiles shown as shaded regions. For \texttt{BayesPIM} the median of the posterior predictive CIF is used. For \texttt{Pimixture} and \texttt{em\_mixed} the corresponding maximum likelihood estimates are used, rescaled to the non-prevalent population.}
	\label{fig:sim2_cdfs_x}  
\end{figure}
\clearpage

\begin{figure}[h]  
	\centering  
	\includegraphics[width=\textwidth, trim=60pt 70pt 70pt 85pt, clip, page = 1]{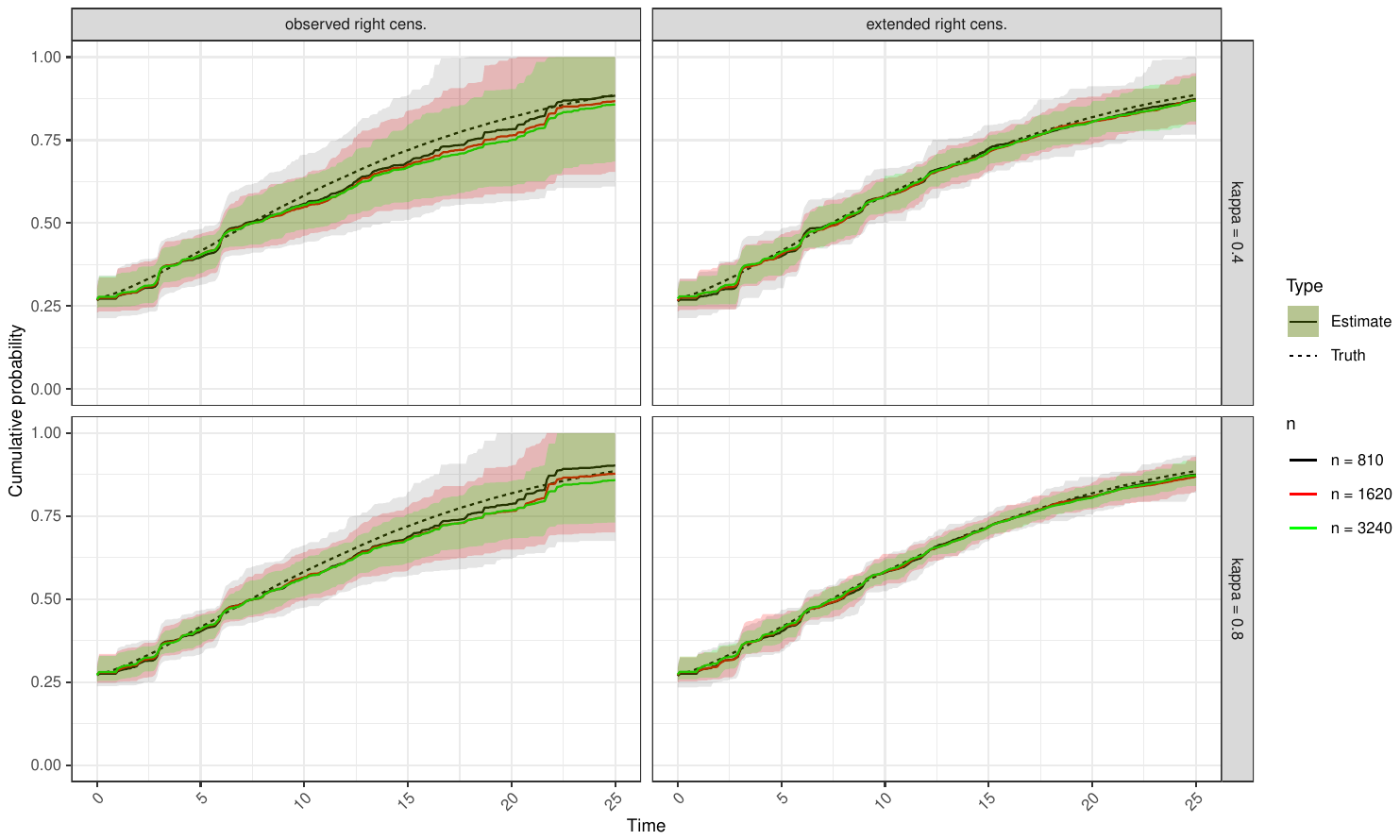}
	\caption{Marginal mixture CIFs, $F_{t^*}(t \mid \betaf, \sigma, \thetaf)$, estimated by \texttt{em\_mixed}, point-wise averaged over 200 Monte Carlo simulation runs with 95\% quantiles shown as shaded regions.}
	\label{fig:sim2_cdfs_emmixed}  
\end{figure}
\clearpage

\begin{figure}[h]  
	\centering  
	\includegraphics[scale = 0.8, trim=120 120pt 120pt 120pt, clip, page = 1]{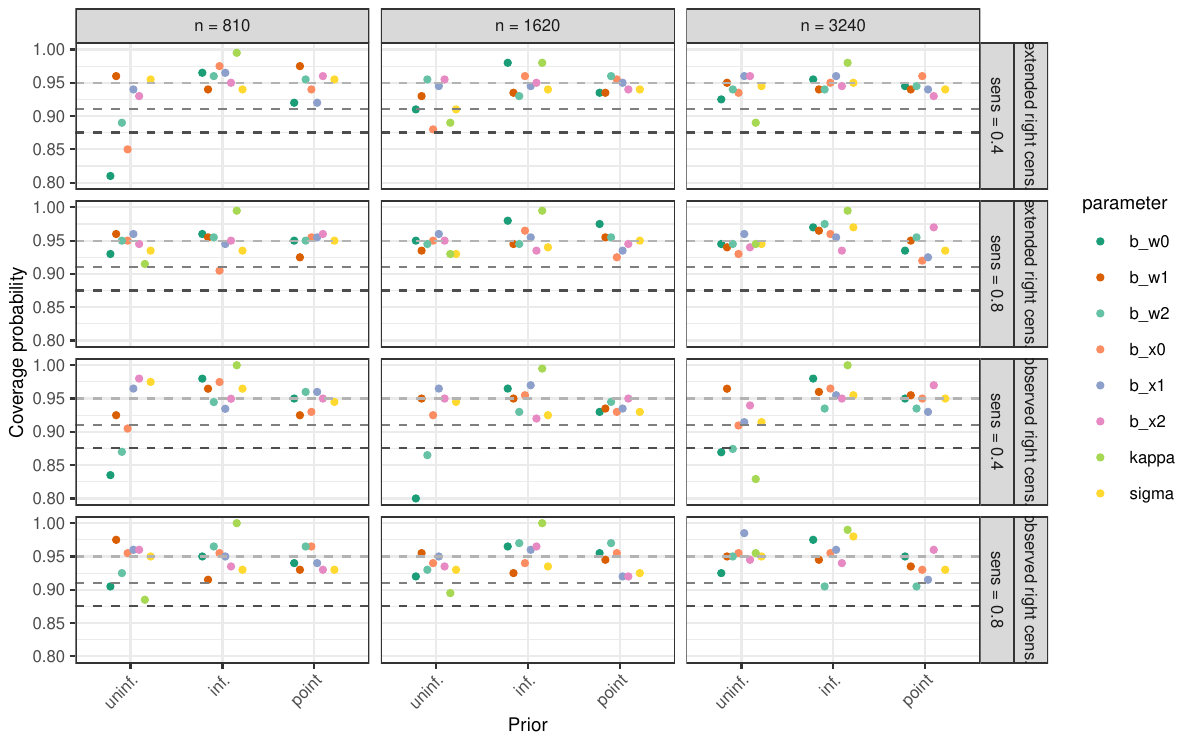}
	\caption{Frequentist coverage probability of the Bayesian 95\% posterior credible intervals for the 36 simulation conditions in Simulation 2 (estimated from 200 Monte Carlo data sets per condition by the proportion of intervals covering the true parameter value). The gray dotted lines in each panel denote, from top to bottom: (a) the nominal 95\% level, (b) the value of a point estimate whose 95\% confidence upper bound is equal to 95\%, (c) the value of a point estimate whose 95\% confidence upper bound is equal to 95\% with a Bonferroni adjustment for 36 repeated tests.}
	\label{fig:sim2_coverage}  
\end{figure}

\clearpage
\newpage
\section{Additional results from the CRC application} \label{sec:supplement_application}

\begin{figure}[h]  
	\centering  
	\includegraphics[scale=0.65, trim=140pt 160pt 140pt 160pt, clip, page = 4]{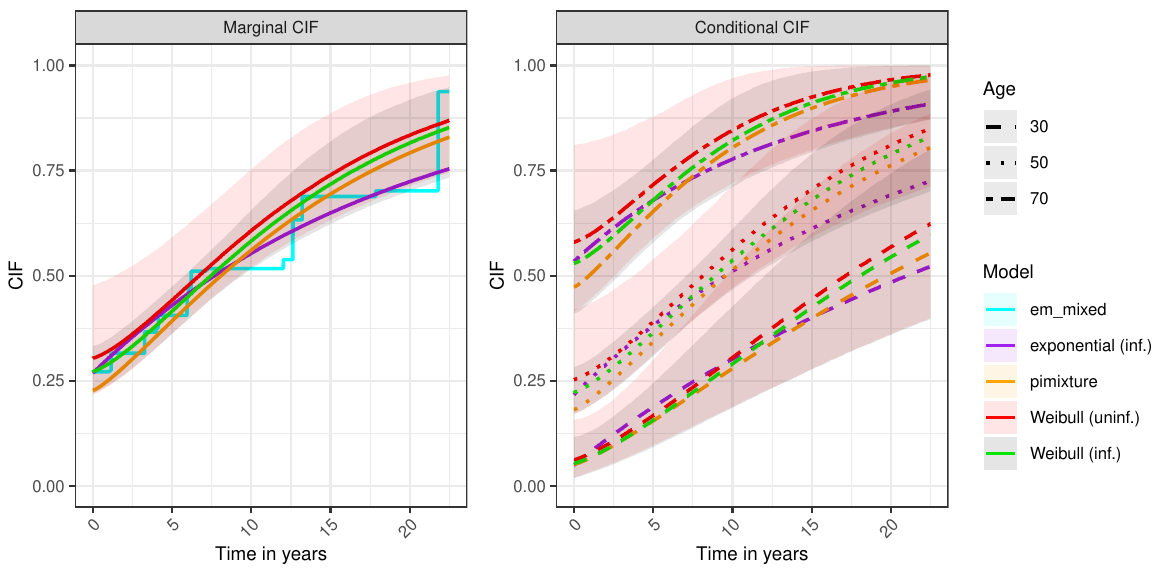}
	\caption{Marginal and conditional mixture cumulative incidence functions (CIF), $F_{t^*}(t \mid \betaf, \sigma, \thetaf)$ and $F_{t^*}(t \mid \tilde{\x}, \betaf, \sigma, \thetaf)$, for Weibull \texttt{BayesPIM} with uninformative (uninf.) prior and informative (inf.) priors. For comparison, CIF from the Weibull \texttt{PIMixture} model, the exponential \texttt{BayesPIM} model (inf.), and \texttt{em\_mixed} are given. The lines represent posterior median estimates and the shaded regions indicates the 95\% credible interval of the Weibull (inf. in gray) and Weibull (uninf. in red) models. }  
	\label{fig:CIF_margcond_complete}  
\end{figure}

\begin{figure}[h]  
	\centering  
	\includegraphics[scale=0.65, trim=140pt 160pt 140pt 160pt, clip, page = 6]{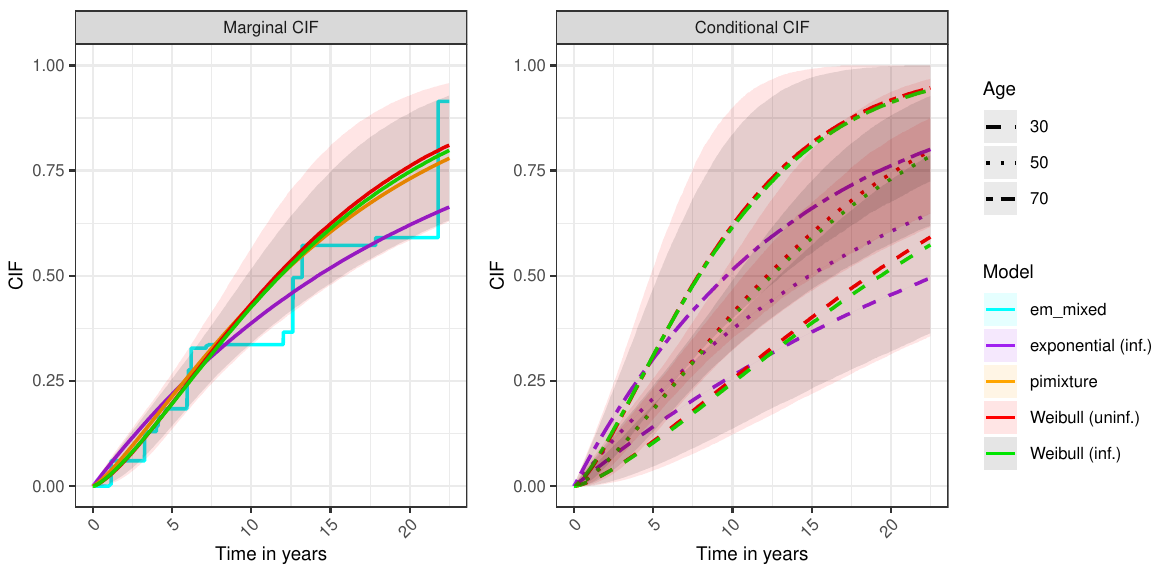}
	\caption{Marginal and conditional cumulative incidence functions (CIF) for the non-prevalent population, $F_{t}(t \mid g=0, \betaf, \sigma)$ and $F_{t}(t \mid g=0, \tilde{\x}, \betaf, \sigma, \thetaf)$, for Weibull \texttt{BayesPIM} with uninformative (uninf.) prior and informative (inf.) priors. For comparison, CIF from the Weibull \texttt{PIMixture} model, the exponential \texttt{BayesPIM} model (inf.), and \texttt{em\_mixed} are given. The lines represent posterior median estimates and the shaded regions indicates the 95\% credible interval of the Weibull (inf. in gray) and Weibull (uninf. in red) models. }  
	\label{fig:CIF_margcond_healthy}
\end{figure}

\end{document}